\documentclass[12pt,a4paper]{article}
\usepackage{authblk}
\usepackage{indentfirst}
\usepackage{graphicx}
\usepackage{epsfig}
\usepackage{bm}
\usepackage{latexsym}
\usepackage{amsmath}
\usepackage{amssymb}
\usepackage{amsfonts}
\usepackage{mathrsfs}
\usepackage{pifont}

\usepackage{amsthm}
\usepackage{float}
\usepackage{subfigure}
\usepackage{color}
\usepackage{multicol}
 \usepackage[flushleft]{threeparttable} 
\usepackage[colorlinks,linkcolor=blue,filecolor=black,citecolor=blue]{hyperref}

\newtheorem{theorem}{Theorem}[section]

 \newtheorem{remark}[theorem]{Remark}

\usepackage{ulem}
\title{Optic Nerve Microcirculation: Fluid Flow and Electrodiffusion
\author[1]{Yi Zhu}
\author[2]{Shixin Xu\thanks{Corresponding author: shixin.xu@dukekunshan.edu.cn}}
\author[3]{Robert.S. Eisenberg}
\author[5,1,6]{Huaxiong Huang}

\affil[1]{\small Department of Mathematics and Statistics, York University, Toronto, Ontario, Canada.} 
\affil[2]{\small Duke Kunshan University, 8 Duke Ave, Kunshan, Jiangsu, China.}
\affil[3]{\small  Department of Applied Mathematics, Illinois Institute of Technology, Chicago IL 60616 USA.}
\affil[4]{\small Computer Science, University of Toronto, Toronto, Ontario, Canada.}
\affil[5]{\small Joint Mathematical Research Centre of Beijing Normal University and BNU-HKBU United International College, Zhuhai, China}
\affil[6]{\small Division of Science and Technology, BNU- HKBU United International College, Zhuhai, 519087,	China}}

\date{}
\begin{document}
\maketitle

\section*{Abstract}
Complex fluids flow in complex ways in complex structures. 
Transport of water and various organic and inorganic molecules in the central nervous system are important in a wide range of biological and medical processes [C. Nicholson, and S. Hrabětová,  Biophysical Journal, 113(10), 2133(2017)]. However, the exact driving mechanisms are often not known. In this paper, we investigate flows induced by action potentials in an optic nerve as a prototype of the central nervous system (CNS). Different from traditional fluid dynamics problems, flows in biological tissues such as the CNS are coupled with ion transport. It is driven by osmosis created by concentration gradient of ionic solutions, which in term influence the transport of ions. Our mathematical model is based on the known structural and biophysical properties of the experimental system used by the Harvard group Orkand et al [R.K. Orkand, J.G. Nicholls, S.W. Kuffler, Journal of Neurophysiology, 29(4), 788(1966)]. Asymptotic analysis and numerical computation show the significant role of water in convective ion transport. The full model (including water) and the electrodiffusion model (excluding water) are compared in detail to reveal an interesting interplay between water and ion transport. In the full model, convection due to water flow dominates inside the glial domain. This water flow in the glia contributes significantly to the spatial buffering of potassium in the extracellular space. Convection in the extracellular domain does not contribute significantly to spatial buffering. Electrodiffusion is the dominant mechanism for flows confined to the extracellular domain.

\section{\label{sec:Introduction}	Introduction}

The theory of complex fluids deals with complex fluids in complex structures \cite{gelbart1996new,krishnan2010rheology,spagnolie2015complex,fuller2012complex}. Here we deal with the complex fluid of an ionic solution \cite{eisenberg2010energy}  in a complex structure typical of biological systems in particular the central nervous system. These structures are known in some detail—both structure and function—because of the work of generations of neuroanatomists, histologists and neurobiologists \cite{kandel2000principles, nicholls2001neuron}. The biophysical properties of membranes are also well known \cite{boron2016medical}. 
So we can formulate a biologically significant problem in the language of theory of complex fluids and use the methods of computational fluid mechanics to analyze the system, here the optic nerve of an amphibian. The results are of interest biologically because of the importance of the central nervous system: the optic nerve of amphibian is an experimentally accessible part of the central nervous system.

The analysis used here may also serve as a bridge, and archetype, of how the theory of complex fluids can deal with what at first may seem formidable challenges of structured biological systems in other biological systems, e.g., kidney, blood brain barrier, and epithelial in general.

The rest of the paper is organized as follows. In Section \ref{background and model}, we present the biological background about the optic nerve and the tridomain mathematical model in detail. 
The three domains, axon, glial and extracellular ones,  are coupled via transmembrane fluxes for three major ions, namely sodium, potassium and chloride, treated as reaction terms. Model calibration is discussed in Section \ref{Model_Calibration} by matching extracellular potassium concentration accumulation after the optic nerve is stimulated by a train of electric current pulses. In Section \ref{Effects_of_Water_Flow}, we present estimates using order of magnitude analysis of transport of ionic and water fluxes cross membranes. They provide useful insight into the mechanisms for potassium clearance.   Then  in Section \ref{sec: Numerical simulation},  numerical simulations are carried out. We investigate the role of water flow (convection) in ionic transport during and after stimulus of the optic nerve.
Our analysis shows that convection is very important within the glia. 
Water flow in glia has an indirect but significant effect in clearing potassium from the narrow extracellular space. This may be an important role for glia wherever they are found in the central nervous system, and even in structures of the peripheral nervous system. A discussion on the parameters in the compartment models and field models are presented in Section \ref{Discussion}.
In Section \ref{Conclusion}, we provide concluding remarks on the limitation of our study and directions for future research.

\section{Biological Background and Model}\label{background and model}
\subsection{Biological Background}
Recent experimental studies \cite{nedergaard2020glymphatic} suggest that transport in the central nervous system during sleep plays a critical role in maintaining the health of brain tissue. Since the nervous system is densely packed with neurons communicating with each other, question arises:  how is the state of steady internal conditions—known as ``homeostasis'' in the biological literature—maintained.
A few action potentials are known to significantly alter ion concentration in the immediate vicinity of peripheral and optic nerve cells \cite{orkand1966effect,frankenhaeuser1956after} and that change in concentration acts on more than one axon, producing “cross talk”. The question is then how does the central nervous system deal with changes in ion concentration produced by hundreds or thousands of action potentials and maintain a healthy environment? 
How does the central nervous system maintain concentrations in its narrow extracellular space? What are the roles played by of glial cells and extracellular space?

Complex flows in complex structures cannot be understood unless the structure is understood. 
The central nervous system contains nerve fibers and glia, separated by a narrow extracellular space.
We use three domains to describe the flow and diffusion of ions and water in the optic nerve bundle of the central nervous system, hoping to glimpse general properties by which the central nervous system controls the concentration of ions in such narrow confines. 
The optic nerve bundle contains paired cranial nerve bundled with cell bodies in the retina. It reaches from the eye through the optic chiasma to the cortex and transfers visual information from the retina to the vision centers of the brain using digital (actually binary) electrical signals (action potentials).
The optic nerve is customarily separated into four main regions \cite{salazar2018anatomy,selhorst2009optic}: (1) intraocular nerve head, (2) intraorbital region, (3) intracanalicular and (4) intracranial \cite{salazar2018anatomy,hayreh2009ischemic}. In this paper, we mainly focus on the intraorbital region, which occupies more than half of the optic nerve.

There are about one million optic nerve fibers in the optic nerve bundle. The ganglion cells that are the cell bodies of the axons are scattered on the retina and form into a bundle at the optic disc. The bundle passes through the mesh-like lamina cribrosa region into the intraorbital region. 
Like almost all nerve cells, optic nerve fibers are functionally isolated, nearly insulated one from another , without connexins between them, so neither ions nor electrolytes can flow directly from the interior of one nerve cell to another. Current flow down one axon cannot flow into the adjacent axon or glia \cite{atchison2000optics,kuffler1966physiological}. The ‘ephaptic communication’ of concern to pioneers in electrophysiology rare occurs. 

Glial cells wrap the nerve fiber bundles producing a narrow cleft of extracellular space between nerve fiber and glia. 
Glial cells are connected to each other through connexin proteins, called `gap junctions’,
and form an electrical syncytium (as do so many other cells, e.g., epithelia, cardiac muscle, lens of the eye, liver, etc.) in which current flow in one cell spreads into another with little extra resistance. In syncytia like this, inorganic ions, and many organic molecules (typically less than 2 nanometer diameter) can diffuse from cell to cell with hardly any restriction and thus with mobility and ionic conductance similar to that in cytoplasm. 
Thus, glial cells are thought to play an important role in accelerating $\mathrm{K^+}$ clearance from the extracellular space \cite{bellot2017astrocytic,wallraff2006impact}. Sometimes, central retinal blood vessels (CRV, arterioles in fact) are found in the center of the optic nerve bundle in the intraorbital region. Here we consider the case where the blood vessel is not present, as in the optic nerve of the mud puppy, the amphibian salamander Necturus used in the experiments of Orkand et al. \cite{orkand1966effect,kuffler1966physiological}.


\begin{figure}[hpt]
	\centering
	\includegraphics[bb=50 55 420 380, clip=true,width=3.25in,height=6.5cm ]{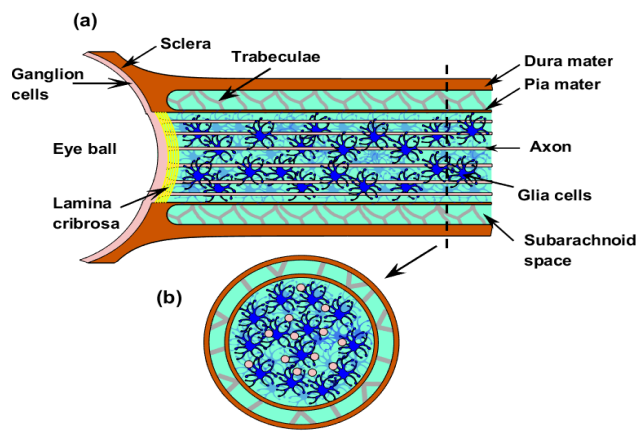}
	\caption{Optic nerve structure. (a) Longitudinal section of the optic nerve; (b) Cross section of the optic nerve. \label{fig:schmatic}}
\end{figure}

The optic nerve bundles are surrounded by the meningeal sheath which consists of dura mater, arachnoid mater and pia mater, and cerebrospinal fluid (CSF) in the subarachnoid space (SAS) \cite{hayreh2009ischemic,hayreh1984sheath}. Also see Fig. \ref{fig:schmatic}a. The pia mater and dura mater are thin deformable shells, with mechanical properties important in glaucoma \cite{hayreh1984sheath,killer2003architecture,pache2006morphological,hua2018cerebrospinal}. Andrew et. al \cite{andres1987nerve} and Killer et. al  \cite{killer2003architecture,killer1999lymphatic}show that the dura mater contains lymphatic vessels that drain CSF out of SAS \cite{hua2018cerebrospinal,morgan2016cerebrospinal}. Pia mater forms a macroscopic semipermeable membrane made of many cells, not just one lipid bilayer \cite{filippidis2012permeability}. Many layered epithelia have been characterized as ``semipermeable membranes'' in low resolution studies of epithelia for more than a century. Filipidis et. al. \cite{vogiatzidis2006mu} have written a most helpful review that identifies analogous leptomeningeal structures important in the physiology of ``like pleura \cite{hatzoglou2001effects,payne1988comparative,sarkos2002effect,zarogiannis2007comparison,zarogiannis2007adrenergic,zarogiannis2009dexamethasone}, peritoneum \cite{li2001electrophysiology,simon1996peritoneal,stefanidis2007amiloride,zarogiannis2005influence,zarogiannis2007effect,zarogiannis2007mu}, pericardium \cite{vogiatzidis2006mu}, fetal membranes  \cite{verikouki2008rapid,adams2005comparison}, and leptomeninges \cite{filippidis2010transmembrane},''  We imagine that a general tridomain model may help understand many of these tissues.

\subsection{Mathematical Model\label{Math_model}}

The model is first proposed in Ref.~\cite{zhu2020tridomain}. Here in order to make  this paper self-contained, we summarize the model.  The model deals with two types of flow: the circulation of water (hydrodynamics) and the circulation of ions (electrodynamics) in the glial compartment $\Omega_{gl}$, axon compartment $\Omega_{ax}$ and extracellular space $\Omega_{ex}$.

\begin{figure}[hpt]
	\centering
	\includegraphics[bb=0 0 920 430, clip=true,width=3.25in,height=4cm ]{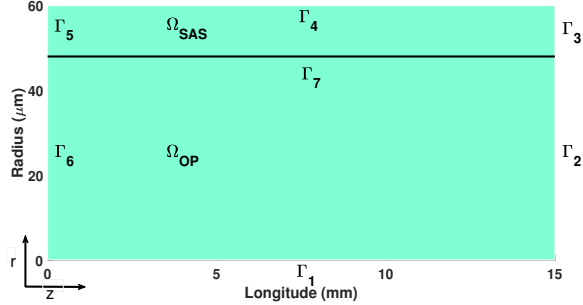}
	\caption{Domain of axial symmetry model. The optic nerve  $\Omega_{OP}$ consist of axon compartment $\Omega_{ax}$, glial compartment $\Omega_{gl}$ and extracellular space $\Omega_{ex}^{OP}$.  The subarachnoid space $\Omega_{SAS}$ only has extracellular space. \label{fig:model_domain}}
\end{figure}

The glial compartment and axon compartment are limited to the optic nerve bundle, while extracellular space exists both in the optic nerve bundle $\Omega_{ex}^{OP}$ and in the  subarachnoid space $\Omega_{ex}^{SAS}$, (See Fig. \ref{fig:model_domain})

\begin{equation*}
\Omega_{OP}=\Omega_{ax} \cup \Omega_{gl} \cup \Omega_{ex}^{OP}, \quad \Omega_{SAS}=\Omega_{ex}^{SAS}.
\end{equation*}
The model is mainly based on the law of mass conservation \cite{nicholson2001diffusion}, in $\Omega_l,~l = ax, ~gl, ~ex$
\begin{equation}
\frac{\partial }{\partial t}(\eta_l f_l) +\nabla\cdot (\eta_l \mathbf{J}_l) + S= 0,
\end{equation}
where $\eta_l$ is the volume fraction of $l$ compartment, $f_l$ is the concentration of  given substance, $ \mathbf{J}_l$ is the flux inside compartment, and $S$ is the source term induced by the pumps and channels on the membranes. 

We first introduce the following notations used in the paper, where $i=\mathrm{Na^+,K^+,Cl^-}$ for ion species, $l=ex,gl,ax$ for extracellular space, glial compartment and axon compartment, and $k=gl,ax$ for glial or axon membrane in the optic nerve. The summary of notations is listed in Appendix A1.

In each domain, we assume that electroneutrality such that
\begin{subequations}
	\label{Charge_nuetrality}
	\begin{align}
	\eta_{gl} \sum_{i} z^{i} c_{gl}^{i}+z^{gl} \eta_{gl}^{re}A_{gl}   &=0, \\
	\eta_{ax} \sum_{i} z^{i} c_{ax}^{i}+z^{ax}  \eta_{ax}^{re}A_{ax}  &=0, \\
	\sum_{i} z^{i} c_{ex}^{i} &=0,
	\end{align}
\end{subequations}
where $A_{l}>0$ with $l=ax,gl$ is the density of proteins in axons or glial cells with valence $z^{l}$, $l=gl,ax$. The $\eta_{ax}$ and $\eta_{gl}$ are the volume fraction of axon and glial compartments in the optic nerve and $\eta_{ax}^{re}$ and $\eta_{gl}^{re}$ are the resting state volume fractions.

\subsubsection{Water Circulation\label{Water_circulation}}
The conservation of mass in each domain yields  

\begin{subequations}\label{water_eq} 
	\begin{align}
	\frac{\partial \eta_{gl}}{\partial t}+\mathcal{M}_{gl} U^{m}_{gl}+\nabla \cdot\left(\eta_{gl} \mathbf{u}_{gl}\right) =0, ~\text {in} ~ \Omega_{OP}, \label{water_gl} \\
	\frac{\partial \eta_{ax}}{\partial t}+\mathcal{M}_{ax}U^{m}_{ax}+\frac{\partial}{\partial z}\left(\eta_{ax} u_{a x}^{z}\right) =0, ~\text {in} ~\Omega_{OP},  \label{water_ax} \\
	\nabla \cdot\left(\eta_{gl} \mathbf{u}_{gl}\right)+\nabla \cdot\left(\eta_{ex} \mathbf{u}_{e x}\right)+\frac{\partial}{\partial z}\left(\eta_{ax} u_{ax}^{z}\right)=0, ~\text {in}~ \Omega_{OP}, \label{water_ex}  \\  
	\eta_{gl}+\eta_{ax}+\eta_{ex}=1, ~\text{in}~ \Omega,  
	\end{align}
\end{subequations}
where the transmembrane water flux is proportional to the intracellular/extracellular hydrostatic pressure and osmotic pressure differences, i.e., Starling’s law on the membrane, 
\begin{equation*}
\begin{aligned}
U^{m}_{gl}&= L_{g l}^{m}\left(p_{gl}-p_{ex}-\gamma_{gl} k_{B} T\left(O_{gl}-O_{e x}\right)\right),\\
U^{m}_{ax}&= L_{ax}^{m}\left(p_{ax}-p_{ex}-\gamma_{ax} k_{B} T\left(O_{ax}-O_{e x}\right)\right).
\end{aligned}
\end{equation*}

The glial cells are connected to each other by connexins and form a syncytium; While the axons are separate, more or less parallel cylindrical cells that do not form a syncytium. (See Fig. \ref{fig:schmatic}) Then we 
assume that glial cells are isotropic and  axons are anisotropic. Here $\mathbf{u}_{l}$ and $p_{l}$ with $l=gl,ax,ex$ are the velocity and pressure in the glial cells and axons and extracellular space, respectively. And $k_{B}TO_{l}$, is the osmotic pressure \cite{xu2018osmosis,zhu2019bidomain} defined by 

\begin{equation*}
O_{ex}=\sum_{i} c_{ex}^{i}, \quad O_{l}=\sum_{i} c_{l}^{i}+ A_{l}\frac{\eta_{j}^{re}}{\eta_{l}}, \quad l=gl, ax,
\end{equation*}
where $A_{l}\frac{\eta_{l}^{re}}{\eta_{l}} >0 \  (l=gl,ax)$ is the density of the permanent negatively charged protein in glial cell and axons that varies with the volume (fraction) of the region. 

The relation between the hydrostatic pressure $p_{l}$ and volume fraction $\eta_{l}$  $(l=ex,gl,ax)$ is connected by the force balance on the membrane $k (=gl,ax)$ \cite{mori2015multidomain,xu2018osmosis}.   

\begin{subequations}
	\begin{align}
	K_{gl}\left(\eta_{gl}-\eta_{gl}^{re}\right) &=p_{gl}-p_{ex}-\left(p_{gl}^{re}-p_{ex}^{re}\right), \text { in }\Omega_{OP}, \label{hydro_relation} \\
	K_{ax}\left(\eta_{ax}-\eta_{ax}^{re}\right) &=p_{ax}-p_{ex}-\left(p_{ax}^{re}-p_{e x}^{re}\right), \text { in } \Omega_{OP}, \label{hydro_relation_ax}   
	\end{align}
\end{subequations}
where $K_{k}\ (k=gl,ax)$ is the stiffness constant related to Young's modules and Poisson's ratio. The $p_{l}^{re} \ (l=gl,ax,ex)$ is the resting state  hydrostatic pressure.

\begin{remark}
	\normalfont
	If we introduce the characteristic velocities $u^*_l$ in $l$ compartment, the characteristic transmembrane velocity $U^*_l$,  the characteristic time $t^*$, the characteristic lengths $r^*$ in radius direction and $z^*$ in longitude direction,  Eqs. (\ref{water_gl}),  (\ref{water_ax}) and (\ref{water_ex}) could be written as 
	\begin{subequations}
		\begin{align}
		\frac{\partial
			\eta_{gl}}{\partial \tilde{t}}+ \delta_{1}\tilde{U}^{m}_{gl}+  \delta_{2}\tilde{\nabla} \cdot \left(\eta_{gl}\tilde{\mathbf{u}}_{gl}\right)=0, &\label{fluid_nd_gl}\\
		\frac{\partial
			\eta_{ax}}{\partial \tilde{t}}+ \delta_{3} \tilde{U}^{m}_{ax}+\delta_{4} \frac{\partial \left(\eta_{ax} \tilde{u}^{z}_{ax}\right)}{\partial \tilde{z}}=0,  &\label{fluid_nd_ax}\\
		\tilde{\nabla} \cdot \left(\eta_{ex}\tilde{\mathbf{u}}_{ex}\right)+\delta_{5}\tilde{\nabla} \cdot \left(\eta_{gl}\tilde{\mathbf{u}}_{gl}\right) +\delta_{6}\delta_{0}\frac{\partial \left(\eta_{ax}\tilde{u}^{z}_{ax}\right)}{\partial \tilde{z}}=0,&  
		\end{align}\end{subequations}
	where 
	\begin{eqnarray*}
		&&\tilde{\nabla} \cdot \left(\eta_{l}\tilde{\mathbf{u}}_{l}\right)= \frac{1}{\tilde{r}}\frac{\partial \left( \tilde{r}\eta_{l} \tilde{u}^{r}_{l}\right)}{\partial \tilde{r}} + \delta_{0}\frac{\partial \left(\eta_{l} \tilde{u}^{z}_{l}\right)}{\partial \tilde{z}}, \ \ l=gl,ex,  
	\end{eqnarray*}
	and
	\begin{eqnarray*}
		&& \delta_0 = \frac{r^*}{z^*}, \delta_{1} =  \mathcal{M}_{gl}U^{*}_{gl} t^*,  \ \  \delta_{2}=\frac{u^{*}_{gl}t^*}{r^*},   \ \  \delta_{3}= \mathcal{M}_{ax}U^{*}_{ax} t^*, \\
		&& \delta_{4}=\frac{u^{*}_{ax}t^*}{z^*},  \ \   \delta_{5}= \frac{u^{*}_{gl}}{u^{*}_{ex}},  \ \  \delta_{6}=\frac{u^{*}_{ax}}{u^{*}_{ex}}.
	\end{eqnarray*}
	Further scaling can be applied for velocity components in the r and z directions when the cross membrane flux is absent due to incompressibility. However, no such scaling is considered due to significant cross membrane flux.
\end{remark}

The water flows in glial, axon compartments and extracellular space are low Reynold number flows and the characteristic velocity is around $1\sim 10\ \mathrm{nm/s}$ due to the existence of connexin and high  tortuosity.
\textcolor{black}{Then the stationary Stokes equation is used}  
\textcolor{black}{\begin{equation*}
	-\nabla\cdot(\mu\nabla\bm{u}_l)+\nabla p_l =f_l,
	\end{equation*}
	where $f_l$ is the body force density in different compartments, for example, Lorentz force in the extracellular space \cite{xu2014energetic}. }
Next, \textcolor{black}{since the tissues have similar property as the porous media,  The rigorous homogenization theories \cite{allaire2010homogenization,ray2012multiscale} or the control volume average methods \cite{malcolm2006computational, benedek2000physics} yield Darcy's Law is a good macro-scale approximation for the Stokes flow in the porous media. }
For the sake of simplicity, we model flows in the following as porous media flows by using Darcy's Law \cite{mori2015multidomain,zhu2019bidomain}.    

\textbf{Fluid Velocity in the Glial Compartment.} As we mentioned before, the glial space is a connected space, where water can flow from cell to cell through connexin proteins joining membranes of neighboring cells. 

The velocity of fluid in glial syncytium $\mathbf{u}_{gl}$ depends on the gradients of hydrostatic pressure and osmotic pressure:  

\begin{subequations} \label{def_u_gl}
	\begin{align}
	u_{gl}^{r}=-\frac{\kappa_{gl} \tau_{g l}}{\mu}\left(\frac{\partial p_{gl}}{\partial r}-\gamma_{gl} k_{B} T \frac{\partial O_{gl}}{\partial r}\right), \\
	u_{gl}^{z} =-\frac{\kappa_{gl} \tau_{g l}}{\mu}\left(\frac{\partial p_{gl}}{\partial z}-\gamma_{gl} k_{B} T \frac{\partial O_{gl}}{\partial z}\right).   
	\end{align}
\end{subequations}

The boundary conditions of fluid in the glial syncytium are as follows
\begin{equation}
\left\{
\begin{aligned}
&\mathbf{u}_{gl} \cdot \hat{\mathbf{r}}=0, & \text { on } \Gamma_{1}, \\
&\nabla p_{gl} \cdot \hat{\mathbf{z}}=0, & \text { on } \Gamma_{2}, \\
&\nabla p_{gl} \cdot \hat{\mathbf{z}}=0, & \text { on } \Gamma_{6}, \\
&\mathbf{u}_{gl} \cdot \hat{\mathbf{r}}=0, & \text { on } \Gamma_{7}.
\end{aligned}
\right.
\end{equation}

\textbf{Fluid Velocity in the Axon Compartment}. Since the axons are only connected in the longitudinal direction and the fluid velocity in axons region is defined along $z$ direction as 
\begin{subequations}\label{def_u_ax}
	\begin{align}
	u_{ax}^{r} =0, \\
	u_{ax}^{z} =-\frac{\kappa_{a x}}{\mu} \frac{\partial p_{a x}}{\partial z}.    
	\end{align}
\end{subequations}

Dirichlet boundary conditions are used to the fluid velocity in axons 
\begin{equation}
\nabla p_{ax} \cdot \hat{\mathbf{z}}=0,   \quad \text { on } \Gamma_{2} \cup  \Gamma_{6}.
\end{equation}

\textbf{Fluid  Velocity in the Extracellular Space.} The extracellular space is narrow,  and the extracellular velocity is determined by the gradients of hydro-static pressure and electric potential  
\begin{subequations}\label{def_u_ex}
	\begin{align}
	u_{ex}^{r} =-\frac{\kappa_{ex} \tau_{ex}}{\mu} \frac{\partial p_{ex}}{\partial r}-k_{e} \tau_{ex} \frac{\partial \phi_{e x}}{\partial r}, \\
	u_{ex}^{z} =-\frac{\kappa_{ex} \tau_{ex}}{\mu} \frac{\partial p_{ex}}{\partial z}-k_{e} \tau_{ex} \frac{\partial \phi_{ex}}{\partial z},    
	\end{align}
\end{subequations}
where $\phi_{ex}$ is the electric potential in the extracellular space, $\tau_{ex}$ is the tortuosity of extracellular region \cite{nicholson2001diffusion,perez1995extracellular} and $\mu$ is the viscosity of water, $k_{e}$ is introduced to describe the effect of electro-osmotic flow \cite{mclaughlin1985electro,vaghefi2012development,wan2014self}, $\kappa_{ex}$ is the permeability of extracellular space. Here the hydro permeability $\kappa_{ex}$,  tortuosity $\tau_{ex}$ and electric-osmotic parameter $k_{e}$ have two distinguished values in the region $\Omega_{ex}^{OP}$ and $\Omega_{ex}^{SAS}$,
\begin{equation*}
\begin{aligned}
\kappa_{ex}&=\left\{\begin{array}{l}
\kappa_{ex}^{OP}, \text { in } \Omega_{OP}, \\
\kappa_{ex}^{SAS},\text { in } \Omega_{SAS}, 
\end{array}\right.
\tau_{ex}=\left\{\begin{array}{l}
\tau_{ex}^{OP}, \text { in } \Omega_{OP}, \\
\tau_{ex}^{SAS}, \text { in } \Omega_{SAS},
\end{array}\right.  \\
k_{e}&=\left\{\begin{array}{l}
k_{e}^{OP}, \text { in } \Omega_{OP}, \\
k_{e}^{SAS}, \text { in } \Omega_{SAS},
\end{array}\right. .
\end{aligned}
\end{equation*}
Since $\Gamma_{2} \cup \Gamma_{3}$ are the far end of optic nerve away from eyeball and next to the optic canal, we assume the hydro-static pressure of extracellular is equal to the cerebrospinal fluid (CSF) pressure. On the other hand, the intraocular pressure (IOP) is imposed at $\Gamma_{6}$ where the extracellular space is connected to the retina. At boundary $\Gamma_{5}$, we assume a non-permeable boundary. We are aware of the significance of the pressures and flows at these boundaries for clinical phenomena including glaucoma \cite{band2009intracellular,norman2011finite,gardiner2010computational} and will return to that subject in later publications. 

The water flow across the semi-permeable membrane $\Gamma_4$  is  produced by the lymphatic drainage on the dura membrane, which depends on the difference between extracellular pressure and orbital pressure (OBP). We assume the velocity across the pia membrane $\Gamma_4$, is continuous and determined by the combination of hydrostatic and osmotic pressures. To summarize, the boundary conditions of the extracellular fluid are

\begin{equation} 
\label{P_ex_boundary_cd}
\left\{
\begin{aligned}
& \mathbf{u}_{ex}  \cdot \hat{\mathbf{r}}=0,  &  &\mbox{on $\Gamma_{1}$},\\
&p_{ex}=p_{CSF}, &  &\mbox{on $\Gamma_{2}\cup \Gamma_{3}$},\\
&\mathbf{u}^{SAS}_{ex} \cdot \ \hat{\mathbf{r}}=L^{m}_{dr}\left(p^{SAS}_{ex}-p_{OBP}\right),&  &\mbox{on $\Gamma_{4}$},\\
&\mathbf{u}_{ex} \cdot  \hat{\mathbf{r}}=0, &  &\mbox{on $\Gamma_{5}$},\\
&p_{ex}=p_{ICP},&  &\mbox{on $\Gamma_{6}$},\\
&\mathbf{u}^{OP}_{ex} \cdot  \hat{\mathbf{r}}=\mathbf{u}^{SAS}_{ex} \cdot  \hat{\mathbf{r}}\\
&=L^{m}_{pia} \left(p^{OP}_{ex}-p^{SAS}_{ex}-\gamma_{pia}k_{B}T \left(O^{OP}_{ex}-O^{SAS}_{ex} \right) \right),&  &\mbox{on $\Gamma_{7}$},
\end{aligned}
\right.
\end{equation}
where $p_{CSF}$ is the cerebrospinal fluid pressure \cite{band2009intracellular} and $p_{ICP}$ is the pressure in the eye and $p_{OBP}$ is the orbital pressure on the dura mater.
\begin{remark}
	\textcolor{black}{Substituting velocities \eqref{def_u_gl}, \eqref{def_u_ax} and \eqref{def_u_ex} into conservation law Eq. \eqref{water_eq} yields Poisson Equations of hydrostatic pressures in different compartments. Eqs.  \eqref{def_u_gl}, \eqref{def_u_ax} and \eqref{def_u_ex}  mean that velocities vary in both $r$ and $z$ direction, which depend on the gradient of the hydrostatic pressure, osmotic pressure, or electric field. The distribution of velocity in radius direction during and after a train of stimuli is shown in Appendix Fig. \ref{fig: velocityr}. }
\end{remark}

\subsubsection{Ion Transport}\label{Ion_transport}
The conservation of chemical species  implies the following system of partial differential equations to describe the dynamics of ions in each region, for $i=\mathrm{Na^+},\mathrm{K^+},\mathrm{Cl^-}$  

\begin{eqnarray}
\label{ion_governing}
&&\frac{\partial\left(\eta_{gl} c_{gl}^{i}\right)}{\partial t}+\mathcal{M}_{gl}J^{m,i}_{gl}+\nabla \cdot\left(\eta_{gl} \mathbf{j}_{gl}^{i}\right)=0, \text { in } \Omega_{OP},  \label{C_i_gl} \\
&&\frac{\partial\left(\eta_{ax} c_{ax}^{i}\right)}{\partial t}+\mathcal{M}_{ax}J^{m,i}_{ax}+\frac{\partial}{\partial z}\left(\eta_{ax} j_{ax,z}^{i}\right)=0, \text { in } \Omega_{OP},   \label{C_i_ax} \\
&&\frac{\partial\left(\eta_{ex} c_{ex}^{i}\right)}{\partial t}-\mathcal{M}_{ax}J^{m,i}_{ax}-\mathcal{M}_{gl}J^{m,i}_{gl}+\nabla \cdot\left(\eta_{ex} \mathbf{j}_{e x}^{i}\right)=0, \text {in } \Omega_{OP}, \nonumber \\
\label{C_i_ex}
\end{eqnarray}
where the last equation reduces to the following in the $\Omega_{SAS}$ region, 
\begin{equation}
\frac{\partial c_{ex}^{i, SAS}}{\partial t}+\nabla \cdot \mathbf{j}_{ex}^{i, SAS}=0.
\end{equation}
The transmembrane ion flux  $J_{k}^{m,i} \ (k=gl,ax)$   consists of  active ion pump source  $J_{p,k}^{i}$  and  passive ion channel source $J_{c,k}^{i}$, on the $k$ membrane,
\begin{equation*}
J_{k}^{m, i}=J_{p,k}^{i}+J_{c,k}^{i}, \quad k=gl,ax, \quad i=\mathrm{Na}^{+}, \mathrm{K}^{+}, \mathrm{Cl}^{-}.
\end{equation*}
On the glial cell membranes, $J_{c,gl}^{i}$ is defined as 
\begin{equation}
J_{c,gl}^{i}=\frac{g_{g l}^{i}}{z^{i} e}\left(\phi_{g l}-\phi_{e x}-E_{g l}^{i}\right),    \ \  i=\mathrm{Na^+,K^+,Cl^-},
\end{equation}
where the Nernst potential is used to describe the gradient of chemical potential $E_{gl}^{i}= \frac{k_{B}T}{ez^{i}} \log\left(\frac{c_{ex}^i}{c_{gl}^i}\right)$ and the conductance $g_{gl}^{i}$ for $i$th ion specie  on the glial membrane is a fixed constant, independent of voltage and time. On the axon's membrane, $J_{c,ax}^{i}$ is defined as 
\begin{equation*}
J_{c,ax}^{i}=\frac{g_{a x}^{i}}{z^{i} e}\left(\phi_{a x}-\phi_{ex}-E_{ax}^{i}\right), \ \ i=\mathrm{Na^+,K^+,Cl^-},
\end{equation*}
where 
\begin{equation*}
g_{ax}^{Na}=\bar{g}^{Na} m^{3} h+g_{leak}^{Na} , \ \   
g_{ax}^{K}=\bar{g}^{K} n^{4}+g_{leak}^{K},  \ \ g_{ax}^{Cl}=g_{leak}^{Cl}.
\end{equation*}
The time dependent dynamic of open probability, often loosely called `gating' is governed by the Hodgkin-Huxley model \cite{fitzhugh1960thresholds,gabbiani2017mathematics}
\begin{equation}
\label{HHMODEL}
\begin{aligned}
\frac{dn}{dt} &=\alpha_{n}(1-n)-\beta_{n} n, \\
\frac{dm}{dt} &=\alpha_{m}(1-m)-\beta_{m} m, \\
\frac{dh}{dt} &=\alpha_{h}(1-h)-\beta_{h} h,
\end{aligned}
\end{equation}
where $n$ is the open probability of $\mathrm{K}^+$ channel, $m$ is the open probability of the  $\mathrm{Na}^+$ activation gate, and $h$ is the open probability of the $\mathrm{Na}^+$ inactivation gate.

We assume that the only pump is the Na/K active transporter. We are more than aware that other active transport systems can and likely do move ions and thus water in this system. They will be included as experimental information becomes available.

In the case of the Na/K pump $J_{p,k}^i$ $(k=ax,gl)$, the strength of the pump $I_{k}$ depends on the concentration in the intracellular and extracellular space \cite{gao2000isoform,fitzhugh1960thresholds}, i.e. 
\begin{equation}
\label{pump_eq}
J_{p,k}^{Na}=\frac{3I_{k}}{e}, \quad J_{p,k}^{K}=- \frac{2I_{k}}{e}, \quad J_{p,k}^{Cl}=0, \quad k=gl,ax, 
\end{equation}
where 
\begin{equation}
\begin{aligned}
I_{k}&=I_{k,1}\left(\frac{c_{k}^{N a}}{c_{k}^{Na}+K_{Na1}}\right)^{3}\left(\frac{c_{ex}^{K}}{c_{ex}^{K}+K_{K1}}\right)^{2} \\
&+I_{k,2}\left(\frac{c_{k}^{Na}}{c_{k}^{Na}+K_{Na2}}\right)^{3}\left(\frac{c_{ex}^{K}}{c_{ex}^{K}+K_{K2}}\right)^{2}, \quad k=ax,gl.
\end{aligned}
\end{equation}
$I_{k,1}$ and $I_{k,2}$ are related to the maximum current of  $\alpha_{1}-$ and $\alpha_{2}-$ isoform of $\mathrm{Na/K}$ pump on the glial membrane ($k=gl$) or axon membrane ($k=ax$).

The definitions of ion flux in each domain are as follows, for $i=\mathrm{Na^+,K^+,Cl^-}$,  
\begin{equation*}
\begin{aligned}
&\mathbf{j}_{l}^{i} =c_{l}^{i} \mathbf{u}_{l}-D_{ l}^{i} \tau_{l}\left(\nabla c_{l}^{i}+\frac{z^{i} e}{k_{B} T} c_{l}^{i} \nabla \phi_{l}\right), \quad l=gl, ex, \\
&j_{ax,z}^{i} = c_{ax}^{i} u_{ax}^{z}-D_{a x}^{i}\left(\frac{\partial c_{ax}^{i}}{\partial z}+\frac{z^{i} e}{k_{B}T} c_{ax}^{i} \frac{\partial \phi_{ax}}{\partial z}\right).
\end{aligned}
\end{equation*}
For the axon compartment and glial compartment boundary condition, we have 
\begin{equation}
c_{ax}^{i}=c_{ax}^{i, re}, \quad \text { on } \ \Gamma_{2} \cup \Gamma_{6},
\end{equation}
and 
\begin{equation}
\left\{\begin{array}{ll}
\mathbf{j}_{g l}^{i} \cdot \hat{\mathbf{r}}=0, & \text { on } \Gamma_{1}, \\
c_{gl}^{i}=c_{gl}^{i, re}, & \text { on } \Gamma_{2} \cup \Gamma_{6}, \\
\mathbf{j}_{gl}^{i} \cdot \hat{\mathbf{r}}=0, & \text { on } \Gamma_{7},
\end{array}\right.
\end{equation}
where the Dirichlet boundary conditions are used at locations $\Gamma_{2}\cup \Gamma_{6} $ for axons and glial cell, and a non-flux boundary condition is used for glial cells ions flux on pia mater $\Gamma_{7}$.

For the extracellular space boundary condition, similar boundary conditions are imposed except on the pia mater $\Gamma_{7}$. The flux across the pia mater is assumed continuous and Ohm’s law is used \cite{zhu2019bidomain}. Additionally, a non-permeable boundary condition is used at location $\Gamma_{5}$ and a homogeneous Neumann boundary condition is applied at the location of the dura mater $\Gamma_{4}$, 
\begin{equation}
\label{C_ex_bc}
\left\{\begin{array}{ll}
\mathbf{j}_{ex}^{i} \cdot \hat{\mathbf{r}}=0, & \text { on } \Gamma_{1}, \\
c_{ex}^{i}=c_{csf}^{i}, & \text { on } \Gamma_{2} \cup \Gamma_{3}, \\
\nabla c_{ex}^{i} \cdot \hat{\mathbf{r}}=0, & \text { on } \Gamma_{4}, \\
\mathbf{j}_{ex}^{i}  \cdot \hat{\mathbf{z}}=0, & \text { on } \Gamma_{5}, \\
c_{ex}^{i}=c_{eye}^{i},  & \text { on } \Gamma_{6}, \\
\mathbf{j}_{ex}^{i, OP} \cdot \hat{\mathbf{r}}=\mathbf{j}_{ex}^{i,SAS} \cdot \hat{\mathbf{r}}=\frac{G_{pia}^{i}}{z^{i} e}\left(\phi_{e x}^{OP}-\phi_{ex}^{SAS}-E_{pia}^{i}\right), & \text { on } \Gamma_{7}.
\end{array}\right.
\end{equation}

\begin{remark}
	\normalfont
	Suppose the $c^{i,*}_{l}$ is the scale of $i$ ion specie in the $l$ space and $\Delta c^{i,*}_{l}$ is the scale of $r$ and $z$ direction $i$ ion specie concentration variation in the $l$ space. If We define 
	\begin{equation*}
	\delta^{i}_{7,l}=\frac{\Delta c^{i,*}_{l}}{c^{i,*}_{l}} , \ \ i=\mathrm{Na}^{+},\mathrm{K}^{+},\mathrm{Cl}^{+} , \ \  l=ax,gl,ex. 
	\end{equation*}
	the ion fluxes could be written as 
	\begin{eqnarray*}
		\tilde{\mathbf{j}}^{i}_{l} &&= Pe^{i}_{l}\delta^{i}_{7,l}\tilde{c}^{i}_{l} \tilde{\mathbf{u}}_{l} - \left( \delta^{i}_{7,l} \tilde{\nabla} \tilde{c}^{i}_{l} + z^{i}\tilde{c}^{i}_{l} \tilde{\nabla} \tilde{\phi}_{l} \right), \ \ \ l=gl,ex,\\
		\tilde{j}^{i}_{ax,z}&&= Pe^{i}_{ax} \delta^{i}_{7,l} \tilde{c}^{i}_{l} \tilde{u}^{z}_{ax} - \left( \delta^{i}_{7,l} \frac{ \partial \tilde{c}^{i}_{l}}{\partial \tilde{z} } + z^{i}\tilde{c}^{i}_{l} \frac{\partial \tilde{\phi}_{l}}{\partial \tilde{z}} \right),
	\end{eqnarray*}
	with Peclet numbers 
	\begin{eqnarray}
	Pe^{i}_{ax}= \frac{ u^{*}_{ax}z^{*}c^{i,*}_{ax}}{ D^{i}_{ax}\Delta c^{i,*}_{ax}}, \ \ Pe^{i}_{l}=\frac{ u^{*}_{l}r^{*}c^{i,*}_{l}}{ D^{i}_{l}\tau_{l}\Delta c^{i,*}_{l}},  \ l=gl,ex. \label{Peclet}
	\end{eqnarray}
	If we let $g_l^*,~l=ax,gl$ be the characteristic membrane conductance, $\frac{k_BT}{e}$ be the characteristic electric potential,  the dimensionless form of transmembrance flux   is 
	\begin{equation*}
	\tilde{J}^{m,i}_{l} = \tilde{J}^{i}_{c,l}+ \tilde{J}^{i}_{p,l},
	\end{equation*}
	where for $i=\mathrm{Na^+,K^+,Cl^-}, \ l=gl,ax,$
	\begin{equation*}
	\begin{aligned}
	\tilde{J}_{c,l}^{i}=\frac{ \tilde{g}_{l}^{i}}{z^{i} }\left(\tilde{\phi}_{k}-\tilde{\phi}_{ex}-\tilde{E}_{gl}^{i}\right), \ \ \ 
	\tilde{J}_{p,l}^{i} = \frac{J_{p,l}^{i}e^2}{k_BTg^{*}_{l}}.
	\end{aligned}
	\end{equation*}
	The governing equations for ions become
	\begin{eqnarray}
	&&\frac{\partial\left(\eta_{gl} \tilde{c}_{gl}^{i}\right)}{\partial \tilde{t} }+ \delta^{i}_{8} \tilde{J}^{m,i}_{gl}+ \delta^{i}_{9} \tilde{\nabla} \cdot \left(\eta_{gl}\tilde{\mathbf{j}}^{i}_{gl}\right) =0, \\
	&&\frac{\partial\left(\eta_{ax} \tilde{c}_{ax}^{i}\right)}{\partial \tilde{t}}+\delta^{i}_{10}\tilde{J}^{m,i}_{ax}+\delta^{i}_{11}\frac{\partial}{\partial \tilde{z}}\left(\eta_{ax} \tilde{j}_{ax,z}^{i}\right)=0, \\
	&&\frac{\partial\left(\eta_{ex} \tilde{c}_{ex}^{i}\right)}{\partial \tilde{t}}-\delta^{i}_{12}\delta^{i}_{10}\tilde{J}^{m,i}_{ax}-\delta^{i}_{13}\delta^i_{8}\tilde{J}^{m,i}_{gl} + \delta^{i}_{14} \tilde{\nabla} \cdot \left(\eta_{ex}\tilde{\mathbf{j}}^{i}_{ex}\right)=0, \quad   \quad   \quad  \label{ion_ex_nd}
	\end{eqnarray}
	where 
	\begin{eqnarray*}
		\tilde{\nabla} \cdot &&\left(\eta_{l}\tilde{\mathbf{j}}^{i}_{l}\right) = \frac{1}{\tilde{r}} \frac{\partial \left(\tilde{r}\eta_{l} \tilde{j}_{l}^{r,i}\right)}{\partial \tilde{r}}+( \delta_{0})^2 \frac{\partial\left(\eta_{l} \tilde{j}_{l}^{z,i}\right)}{\partial \tilde{z}},\ \ l=gl,ex,\\
		\delta^{i}_{8}&&=\frac{t^*\mathcal{M}_{gl}g^{*}_{gl}k_{B}T}{c_{gl}^{i,*}e^2},  \ \  \delta^{i}_{9}=\frac{D^{i}_{gl}\tau_{gl}t^*}{(r^*)^2}, \\
		\delta^{i}_{10}&&=\frac{t^*\mathcal{M}_{ax}g^{*}_{ax}k_{B}T}{c_{ax}^{i,*}e^2}, \ \ \delta^{i}_{11}=\frac{D^{i}_{ax}t^*}{(z^*)^2},\\
		\delta^{i}_{12}&&= \frac{c_{ax}^{i,*}}{c_{ex}^{i,*}}, \ \ \delta^{i}_{13}=\frac{c_{gl}^{i,*}}{c_{ex}^{i,*}}, \ \ \delta^{i}_{14} = \frac{D^{i}_{ex}\tau_{ex}t^*}{(r^*)^2}.
	\end{eqnarray*}
	
\end{remark}
\begin{remark}
	In the rest of this paper, the symbol $\Delta f$ is used to denote the variation of the variable $f$ from its resting state value.
\end{remark}

Multiplying Eqs. in (\ref{C_i_gl}-\ref{C_i_ex}) with $z_{i} e$ respectively, summing up, and using the charge neutrality condition,  we have the following system for the electric fields in $ax,gl,ex$,

\begin{eqnarray}
\label{phi_governing}
&\sum_{i} z^{i} e \mathcal{M}_{gl}J^{m,i}_{gl}+\sum_{i}  \nabla \cdot\left(z^{i} e\eta_{g} \mathbf{j}_{gl}^{i}\right) =0, \\
&\sum_{i} z^{i} e \mathcal{M}_{a x}J^{m,i}_{ax}+\sum_{i}  \frac{\partial}{\partial z}\left(z^{i} e\eta_{ax} j_{ax, z}^{i}\right) =0, \\
&\sum_{i} z^{i} e \nabla \cdot\left(\eta_{gl} \mathbf{j}_{gl}^{i}\right)+\sum_{i}  \frac{\partial}{\partial z}\left(z^{i} e\eta_{a x} j_{ax,z}^{i}\right)+\sum_{i}  \nabla \cdot\left(z^{i} e\eta_{ex} \mathbf{j}_{ex}^{i}\right) =0,\nonumber\\
& 
\end{eqnarray}
In the subarachnoid space $\Omega_{SAS}$, the extracellular equations reduce to 
\begin{equation}
\sum_{i}  \nabla \cdot\left(z^{i} e\sum_{i} \mathbf{j}_{ex}^{i,SAS}\right)=0.
\end{equation}
The boundary conditions for electric fields $\phi_{ax}$, $\phi_{gl}$ and $\phi_{ex}$ are given below.\\
In the axon compartment: 
\begin{equation}
\left\{
\begin{aligned}
\nabla \phi_{ax} \cdot \hat{\mathbf{z}}=0, & \text { on } \Gamma_{2}, \\
\nabla \phi_{ax} \cdot \hat{\mathbf{z}}=0, & \text { on } \Gamma_{6},
\end{aligned}\right.
\end{equation}
In the glial compartment: 
\begin{equation}
\left\{
\begin{aligned}
&\nabla \phi_{gl} \cdot \hat{\mathbf{r}}=0, & \text { on } \Gamma_{1}, \\
&\nabla \phi_{gl} \cdot \hat{\mathbf{z}}=0, & \text { on } \Gamma_{2}, \\
&\nabla \phi_{gl} \cdot \hat{\mathbf{z}}=0, & \text { on } \Gamma_{6}, \\
&\nabla \phi_{gl}  \cdot \hat{\mathbf{r}}=0, & \text { on } \Gamma_{7},
\end{aligned}
\right.
\end{equation}
and in the extracellular space: 
\begin{equation}
\label{phi_bd}
\left\{
\begin{aligned}
&\nabla\phi_{ex} \cdot \hat{\mathbf{r}}=0,  &&\text {on } \Gamma_{1}, \\
&\nabla \phi_{ex} \cdot \hat{\mathbf{z}}=0, &&\text {on } \Gamma_{2} \cup \Gamma_{3}, \\
&\nabla \phi_{ex} \cdot \hat{\mathbf{r}}=0, &&\text {on } \Gamma_{4}, \\
&\nabla\phi_{ex}  \cdot \hat{\mathbf{z}}=0, &&\text {on } \Gamma_{5}, \\
&\nabla \phi_{ex} \cdot \hat{\mathbf{z}}=0, &&\text {on } \Gamma_{6}, \\
&\sum_{i} z^{i} e \mathbf{j}_{ex}^{i, OP} \cdot \hat{\mathbf{r}}=\sum_{i} z^{i} e \mathbf{j}_{e x}^{i, SAS} \cdot \hat{\mathbf{r}} &&\\
&=\sum_{i} G_{pia}^{i}\left(\phi_{ex}^{OP}-\phi_{e x}^{SAS}-E_{pia}^{i}\right), &&\text {on } \Gamma_{7}.
\end{aligned}
\right.
\end{equation}

In the rest of this paper, the full electric-diffusion-convection model is defined by Eqs.  (\ref{water_gl}) through (\ref{phi_bd}). The electric-diffusion model is defined by Eqs.  (\ref{ion_governing})-(\ref{phi_bd}). The electric diffusion model is a reduced version of the full model in which water is neglected.

\section{Model Calibration   and Validation \label{Model_Calibration}}
In this section, we use the physiological and anatomical data in Orkand et al. \cite{orkand1966effect} to calibrate the value of parameters, like membrane conductance, capacitance, and structural parameters.   We then validate our model by computing results with these parameters and comparing the computation with the experiment, which are designed to measure the change in potential across the glial membrane produced by a train of action potentials.

In the Orkand experiment, optic nerve has been put in bathing solutions with three different $\mathrm{K^+}$ concentration $(1.5 \   \mathrm{mM}, 3 \ \mathrm{mM}, 4.5 \  \mathrm{mM})$ and the resting potential across the glia membrane was measured. Then the axon was stimulated simultaneously at both ends (see lines 5-6 of the Methods section of Orkand paper) to give a train of action potentials. The action potentials increased $\mathrm{K^+}$ in extracellular space (ECS). The accumulated $\mathrm{K^+}$ then made the glia membrane potential more positive.

\begin{figure}[hpt]
	\centering
	\includegraphics[bb=65 0 880 430, clip=true,width=3.25in,height=5.5cm ]{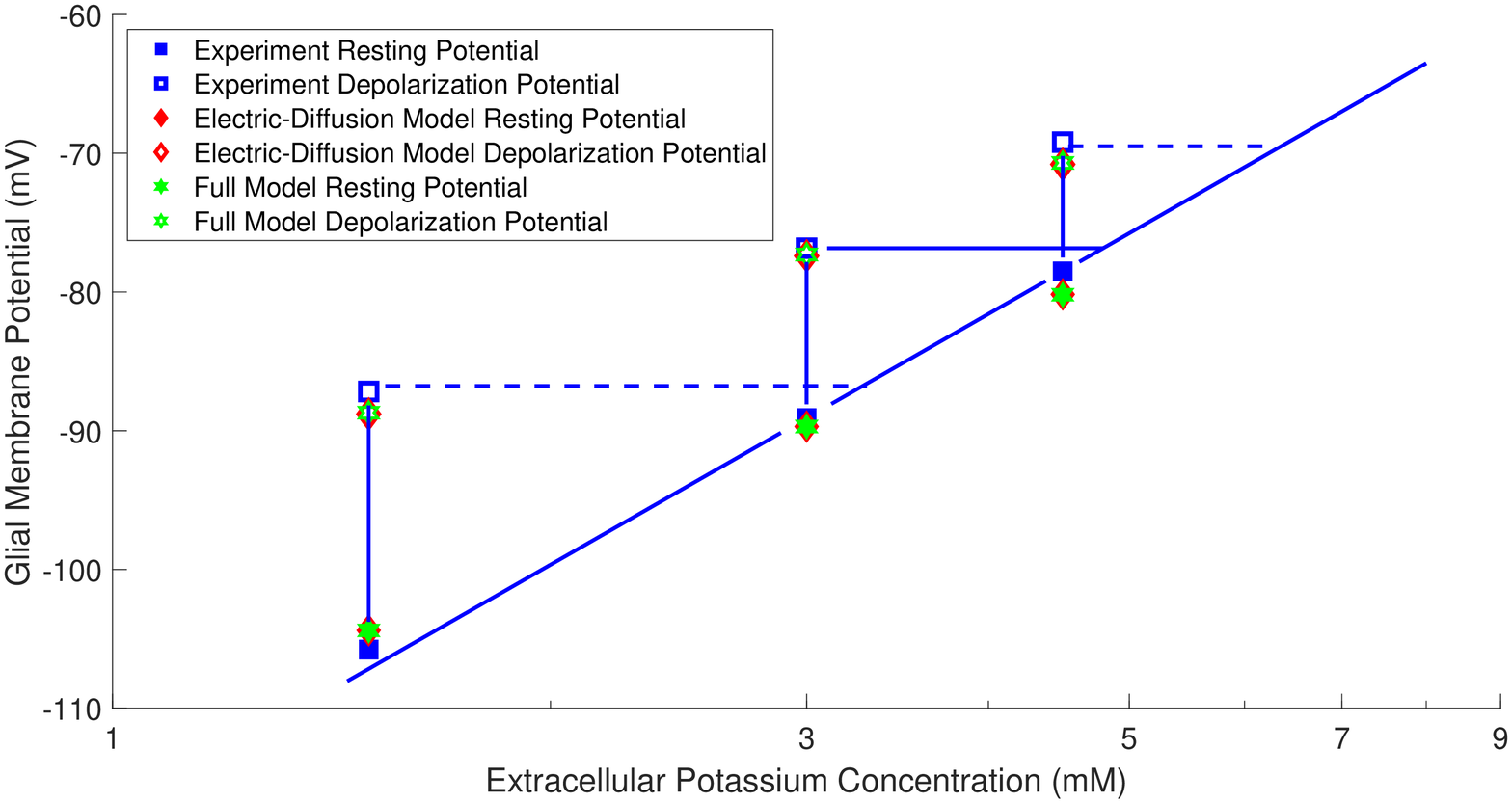}
	\caption{ The comparison between the experiment  \cite{orkand1966effect}  and simulation on the effect of nerve impulses on the membrane potential of glial cells. The solid symbols are resting potentials and the open symbols are depolarization potentials with different ECS $\mathrm{K^+}$ concentrations. \label{fig:Calibration}}
\end{figure}

In the simulation, we applied a train of stimuli with frequency $17/ \mathrm{s}$ for $1 \mathrm{s}$ to the axon membrane at $z=\textcolor{black}{2.25} \mathrm{mm}, 13.5 \mathrm{mm}, 0<r<R_{a}=48 \mathrm{\mu m}$.  Each individual stimulus in the train lasted $3 \ \mathrm{ms}$ (as Orkand's paper indicated) and had strength $3 \ \mathrm{mA/m^2}$.   The stimulus was large enough to exceed threshold and generate action potentials. We set the ECS $\mathrm{K^+}$ to be $1.5 \ \mathrm{mM}, 3 \ \mathrm{mM}$, or  $4.5\ \mathrm{mM}$ and record the largest absolute value of the change in glial membrane potential in each case as  in the Fig. \ref{fig:Calibration} . This number is loosely called `the depolarization' in most laboratories. The blue symbols show experimental data, red ones are the simulations results of electrodiffusion model and the green ones are the full model.  Fig. \ref{fig:Calibration} shows that both the full model and electrodiffusion model could match the experimental resting potentials (solid symbols) and depolarizations (open symbols) very well for the different ECS $\mathrm{K^+}$ concentrations. 

\begin{figure}[hpt]
	\centering
	\includegraphics[bb=0 0 1120 1220, clip=true,width=3.25in,height=6.5cm ]{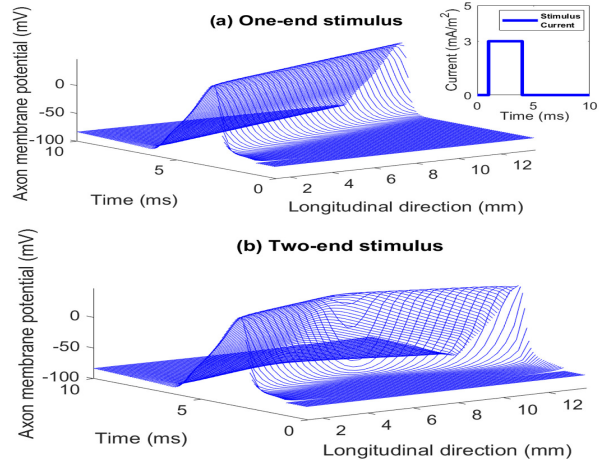}
	\caption{(a) axon membrane potential profile when eye-end axon stimulated. The built-in figure is the stimulus current profile.  (b) axon membrane potential profile when two-end axon simulated.  \label{fig:axon_potential}}
\end{figure}

Fig. \ref{fig:axon_potential} shows the propagation of the axon action potential. The membrane potential from axons at the center of the optic nerve bundle is shown when different locations of the axon had been stimulated. In both eye-end and two-end cases, the stimulus current was applied  from  $t=1 \ \mathrm{ms}$ to   $t=4 \ \mathrm{ms}$.  In Fig. \ref{fig:axon_potential}a, the stimulus was applied near to the optic nerve near the eye-end $(z=\textcolor{black}{2.25} \  \mathrm{mm})$. At $t=1 \ \mathrm{ms} $, the discontinuity of stimulus current induces jumps of the axon membrane potential in Fig. \ref{fig:axon_potential}.  At $t=10 \ \mathrm{ms}$, the action potential completely has propagated and left the location near far-eye-end $(13.5 \ \mathrm{mm})$.  The axon in the optic nerve of the mud puppy is unmyelinated. This speed of action potential propagation in the model lies in the range of the action potential  speeds typical of unmyelinated axons,  i.e., between $0.5 \ \mathrm{m/s}$ and $2.0 \ \mathrm{m/s}$ \cite{susuki2010myelin}.  In the Fig. \ref{fig:axon_potential}b, when the two-ends of the axon stimulated, the axon membrane potential has is more uniform spatially at each time point in compare to the single side stimulus case.  Orkand et al used the dual stimulation to more closely approximate a `space clamp'.

\section{Effects of Water Flow \label{Effects_of_Water_Flow}}

In this section, when part of the nerve is stimulated, we estimate the transmembrane fluxes and the resulting accumulation of ions in the extracellular space and glial cells.  
Our main conclusion is  that the variation of osmotic pressure between extracellular space and glial cells is the dominant mechanism that drives water flow.  And water flows are significant and many important flows occur in the glial region. It is important to note that these flows can occur in the glia because it is a syncytium of irregular but finite cells (i.e., not long cylinders) that allows easy flow from cell to cell.
The circulation pattern and strength of water flow in optic nerve are also presented. 

\begin{figure}[hpt]
	\centering
	\includegraphics[bb=10 50 400 300, clip=true,width=3.25in,height=4.5cm ]{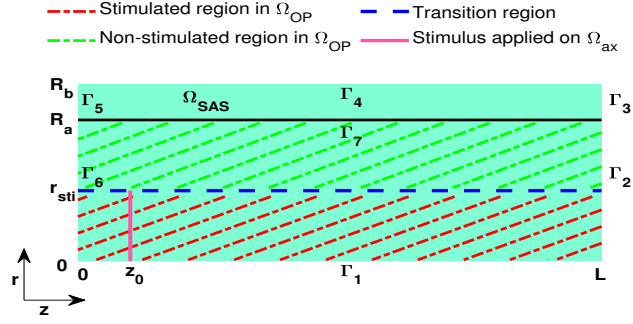}
	\caption{Stimulated region and non-stimulated region in the optic nerve $(\Omega_{OP})$. The stimulus is applied in the axon compartment where  $0<r<r_{sti}$ at a given location $z=z_{0}$. \label{fig:stimulated_region}}
\end{figure}

To simplify our discussions, we focus our analyses on an idealized setting where the stimulus is applied at an inner part of the axon compartment. As shown in Fig. \ref{fig:stimulated_region}, the stimulus was applied at $0<r<r_{sti}$  at a given location $z=z_{0}$. This stimulus is within the optic nerve, so $r_{sti}<R_{a}=r^{*}$  shown in Fig. \ref{fig:stimulated_region}. We distinguish the stimulated region and the non-stimulated region in the optic nerve $\Omega_{OP}$ shown in the Fig. \ref{fig:stimulated_region}, since the electrical signal propagates in the $z$ direction in the axon compartment. We do not put the stimulus everywhere in this region, rather we only apply the stimulus at the location $(z_{0})$ within a radial. 

To understand the mechanism inducing the water circulation, we first estimate the variations of ion concentrations from axon to the extracellular space during a single action potential. Then we analyze the different transmembrane current on the glial cells and identify the dominant $\mathrm{K^+}$  current.  Finally, we study osmotic pressure change after a train of action potentials on axon.

\subsection{Single action potential estimation}
We first estimate the amount of ion exchange between axon and extracellular space during a single action potential. We assume that during the single action potential, the volume fraction $\eta_l,\ l=ax,gl,ex$, does not differ from their resting state. We find then that the variation of $\mathrm{Na^{+}}$ and $\mathrm{K^{+}}$   in the stimulated extracellular region is the same to leading order, and that agrees with experimental observations \cite{ostby2009astrocytic,keynes1951ionic,dietzel1982stimulus}. Although our estimation is based on the classic Hodgkin-Huxley model, the methods are general and can be applied to systems with other channels and transporters.  

When an action potential occurs in the nerve, the equilibrium (or steady state) balance between the ions and electric fields is lost and resting state changes. We introduce notations to separate the resting state variables (with superscript `$re$') before the action potentials from the variables  during the action potentials (with superscript `$dy$'). 

We introduce the current of $i$th ionic species through axon and glial membrane as
\begin{eqnarray*}
	&I_{k}^{i,j}=z^{i}eJ_k^{m,i,j}=z^{i}eJ_{p,k}^{i,j}+z^{i}eJ_{c,k}^{i,j},\ i=\mathrm{Na}^{+},\mathrm{K}^{+},\mathrm{Cl}^{-}, \\
	&~~~~~~~~~~~\ 
	j=re,dy, \ k=gl,ax, 
\end{eqnarray*}
where $J_{k}^{m,i,j}$ consists of the active $\mathrm{Na/K}$ pump source $(J_{p,k}^{i,j})$ and  passive ion channel source $(J_{c,k}^{i,j})$  for $i$th ionic species on the axons $(k=ax)$ or glial cells membranes $(k=gl)$ at resting state $(j=re)$ before the action potentials or during the action potentials $(j=dy)$.

At the resting state,  $\mathrm{Na/K}$ pump source $J_{p,k}^{i,re}$ and ion channels source $J_{c,k}^{i,re}$ on the axon membrane $(k=ax)$ and glial membrane $(k=gl)$ satisfy
\begin{eqnarray*}
	&J_{p,k}^{Na,re}=\frac{3I_{k}^{re}}{e},\ \ \ J_{p,k}^{K,re}= -\frac{2I_{k}^{re}}{e},\ \ \ J_{p,k}^{Cl,re}=0, \\ 
	&J_{c,k}^{i,re}=\frac{g_{k}^{i,re}}{z^{i} e} \left(V_{k}^{re}-E_{k}^{i,re} \right),i=\mathrm{Na^+,K^+,Cl^-},  k=gl,ax
\end{eqnarray*}
where the membrane potential $V^{re}_{k}$ at the resting state is
\begin{equation*}
V_{k}^{re}=\phi_{k}^{re}-\phi_{ex}^{re}, \ \  k=gl,ax.
\end{equation*}
The ion channel conductance on the glial membrane is a fixed constant,
\begin{equation*}
g_{gl}^{i,re}=g_{gl}^{i} ,\quad i=\mathrm{Na^+,K^+,Cl^-}.
\end{equation*}
and the ion channel conductance on the axon membrane is defined as in the classical Hodgkin-Huxley model
\begin{eqnarray*}
	&g_{ax}^{Na,re}=\bar{g}^{Na} \left(m^{re}\right)^{3}h^{re}+ g_{leak}^{Na}, \ \ g_{ax}^{K,re}=\bar{g}^{K} \left(n^{re} \right)^{4} +g_{leak}^{K},  \\ &g_{ax}^{Cl,re}=g_{leak}^{Cl}, 
\end{eqnarray*}
The kinetic variables $m^{re}$, $h^{re}$ and $n^{re}$ are measures of the resting state open probability for the voltage-gated $\mathrm{Na^+}$ and $\mathrm{K^+}$ channel on the axon membrane. In addition, in the resting state, the ion fluxes through the active Na/K pump $J_{p,k}^{i,re}$ and  ion channel $J_{c,k}^{i,re}$  in the  glial membrane ($k=gl$) or axon membrane ($k=ax$) are balanced in magnitude 
\begin{equation*}
O\left( |J_{p,k}^{i,re}| \right) = O\left( |J_{c,k}^{i,re}| \right), \ i=\mathrm{Na^+,K^+,Cl^-}, \ k=gl,ax.
\end{equation*}
During action potentials, the ion fluxes through active $\mathrm{Na/K}$ pump are
\begin{equation*}
J_{p,k}^{Na,dy}=\frac{3\left(I_{k}^{re}+\Delta I_{k}\right)}{e},    \ \ \   J_{p,k}^{K,dy}=-\frac{2\left( I_{k}^{re}+\Delta I_{k}\right)}{e} ,  \ \ \ k=gl,ax,
\end{equation*}
where $\Delta I_{k}$ is the variation of current through Na/K pump in the membrane due to the ion concentration changes. The ion fluxes through ion channels can be written as 
\begin{eqnarray*}
	J_{c,k}^{i,dy}=&\frac{g_{k}^{i,dy}}{z^{i} e} \left( V_{k}^{re}-E_{k}^{i,re} \right)+ \frac{g_{k}^{i,dy}}{z^{i} e} \left(\Delta V_{k} -\Delta E_{k}^{i}  \right),  \ k=gl,ax,
\end{eqnarray*}
where $\Delta \mathrm{X}_{k}= \mathrm{X}_{k}^{dy}-\mathrm{X}_{k}^{re}$ is the deviation of $\mathrm{X}$ away from the resting state value with $\mathrm{X}=V,E,I$ on the membrane $k$. For the conductance on  membranes, we have
\begin{eqnarray*}
	&g_{ax}^{Na,dy}=\bar{g}^{Na}\left( m^{dy} \right)^{3} h^{dy}+g_{leak}^{Na}, \ \ g_{ax}^{K,dy}=\bar{g}^{K} \left( n^{dy} \right )^{4}+g_{leak}^{K}, \\ &g_{ax}^{Cl,dy}=g_{ax}^{Cl,re}, \ \
	g_{gl}^{i,dy}=g_{gl}^{i,re},  \ \  i=\mathrm{Na^+,K^+,Cl^-},
\end{eqnarray*}
where $m^{dy}$, $h^{dy}$ and $n^{dy}$ are governed by system (\ref{HHMODEL}).
During a single action potential, we claim that the variation of  ion's Nernst potential is much smaller than changes in the axon membrane potential (see Appendix \ref{nerst_axon}),
\begin{equation*}
\Delta E_{ax}^{i}=o\left( \Delta V^*_{ax} \right), \ \  i=\mathrm{Na^+,K^+,Cl^-},
\end{equation*}
At the same time, we estimate that
\begin{equation*}
J_{p,ax}^{i,dy} = o\left( \frac{g_{ax}^{i,dy}}{z^{i} e} \left( V_{ax}^{re}-E_{ax}^{i,re} \right) \right), \ \  i=\mathrm{Na^+,K^+}.  
\end{equation*}
This is because the voltage-gated $\mathrm{Na^{+}}$ and $\mathrm{K^{+}}$ channels are open during the action potential and satisfy
\begin{equation*}
g_{ax}^{i,re}=o\left( g_{ax}^{i,dy} \right), \quad i=\mathrm{Na^+,K^+}.
\end{equation*}
In addition, the increments of $\mathrm{Na/K}$ pump strength is limited since the ion fluxes through  the $\mathrm{Na/K}$ pump is controlled by its maximum currents $I_{ax,1}$ and  $I_{ax,2}$   in  Eq. (\ref{pump_eq}).

In sum, during action potentials, we can approximate the axon transmembrane current for each ionic species as        
\begin{equation}
\label{current_app_1}
I_{ax}^{i,dy}\approx g_{ax}^{i,dy} \left( V_{ax}^{re}-E_{ax}^{i,re} \right)+g_{ax}^{i,dy} \Delta V_{ax},\quad i=\mathrm{Na^+,K^+,Cl^-}.        
\end{equation}
In the next paragraphs, by using Eq. (\ref{current_app_1}), we estimate the accumulative $\mathrm{Na^+}$ and $\mathrm{K^+}$ fluxes through the axon membrane during a single action potential. This estimation helps us estimate the concentration changes in the stimulated extracellular region. 

\begin{figure}[h]
	\centering
	\includegraphics[width=3.2in,height=5.5cm ]{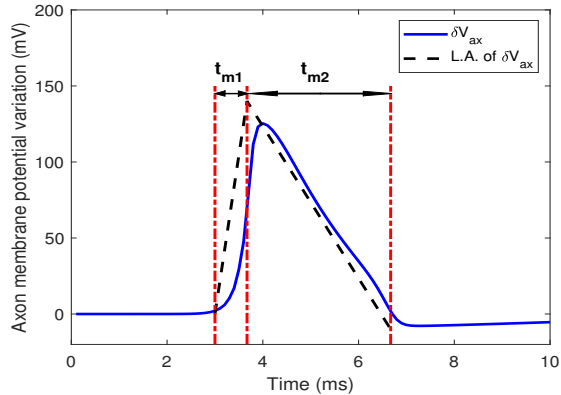}
	\caption{Two distinguished time intervals used in the estimation during a single action potential. The blue line is the axon membrane potential variation $\Delta V_{ax} (= V^{dy}_{ax}-V_{ax}^{re})$  during a single action potential. The dark dash line is the linear approximation of the $\Delta V_{ax}$.  $t_{m1}$ and $t_{m2}$ are the time parameters in Eqs. (\ref{m_eq_1}) and (\ref{m_eq_2}). }
	\label{fig::estimate_time_intervals}
\end{figure}

The governing equation of the open probability  for $\mathrm{Na}^{+}$ channel $m$-gates  in the Hodgkin-Huxley model is
\begin{equation}
\label{HH_1}
\frac{dm^{dy}}{dt}=\alpha_{m} \left(1-m^{dy}\right)-\beta_{m} m^{dy},                      
\end{equation}
where      
\begin{equation}
\label{alpha_m}
\alpha_{m}=\frac{1}{10} \frac{ 25 - \Delta V_{ax} }{ \exp\left( \frac{25-\Delta V_{ax}}{10}\right)-1} ,  \ \  \
\beta_{m}=4\exp\left(-\frac{\Delta V_{ax}}{18} \right),
\end{equation}
and $\Delta V_{ax}=V_{ax}^{dy}-V_{ax}^{re}$. The solution for Eq. (\ref{HH_1}) is 
\begin{eqnarray}
\label{m_1_solution}
m^{dy}&&(t)=m_{0}\exp\left(\int_{0}^{t}\alpha_{m}(s)+\beta_{m} ds\right) \nonumber \\ 
&&+\int_{0}^{t} \alpha_{m}(s)\exp \left(-\int_{s}^{t}\alpha_{m}(u)+\beta_{m}(u)du\right)ds, 
\end{eqnarray}
with initial value $m_{0}$.\\
During a single action potential period $[0,T_{ax}^{*}]$, we define two distinguished time intervals based on the rapidly-responding $m$-gates open probability $m^{dy}$ as shown in Fig. \ref{fig::estimate_time_intervals}.
The first period $[0,t_{m1}]$ is when the  $\mathrm{Na^+}$ channel  becomes fully open, and the action membrane  potential moves positive  from its resting value to its most positive value. The second period $[t_{m1},T_{ax}^{*}=t_{m1}+t_{m2}]$ occurs  when the  $\mathrm{Na^+}$ channel closes and the action potential recovers from the peak value to the hyperpolarization value.

In the first time interval $[0,t_{m1}]$, we estimate that $\Delta V_{ax}$ increases monotonically from $0$ to $E_{ax}^{Na,re}-V_{ax}^{re}$, where we approximate the peak value of action potential by the Nernst potential of $\mathrm{Na^+}$  in  the resting state such that
\begin{equation}
\label{V_ax_app}
\Delta V_{ax}(t)=\frac{E_{ax}^{Na,re}-V_{ax}^{re}}{t_{m1}}t, \ \ \ t\in [0,t_{m1}]. 
\end{equation}
where $E_{ax}^{Na,re}-V_{ax}^{re}\approx 1.4\times 10^{2}$  $\mathrm{mV}$.  In Eq. (\ref{V_ax_app}), the  $t_{m1}$ is an unknown variable. The initial value of Eq. (\ref{m_1_solution}) is chosen when $\Delta V_{ax}=0 \ \mathrm{mV}$ as
\begin{eqnarray*}
	m_{0}=m^{re}=m^{eq}(0),
\end{eqnarray*}
where $m^{eq}$ is the equilibrium state of Eq. (\ref{HH_1}) depending on $\Delta V_{ax}$, 
\begin{equation}
m^{eq}(\Delta V_{ax})=\frac{\alpha_{m}(\Delta V_{ax} )}{\alpha_{m} (\Delta V_{ax} )+\beta_{m}(\Delta V_{ax})} .
\end{equation}
By using Eqs. (\ref{alpha_m}), (\ref{m_1_solution}) and (\ref{V_ax_app}),  we can obtain one equation for $t_{m1}$ as shown in Eq. (\ref{m_eq_1}) (see Appendix \ref{robust}). Without loss of generality, we assume the voltage-gated $\mathrm{Na^{+}}$ channel is almost fully open when $t=t_{m1}$ and $m^{dy}(t_{m1})=0.95$. The estimation from Eq. (\ref{m_eq_1}) gives $t_{m1}\approx 0.67\ \mathrm{ms}$.

In the second time interval, we use the homogeneous property of Eq. (\ref{HH_1}) and move the time interval $[t_{m1},T_{ax}^{*}=t_{m1}+t_{m2}]$ to $[0,t_{m2}]$ to simplify the notation. We assume that $\Delta V_{ax}$  decreases monotonically from  $E^{Na,re}_{ax}-V_{ax}^{re}$ to  $E^{K,re}_{ax}-V_{ax}^{re}$ at second time period such that 
\begin{equation}
\label{V_ax_app2}
\Delta V_{ax}(t) =E_{ax}^{Na,re}-V_{ax}^{re}-\frac{E_{ax}^{Na,re}-E_{ax}^{K,re}}{t_{m2}}t, \ \ t\in[0,t_{m2}],
\end{equation}
where $E_{ax}^{Na,re}-E_{ax}^{K,re}\approx1.5\times10^{2}\ \mathrm{mV}$. We assume that the initial value $m_{0}$ of Eq. (\ref{m_1_solution})  at the second time period is
\begin{equation*}
m_{0}=m^{dy}(t_{m1}).
\end{equation*}
The $\mathrm{Na^{+}}$ channel is in a nearly closed state when the $\Delta V_{ax}$ approaching $E_{ax}^{K,re}-V_{ax}^{re}$ and we estimate $m^{dy}(t_{m2})=0.1$.  In a similar way, by using Eqs. (\ref{alpha_m}), (\ref{m_1_solution}) and (\ref{V_ax_app2}),  we could have another equation for $t_{m2}$ as shown in Eq. (\ref{m_eq_2}) (see Appendix \ref{robust}). Based on Eq. (\ref{m_eq_2}), we get $t_{m2}\approx 3\ \mathrm{ms}$.

In sum, based on estimated $t_{m1}$ and $t_{m2}$ in above, we obtain the approximations for the $\Delta V_{ax}$ and the $h$ during a single action potential period
$(t\in [0,T_{ax}^{*}=t_{m1}+t_{m2}])$ as
\begin{eqnarray*}
	\Delta V_{ax} =\left\{
	\begin{aligned}
		& \frac{E_{ax}^{Na,re}  - V_{ax}^{re} } {t_{m1}}t, &&t\in[0,t_{m1}], \\
		& E_{ax}^{Na,re}- V_{ax}^{re}- \frac{E_{ax}^{Na,re}-E_{ax}^{K,re}}{t_{m2}} (t-t_{m1}),&& t\in[t_{m1},T_{ax}^{*}].
	\end{aligned}
	\right.
\end{eqnarray*}
and 
\begin{eqnarray*}
	h^{dy}(t)&&=h_{0} \exp \left( -\int_{0}^{t} \alpha_{h}(s)+\beta_{h}(s)ds  \right) \\
	&&+\int_{0}^{t}\alpha_{h}(s)\exp \left( -\int_{s}^{t} \alpha_{h}(u)+\beta_{h}(u)du\right)ds,
\end{eqnarray*}
where 
\begin{eqnarray*}
	\alpha_{h}= \frac{7}{100}\exp\left( - \frac{\Delta V_{ax}}{ 20}\right),  \ \ \beta_{h}= \frac{1}{ \exp\left( \frac{30-\Delta V_{ax}}{10}\right)+1 },
\end{eqnarray*}
with the initial value $h_{0}$ 
\begin{equation*}
h_{0}=h^{re}(0)=\frac{\alpha_{h} (0)}{\alpha_{h}(0)+\beta_{h}(0)}.
\end{equation*}
By using Eq. (\ref{current_app_1}),  we estimate the cumulative $\mathrm{Na^+}$ flux Eqs.{black}{through the axon membrane} during a single action potential $[0,T_{ax}^*]$ by 
\begin{eqnarray}
\label{Na_flux}
&&\int_{0}^{T_{ax}^{*}} 
J_{ax}^{m,Na,dy}dt  \nonumber  \\
&&\approx \int_{0}^{T_{ax}^{*}} \frac{\bar{g}^{Na}h^{dy}(m^{dy})^{3} }{z^{Na}e}\left(V_{ax}^{re}-E_{ax}^{Na,re} \right) \nonumber +\frac{\bar{g}^{Na}h^{dy}(m^{dy})^{3} }{z^{Na}e}\Delta V_{ax} dt \nonumber \\
&&\approx-2\times 10^{-9}\ \mathrm{mol/m^{2}}.
\end{eqnarray}   

In the next step, we estimate the cumulative $\mathrm{Cl^{-}}$ flux  through the axon membrane during a single action potential $[0,T_{ax}^*]$ by 
\begin{equation}
\label{Cl_flux}
\int_{0}^{T_{ax}^{*}}  J_{ax}^{m,Cl,dy} dt\approx \int_{0}^{T_{ax}^{*}} \frac{g_{ax}^{Cl}\Delta V_{ax}}{z^{Cl}e}dt\approx -3.7\times 10^{-10} \ \mathrm{mol/m^{2}}.
\end{equation}
In Eq. (\ref{Cl_flux}), we use
\begin{equation*}
I_{ax}^{Cl,dy}=g_{ax}^{Cl} \left(V_{ax}^{re}-E_{ax}^{Cl,re} \right)+g_{ax}^{Cl}\left(\Delta V_{ax} -\Delta E_{ax}^{Cl} \right) \approx g_{ax}^{Cl}\Delta V_{ax},
\end{equation*}
since both $V_{ax}^{re}-E_{ax}^{Cl,re}$ and $ \Delta E_{ax}^{Cl}=o\left(\Delta V_{ax}\right)$. In the next, we provide the estimation of the cumulative $\mathrm{K}^{+}$ flux through axon membrane during a single action potential.
The governing equation of $\phi_{ax}$ yields  
\begin{equation}
\label{phi_eq}
\sum_{i}^{}z^{i}e \frac{\partial }{\partial z} \left( \eta_{ax} j_{ax}^{i} \right) = -\mathcal{M}_{ax} \left( I_{ax}^{Na,dy}+I_{ax}^{K,dy}+I_{ax}^{Cl,dy}\right).    
\end{equation}
At every location of the stimulated region, the duration of a single action potential is $T_{ax}^{*}$. We introduce  $T_{all}^{*}$ for the electrical signal propagation time, during which the signal propagates from one end of the axon (near the the optic nerve head) to the other end (far-eye-side of the optic nerve) as shown in Fig. \ref{fig:axon_potential}. By integrating right-hand side of Eq. (\ref{phi_eq}) over space $[0,L]$ and time $[0,T_{all}^{*}]$, we have 
\begin{eqnarray}
\label{current_integation}
-\mathcal{M}_{ax} &&\int_{0}^{T_{all}^{*}}\int_{0}^{L}I_{ax}^{Na,dy}+I_{ax}^{K,dy}+I_{ax}^{Cl,dy} dzdt \nonumber \\
&&\approx -\mathcal{M}_{ax} L\int_{0}^{T_{ax}^{*}} I_{ax}^{Na,dy}+I_{ax}^{K,dy}+I_{ax}^{Cl,dy} dt.
\end{eqnarray}
where we use the propagation property of the action potential along $z$ direction, and only the axon firing period is taken into consideration. By integrating the left-hand side of Eq. (\ref{phi_eq}), we have   
\begin{equation}
\label{flux_integation}
\int_{0}^{T_{all}^{*}}\int_{0}^{L}\sum_{i}^{}z^{i}e \frac{\partial }{\partial z} \left( \eta_{ax} j_{ax}^{i} \right)dzdt= O\left( T_{all}^{*} e\eta_{ax} j_{ax}^{bd} \right).
\end{equation}
We assume that the characteristic time scale of $T_{all}^*$ equals $O (10^{-3})$.  The scale of ion flux  $j_{ax}^{bd}$ at left and right boundaries $(z=0,L)$ is dominated by  the diffusion term 
\begin{equation*}
j_{ax}^{bd} =O \left( D_{ax}^*  \frac{\Delta c^{*}_{ax} }{z^*} \right),
\end{equation*}
since the boundary conditions are $\frac{\partial \phi_{ax}}{\partial z} \big{\vert}_{z=0,L}=0 $ and  $u_{ax}(0)=u_{ax}(L)=0$. The $\Delta c^{*}_{ax}$  is the characteristic difference between ion concentration at boundary value and the ion concentration inside the axon after a single action potential.  Based on the $\mathrm{Na^+}$ flux estimation in Eq. (\ref{Na_flux}), we estimate  $\Delta c^{*}_{ax}=O(10^{-1})$. From Eqs. (\ref{Na_flux}) and  (\ref{Cl_flux}), we get the following order of  cumulative fluxes through axon membrane during a single action potential time interval
\begin{eqnarray}
\label{comapre_1}
O\left(T_{all}^{*}  \eta_{ax} j_{ax}^{bd*} \right)&\ll O\left( \mathcal{M}_{ax} L \bigg{\vert}  \int_{0}^{T_{ax}^*}  J_{ax}^{m,Cl,dy} dt \bigg{\vert} \right)
\nonumber \\
&\ll O\left(\mathcal{M}_{ax} L \bigg{\vert} \int_{0}^{T_{ax}^*}  J_{ax}^{m,Na,dy}  dt \bigg{\vert} \right).
\end{eqnarray}
In other words, based on Eqs. (\ref{current_integation}),  (\ref{flux_integation}) and  (\ref{comapre_1}), it yields
\begin{equation}
\label{leading_order2}
O\left(\bigg{\vert} \int_{0}^{T_{ax}^{*}}  J_{ax}^{m,K,dy} dt \bigg{\vert}\right)= O\left(\bigg{\vert} \int_{0}^{T_{ax}^{*}}  J_{ax}^{m,Na,dy} dt \bigg{\vert}\right).          
\end{equation}
Based on Eq. (\ref{Na_flux}),  the cumulative axon transmembrane $\mathrm{K^+}$ flux during a single action potential should be
\begin{equation}
\label{leading_order}
\int_{0}^{T_{ax}^{*}}  J_{ax}^{m,K,dy} dt  \approx 2\times 10^{-9} \ \mathrm{mol/m^{2}}.           
\end{equation}
where $[0,T_{ax}^{*} ]$ is the time interval enclosing a single action potential.
\begin{remark}
	\normalfont
	Eq. (\ref{leading_order2})  shows that for a single action potential, the leading order of the cumulative $\mathrm{K}^{+}$ flux out of the axon to the extracellular space equals the leading order of the  cumulative $\mathrm{Na}^{+}$ flux into the axon from the extracellular space. This estimation is consistent with  observations in the literature \cite{ostby2009astrocytic,keynes1951ionic,dietzel1982stimulus}.
\end{remark}
Next, we estimate the concentration variation in the stimulated extracellular region due to a single action potential. The time scale $t^*$ of a single action potential is in milliseconds and during action potential the scale of $g^*_{ax}$ is $\bar{g}^{Na}$. In Appendix \ref{nerst_axon}, the scale of axon membrane potential $\Delta V^{*}_{ax}$ is
\begin{equation*}
\frac{k_{B}T}{\Delta V^{*}_{ax}e}=o(1).
\end{equation*}
Therefore, in Eq. (\ref{ion_ex_nd}) by taking $
\delta^{i}_{10}=\frac{t^*\mathcal{M}_{ax}\bar{g}^{Na}\Delta V^{*}_{ax}}{c_{ax}^{i,*}e}$ ,
we have 
\begin{equation*}
\left\{\frac{\delta^{i}_{13}\delta^{i}_{8}}{\delta^{i}_{12}\delta^{i}_{10}}, \frac{\delta^{i}_{14}}{\delta^{i}_{12}\delta^{i}_{10}} \right\} \subset o(1).
\end{equation*}

Hence, the cumulative ion fluxes through axon transmembrane are the main source changes the ion concentration in the stimulated extracellular region, 
\begin{equation}
\label{variation_C_1}
\eta_{ex} \Delta c_{ex}^{i}= \mathcal{M}_{ax}  \int_{0}^{T_{ax}^{*}}  J_{ax}^{m,i,dy}  dt,   \ \  i=\mathrm{Na^+,K^+},        
\end{equation}
where $\Delta c_{ex}^{i}$ is the $i$th ion's concentration variation from its resting state and $\eta_{ex}$ is unchanged by Eqs. (\ref{fluid_nd_gl}) and (\ref{fluid_nd_ax}) under time scale $t^*=10^{-3}\mathrm{s}$. 
Based on Eqs. (\ref{leading_order2}) and  (\ref{variation_C_1}), the absolute variation of $\mathrm{Na^+}$  and $\mathrm{K^+}$ concentrations in the stimulated extracellular region due to action potentials, can be written as
\begin{equation}
\label{C_variation}
\Delta c_{sti}= O\left( \frac{\mathcal{M}_{ax}}{\eta_{ex} }  \bigg{\vert}\int_{0}^{T_{ax}^{*}}  J_{ax}^{m,i,dy}  dt \bigg{\vert}\right),\ i=\mathrm{Na^+,K^+}.     
\end{equation}
In the following discussion, we use $ \Delta c_{sti}$  describes the concentration changes in the stimulated extracellular space after a single action potential,
\begin{equation}
\Delta c_{sti}=0.12 \ \mathrm{mM}.
\end{equation}

\subsection{Estimation of glial transmembrane potassium flux}
In this section, we estimate the glial transmembrane current  when the $\mathrm{K}^{+}$ and  the $\mathrm{Na}^{+}$ concentration vary by $\Delta c_{sti}$ in the stimulated extracellular region.  We also find that the electric field $\phi_{gl}$ responds immediately to the glial $\mathrm{K}^{+}$  Nernst potential changes. In the stimulated region, the variation of extracellular electric potential  $\Delta \phi_{ex}$ is small  in compare to the variation of glial electric potential $\Delta \phi_{gl}$.

The dominant current through the glial membrane in the stimulated region is through the passive $\mathrm{K}^{+}$  channel, rather than the $\mathrm{Na}^{+}$  channel or the $\mathrm{Na/K}$ pump. At the same time, in the non-stimulated extracellular region, almost the same amount of $\mathrm{K}^{+}$ moves from the glial compartment to extracellular space. In other words,  both the glial cells and extracellular space in the non-stimulated region participate in the spatial buffering process to help potassium clearance \cite{smith2006physiological,dreier2015stroke}.

In the stimulated region, the Nernst potential for $\mathrm{K^+}$ across the glial membrane changes because of the additional potassium $\Delta c_{ex}^K$ in the extracellular space, 
\begin{equation}
\label{Nernst_K}
\Delta E_{gl}^{K}= \frac{k_B T}{z^{K} e} \left( \log\left(1+\frac{\Delta c_{ex}^K}{c_{ex}^{K,re}} \right)- \log\left(1+\frac{\Delta c_{gl}^K}{c_{gl}^{K,re} } \right) \right),   
\end{equation}
where $\Delta c_{l}^{K},l=gl,ex$ are the variations of concentrations in the $l$ compartment. The variation of  $\mathrm{K}^{+}$  concentration in the glial compartment $\Delta c_{gl}^K$ is a result of the $\Delta c_{ex}^K$ produced by the glial transmembrane $\mathrm{K}^{+}$  flux. Recall that the volume fraction $(\eta_{gl})$ of the glial compartment is much larger than the extracellular space $(\eta_{ex})$. At same time, based on Eq. (\ref{C_variation}) and   $\mathrm{K}^{+}$ concentration at resting state, we get
\begin{equation*}
\Delta c_{ex}^{K}= o\left( c_{ex}^{K,re}\right),
\quad
\frac{\Delta c_{gl}^{K} }{c_{gl}^{K,re}} = o\left(\frac{\Delta c_{ex}^{K}}{ c_{ex}^{K,re}}\right).
\end{equation*}
Therefore, $\Delta E_{gl}^{K}$  in Eq. (\ref{Nernst_K}) can be approximated by its Taylor expansion,
\begin{equation}
\label{Nernst_K_variation}
\Delta E_{gl}^K\approx \frac{k_{B} T}{z^{K} e}  \frac{\Delta c_{ex}^K}{c_{ex}^{K,re}}.          
\end{equation}
The variation of $\mathrm{K}^{+}$ Nernst potential in the stimulated region produces  the changes of  glial membrane potential $\Delta V_{gl}$ and  glial compartment electric potential $\Delta \phi_{gl}$. We move on now to estimate the variations of electric potentials in the stimulated extracellular and glial regions. 

From the governing equation for $\phi_{ex}$,
\begin{eqnarray}
\label{phi_ex_governing}
\sum_{i}z^{i}e\nabla \cdot \left(\eta_{ex} \mathbf{j}_{ex}^{i}\right)
&&=\sum_{i} z^{i}e \mathcal{M}_{gl} \left( J_{p,gl}^{i}+ J_{c,gl}^{i} \right) \nonumber \\
&&+\sum_{i}z^{i}e \mathcal{M}_{ax} \left( J_{p,ax}^{i}+ J_{c,ax}^{i}\right),
\end{eqnarray}
where 
\begin{equation*}
\mathbf{j}_{ex}^{i} = c_{ex}^{i} \mathbf{u}_{ex} -D_{ex}^{i}\tau_{ex} \left(\nabla c_{ex}^{i} +\frac{z^{i}e}{k_{B}T} c_{ex}^{i} \nabla\phi_{ex} \right).
\end{equation*}
We claim that after the axon stops firing,  the major current is through glial membrane   $\mathrm{K^+}$ channels (see Appendix \ref{pump_increment}). Therefore, the right-hand side of Eq. (\ref{phi_ex_governing}) can be approximated as 
\begin{eqnarray}
\label{transm_approx}
\sum_{i} z^{i}e \mathcal{M}_{gl} \left( J_{p,gl}^{i}+J_{c,gl}^{i} \right)&&+\sum_{i}z^{i}e \mathcal{M}_{ax} \left(J_{p,ax}^{i}+ J_{c,ax}^{i}\right) \nonumber \\
\approx &&\mathcal{M}_{gl} g_{gl}^{K} \left(\Delta V_{gl} -\Delta E_{gl}^{K} \right).
\end{eqnarray}
Next, we integrate Eq. (\ref{phi_ex_governing}) over the stimulated region $V_{S}=\{(r,z,\theta)\vert r\in [0,r_{sti}], \ z\in[0,L], \  \theta\in [0,2\pi]\}$, through which the action potential propagates as shown in Fig. \ref{fig:stimulated_region}. By Eq. (\ref{transm_approx}), we have the approximation of the total current
\begin{equation}
\label{glia_K_chanel}
\int_{V_{S}} \mathcal{M}_{gl} g_{gl}^{K} \left(\Delta V_{gl} -\Delta E_{gl}^{K} \right) dv \approx \pi r_{sti}^{2} L \mathcal{M}_{gl} g_{gl}^{K} \left(\Delta V_{gl} -\Delta E_{gl}^{K} \right).               
\end{equation}
{black}{In} the left-hand side of Eq. (\ref{phi_ex_governing}), by the charge neutrality assumption in Eq. (\ref{Charge_nuetrality}), we naturally have
\begin{equation*}
\sum_{i}z^{i}e c_{ex}^{i} \mathbf{u}_{ex} =0.
\end{equation*}
Based on Eqs. (\ref{Cl_flux}),  (\ref{leading_order2}) and  (\ref{C_variation}), we know that after a single action potential the leading order of ion concentration variations in the stimulated extracellular region are as follows
\begin{equation}
\label{order_C}
\Delta c_{ex}^{Na}= -\Delta c_{sti}, \ \   \Delta c_{ex}^{K}  = \Delta c_{sti}, \ \ 
\Delta c_{ex}^{Cl} =o\left(\Delta c_{sti}\right).
\end{equation}
Using Eqs. (\ref{order_C}) and (\ref{phi_bd}), the diffusion term in left-hand side of Eq. (\ref{phi_ex_governing}) can be approximated as 
\begin{equation}
\label{sum_ion_flux_diff}
-\int_{V_{S}} \sum_{i}z^{i}e \nabla \cdot \left( \eta_{ex}D_{ex}^{i} \tau_{ex} \nabla c_{ex}^{i} \right) dv \approx  2\pi r_{sti} L e \eta_{ex}  D_{ex}^{\mathrm{diff}} \tau_{ex} \frac{\Delta c_{sti}}{r^{*}},
\end{equation}
where $D_{ex}^{\mathrm{diff}}=D_{ex}^{K}-D_{ex}^{Na}$.  In Eq. (\ref{sum_ion_flux_diff}), we claim that the currents through  the left  $(z=0)$ and right $(z=L)$ boundaries of the stimulated region $V_{S}$ is much smaller than those through the radial transition region $S_{T}$. This is because  (1) the ion concentration  variations are in radial direction (between stimulated region and non-stimulated region) and (2) the length scales in the $z$ and $r$ direction are different. Therefore, the radial transition region $S_{T} =\{(r,z,\theta )\vert r=r_{sti},z\in[0,L], \theta\in [0,2\pi]\}$  has much larger area than the left and right boundaries of $V_{S}$.

Similarly, the integration of the electric drift term  in left-hand side of Eq. (\ref{phi_ex_governing}) yields the approximation,
\begin{eqnarray}
\label{sum_ion_flux_phi}
&-\int_{V_{S}} \sum_{i} z^{i} e \nabla \cdot \left( \eta_{ex} D_{ex}^{i} \tau_{ex}  \frac{z^{i}e}{k_{B}T} c_{ex}^{i} \nabla\phi_{ex} \right)dv \nonumber \\
&\approx  2\pi r_{sti} L  \eta_{ex} \sigma_{ex} \frac{\Delta \phi_{ex}}{r^{*}},
\end{eqnarray}
where $\sigma_{ex}= \frac{ \tau_{ex} e^{2}  }{k_{B}T}\sum_{i} (z^{i})^{2} D_{ex}^{i} c_{ex}^{i}$. From Eqs. (\ref{glia_K_chanel}), (\ref{sum_ion_flux_diff}) and (\ref{sum_ion_flux_phi}), we get 
\begin{equation}
\label{ex_phi_approx}
\frac{2}{r_{sti}} \left( \frac{\eta_{ex} \tau_{ex} e D_{ex}^{\mathrm{diff}} }{\mathcal{M}_{gl} } \frac{\Delta c_{sti} }{r^{*}}  + \frac{\eta_{ex} \sigma_{ex} }{\mathcal{M}_{gl}}  \frac{\Delta \phi_{ex} }{r^{*}}  \right) \approx  g^{K}_{gl} \left( \Delta V_{gl} -\Delta E_{gl}^{K}  \right).  
\end{equation}
At the same time, from the governing equation of  $\phi_{gl}$                      \begin{equation}
\label{phi_gl_governing}
\sum_{i}z^{i}e\nabla \cdot \left(\eta_{gl} \mathbf{j}_{gl}^{i}\right)
=-\sum_{i} z^{i}e \mathcal{M}_{gl} \left( J_{p,gl}^{i}+ J_{c,gl}^{i} \right),
\end{equation}
where 
\begin{equation*}
\mathbf{j}_{gl}^{i} = c_{gl}^{i} \mathbf{u}_{gl} -D_{gl}^{i}\tau_{gl} \left(\nabla c_{gl}^{i} +\frac{z^{i}e}{k_{B}T} c_{gl}^{i} \nabla\phi_{gl} \right),
\end{equation*}
we obtain the following estimation in a similar way
\begin{equation}
\label{glia_phi_approx}
-\frac{2}{r_{sti} } \frac{\eta_{gl}\sigma_{gl} }{\mathcal{M}_{gl}} \frac{\Delta \phi_{gl}}{r^{*}} \approx  g_{gl}^{K} \left( \Delta V_{gl} -\Delta E_{gl}^{K}\right),
\end{equation}
where $\sigma_{gl}= \frac{ \tau_{gl} e^{2}  }{k_{B}T} \sum_{i} (z^{i})^{2} D_{gl}^{i} c_{gl}^{i}$.  We neglect the diffusion and convection terms in Eq. (\ref{phi_gl_governing}) because these terms require much longer time to respond to the extracellular concentration change.  Based on Eq. (\ref{ex_phi_approx}) and Eq. (\ref{glia_phi_approx}), we have 
\begin{equation}
\label{relation_phi}
\Delta \phi_{ex}=- \frac{\eta_{gl}  \sigma_{gl}}{\eta_{ex}  \sigma_{ex}}  \Delta \phi_{gl}-\frac{ \tau_{ex}eD_{ex}^{\mathrm{diff}}}{\sigma_{ex} }\Delta c_{sti}.
\end{equation}
In  Appendix \ref{phi_ex_and_phi_gl}, by matching the orders in both side of Eq. (\ref{glia_phi_approx}),  we claim that $\Delta \phi_{ex}  =o \left(\Delta \phi_{gl} \right)$ in the stimulated region and therefore,
\begin{equation}
\label{v_gland_phi_gl}
\Delta V_{gl} =\Delta \phi_{gl}-\Delta \phi_{ex} = O( \Delta \phi_{gl} ).
\end{equation}
In the next step, we approximate the  $\mathrm{K}^{+}$ current through the leaking  $\mathrm{K}^{+}$ channel  on the glial membrane. Based on Eqs. (\ref{glia_phi_approx}) and  (\ref{v_gland_phi_gl}), we get 
\begin{equation}
\label{deltaE_relation_1}
g_{gl}^{K}\left(\Delta \phi_{gl}-\Delta E_{gl}^{K}\right) \approx g_{gl}^{K} \left(\Delta V_{g l}-\Delta E_{gl}^{K}\right) \approx -  \frac{2\eta_{gl} \sigma_{gl}}{r_{sti} \mathcal{M}_{gl}}  \frac{\Delta \phi_{gl}}{r^{*}}.
\end{equation}
Hence, by Eq. (\ref{deltaE_relation_1}), we obtain the relation between $\Delta E_{gl}^{K}$ and $\Delta \phi_{gl}$  as
\begin{equation}
\label{deltaE_relation}
\Delta E_{gl}^{K} \approx \left(1+h_{\epsilon}\right) \Delta \phi_{gl}, 
\end{equation}
where 
\begin{equation*}
h_{\epsilon}=\frac{2 \eta_{gl}  \sigma_{gl} }{r_{sti}\mathcal{M}_{gl}r^{*}g_{gl}^{K}}.
\end{equation*}
Based on Eq. (\ref{deltaE_relation_1}), it gives us the following approximation
\begin{equation}
\label{K_current_eq}
g_{gl}^{K}\left(\Delta V_{gl}-\Delta E_{gl}^{K}\right)\approx -\frac{g_{gl}^{K} h_{\epsilon}}{1+h_{\epsilon}} \Delta E_{gl}^{K}.
\end{equation}
Furthermore, from Eqs. (\ref{relation_phi}), (\ref{deltaE_relation}) and  (\ref{Nernst_K_variation}), we get the approximation
\begin{equation}
\label{phi_ex_scale}
\begin{aligned}
\Delta \phi_{ex}  \approx - \frac{\eta_{gl} \sigma_{gl}k_{B} T}{\eta_{ex} \sigma_{ex}\left(1+h_{\epsilon}\right)z^{K}e} \frac{\Delta c_{ex}^{K}}{c_{ex}^{K,re}}.
\end{aligned}
\end{equation}
The variations of electric field $\Delta \phi_{gl}$  in both  stimulated and non-stimulated regions are produced without delay by  $\Delta E_{gl}^{K}$ in the stimulated region, as described in the governing equation of $\phi_{gl}$ in Eq. (\ref{phi_governing}). The $\mathrm{K}^{+}$ leaking current is the major current through the glial membrane in the non-stimulated region as it is in the stimulated region because the current through the ion channel is voltage $\phi_{gl}$ dependent and $\mathrm{K}^{+}$ conductance is one dominant ion conductance in the glial membrane
\begin{equation*}
g_{gl}^{i} = o \left( g_{gl}^{K} \right), \quad i=\mathrm{Na^{+},Cl^{-}}.
\end{equation*}
In the next steps, we introduce the superscript notation `$S$' for the stimulated region variables and superscript  `$NS$' for non-stimulated region ones. For the glial transmembrane currents, we have the following approximation
\begin{equation*}
\begin{aligned}
\sum_{i} z^{i}e\mathcal{M}_{gl}\left(J_{p,gl}^{S,i}+J_{c,g l}^{S,i}\right) & \approx  \mathcal{M}_{gl} g_{g l}^{K}\left(\Delta V_{gl}^{S}-\Delta E_{gl}^{S,K}\right), \\
\sum_{i} z^{i}e \mathcal{M}_{gl}\left(J_{p,gl}^{NS,i}+J_{c,g l}^{NS,i}\right) & \approx  \mathcal{M}_{gl} g_{g l}^{K}\left(\Delta V_{gl}^{NS}-\Delta E_{gl}^{NS,K}\right).
\end{aligned}
\end{equation*}
By integration of the $\phi_{gl}$ Eq. (\ref{phi_governing}) over the stimulated region $V_{S}$  and  the non-stimulated region $V_{NS}$ respectively, it yields
\begin{equation}
\label{equal_current}
\left\{\begin{aligned}
&\int_{V_{S}} \sum_{i} z^{i} \mathrm{e} \nabla \cdot\left(\eta_{gl}^{S} \mathbf{j}_{gl}^{S,i}\right) dv  \approx \int_{V_{S}}  \mathcal{M}_{gl} g_{g l}^{K}\left(\Delta V_{gl}^{S}-\Delta E_{g l}^{S, K}\right), \\
&\int_{V_{NS}} \sum_{i} z^{i} \mathrm{e} \nabla \cdot\left(\eta_{gl}^{NS} \mathbf{j}_{gl}^{NS,i}\right) dv  \approx \int_{V_{NS}}  \mathcal{M}_{gl} g_{g l}^{K}\left(\Delta V_{gl}^{NS}-\Delta E_{gl}^{NS, K}\right).
\end{aligned}\right. 
\end{equation}
Most of the current  between region $V_{S}$ and region $V_{NS}$ goes through the radial transition region $S_{T}$. By Eq. (\ref{equal_current}) and boundary conditions for $\phi_{gl}$ we obtain
\begin{eqnarray}
\label{equal_current_2}
\int_{V_{S}} \mathcal{M}_{gl} &&g_{gl}^{K}\left(\Delta V_{gl}^{S}-\Delta E_{gl}^{S,K}\right) dv \nonumber \\
&&\approx-\int_{V_{NS}}  \mathcal{M}_{gl} g_{gl}^{K}\left(\Delta V_{gl}^{NS}-\Delta E_{gl}^{NS, K}\right) dv.
\end{eqnarray}
{black}{Based on Eq. (\ref{equal_current_2}),}the average $\mathrm{K^+}$ flux through the glial membrane in the non-stimulated region leaks out to extracellular space with an approximate strength
\begin{equation}
\label{equal_current_3}
\frac{g_{gl}^{K}}{z^{K} e}\left(\Delta V_{gl}^{NS}-\Delta E_{gl}^{NS, K}\right)=-\frac{r_{sti}^{2}}{r^{* 2}-r_{sti}^{2}} \frac{g_{gl}^{K}}{z^{K} e}\left(\Delta V_{g l}^{S}-\Delta E_{gl}^{S,K}\right).
\end{equation}
In summary, Eq. (\ref{equal_current_2}) and Eq. (\ref{equal_current_3}), show how the glial compartment in the non-stimulated region serve as spatial buffers and help clear potassium from the extracellular space outside the stimulated axons $\cite{chen2000spatial}$.

\begin{remark}
	\normalfont
	The glial compartment serves as an important and quick potassium transport device to remove accumulated potassium during the axon firing as shown in Fig. \ref{fig::Patterns}.
	
	In the stimulated region, the change in the potassium Nernst potential change makes the glial membrane potential more positive and moves potassium through ion channels into the glial compartment. In the non-stimulated region, since glia is an electrical syncytium, the glial membrane potential simultaneously increases as it does in the stimulated region. However, the glia potassium Nernst potential in the non-stimulated region is not very different from that in the resting state. These potentials produce  an outward potassium flux from the glial compartment in the non-stimulated region. 
	
	Interacting regions of this sort depend on spatial variables and the properties of the glia as a syncytium. It is difficult to capture these effects in models that do not include space as an independent variable. Even if such compartment models capture these effects correctly in one set of conditions (because parameters are chose to make the description correct), they are unlikely to describe the effects of changes in conditions consistently, including membrane potential.
\end{remark}

\subsection{The water flow: circulation and estimation}
In this section, we discuss water circulation between the stimulated  and the non-stimulated regions. As  extra $\mathrm{K^+}$ is gradually cleared, it produces an osmotic pressure difference between the intra- and inter- domain, i.e., between the inside the glial compartment and the extracellular space. This osmotic pressure variation drives transmembrane water flow and  water circulation in the optic nerve.

Now we consider a train of stimulus stimulated with the frequency $f_{m}$ in the axon region $( r<r_{sti},z=z_{0})$ during time $[0, T_{sti}]$. The estimation depends on the  $\mathrm{K}^{+}$ and $\mathrm{Na}^{+}$  concentration variations in the extracellular space and charge neutrality condition. The clearance of extra amount of  $\mathrm{K^+}$ $(\Delta c_{ex}^{K})$  in the stimulated extracellular space mostly  goes through glial membrane and extracellular pathway (see Appendix \ref{Estimation_on_transport}), 
\begin{equation}
\label{dynamic_K_ex}
\frac{d\left(\eta_{ex} \Delta c_{ex}^{K}\right)}{d t}=-\left(\lambda_{gl}^{m,K}+\lambda_{ex}^{K}\right) \Delta c_{ex}^{K},
\end{equation}
where
\begin{equation*}
\lambda_{gl}^{m,K} = \frac{\mathcal{M}_{gl} g_{gl}^{K} h_{\epsilon} k_{B} T}{z^{K}\left(1+h_{\epsilon}\right) e^{2} c_{ex}^{K,re}}, \ \ \ \lambda_{ex}^{K}=\frac{2\eta_{ex}D_{ex}^{K} \tau_{ex}  }{r_{sti}r^{*}}. 
\end{equation*}
The $\lambda_{gl}^{m,K}$ presents the effect of  glial transmembrane $\mathrm{K}^{+}$  flux and the $\lambda_{ex}^{K}$ describes the spatial effect of the extracellular $\mathrm{K}^{+}$ transport between the stimulated region and non-stimulated region.  This spatial communication is not negligible since $\lambda_{ex}^{K}$  is  comparable magnitude to the $\lambda_{gl}^{m,K}$.  The initial value of Eq. (\ref{dynamic_K_ex}) starts with the first stimulus on axon as 
\begin{equation*}
\Delta c_{ex}^{K} (0)=\Delta c_{sti},
\end{equation*}
and at the beginning of each period $T$, there is an additional $\Delta c_{sti}$ amount of  $\mathrm{K^+}$  accumulated in the extracellular space due to the axon firing
\begin{equation*}
\Delta c_{ex}^{K}(iT)=\Delta c_{ex}^{K} (iT)+\Delta c_{sti}, \ \  i=1\dots n-1,
\end{equation*}
where $ n \left(=\frac{T_{sti}}{f_m} \right) $  is the total number of periods. 
In the above, we view the extracellular $\mathrm{K^{+}}$ concentration changes due axon firing as a source term $\Delta c_{sti}$.

\begin{remark}
	\normalfont
	The concentration in the stimulated extracellular region changes rapidly because of the transmembrane action potentials, as well as the extracellular electric potential $\phi_{ex}$. The effect of fluid circulation is the cumulative result of the above $\Delta O_{ex}$. The fluid flows from the non-stimulated region to the stimulated region are dominated by the trans-glia-membrane flow. So, the convection in the extracellular reduces (i.e., flattens) the variation of osmotic pressure.  
\end{remark}

\begin{remark}
	\normalfont
	These effects make our spatially inhomogeneous model quite different from existing ODE models \cite{ostby2009astrocytic,murakami2016mechanisms},  since those ODE models either take the extracellular ion concentration as constant or they do not consider the ion exchange between the extracellular space and other compartments at all. In a recent work, Marte J. et al \cite{saetra2020electrodiffusive} introduce a compartment model similar to Eq. \eqref{dynamic_K_ex} by considering ion flux between neuron, glia and extracellular regions in both the dendrite and soma region. It is always possible to take a field theory and approximate its $x$ dependence into compartments. But it is quite difficult to know how to describe the parameter dependence, and compartment inter-dependence in such models consistently.
	And it is probably impossible to describe the parameter dependence and compartment inter-dependence uniquely. These issue are also considered in the Discussion Section.
	
	Field theories show the interdependence as outputs of the analysis. Because field models are consistent, and their solutions are unique, parameter dependence and compartmental interdependence is unique. 
	
	In compartment models, different assumptions are possible and difficult to compare. Analysis with different sets of assumed compartments is likely then to give different results in the hands of different investigators, creating uproductive controversies, and slowing progress. Field models have many fewer assumptions and are more productive. However, they involve considerably more mathematical analysis \cite{xu2018osmosis,zhu2019bidomain} and numerical difficulties. Field models still contain many known parameters (e.g., most structural parameters, capacitance of membranes, conductivity of extra and intraellular solutions) and a number of not well known parameters, like the properties and distributions of membrane channels (and their ensemble properties) and active transport systems. Direct experimentation is the best way to determine these parameters and modern optical methods in particular allow many such measurements on scales much smaller than a cell diameter. But curve fitting to available data is often all that is possible, as in some cases in this paper, with its unavoidable ambiguities.
\end{remark}

The time course of  $\mathrm{Na}^{+}$ variation  $(\Delta c_{ex}^{Na})$ in the stimulated extracellular space is (see Appendix \ref{Estimation_on_transport})
\begin{equation}
\label{Na_dynimic}
\frac{d\left(\eta_{ex} \Delta c_{ex}^{Na}\right)}{d t}=-\lambda_{ex}^{Na, 1} \Delta c_{ex}^{Na}+\lambda_{ex}^{Na, 2} \Delta c_{ex}^{K},
\end{equation}
with the initial condition
\begin{equation*}
\Delta c_{ex}^{Na}(0) =-\Delta c_{sti}.
\end{equation*}
There is  $\Delta c_{sti}$ amount of  $\mathrm{Na^+}$ flux into axon compartment from the extracellular space at the beginning of each period
\begin{equation*}
\Delta c_{ex}^{Na}(iT)= \Delta c_{ex}^{Na} (iT)-\Delta c_{sti}, \ \  i=1 \dots n-1.
\end{equation*}
In Eq. (\ref{Na_dynimic}), the $\lambda_{ex}^{Na,1}$  describes the effect of extracellular diffusion and $\lambda_{ex}^{Na,2}$   presents the extracellular electric drift  between stimulated and non-stimulated regions. In Eq. (\ref{Na_dynimic}), we have
\begin{equation*}
\lambda_{ex}^{Na,1}=\frac{2\eta_{ex} D_{ex}^{Na} \tau_{ex} }{r_{sti}r^{*}}, \quad \lambda_{ex}^{Na, 2}=  \frac{2 \eta_{gl}  \sigma_{gl}D_{ex}^{Na}\tau_{ex} c_{ex}^{Na,re} }{r_{sti}\sigma_{ex}\left(1+h_{\epsilon}\right) r^{*}  c_{ex}^{K,re} }.
\end{equation*}
In  Appendix \ref{Estimation_on_transport}, we present the solution of the coupled linear system of (\ref{dynamic_K_ex}) and (\ref{Na_dynimic}). 
By the charge neutrality condition Eq. (\ref{Charge_nuetrality}), the variation of extracellular osmotic concentration  is 
\begin{equation}
\label{O_ex_nT}
\Delta O_{ex}= 2\left (\Delta c_{ex}^{K}+ \Delta c_{ex}^{Na}\right), 
\end{equation}
where $\Delta c_{ex}^{K}$ and $\Delta c_{ex}^{Na} $ are written in Eqs. (\ref{K_solution}) and  (\ref{Na_solution}).\\
Notice that sodium and potassium behave differently in the extracellular space. In the extracellular space, the electric drift $\mathrm{K^+}$ flux has a much smaller magnitude in comparison to diffusive $\mathrm{K^+}$ flux, since the scale ratio $R_{ex}^K$  between the electric drift term and  diffusion term for $\mathrm{K^+}$  is  (see Appendix \ref{Estimation_on_transport})
\begin{equation}
\label{R_ex_K}
R_{ex}^{K}=  \frac{\eta_{gl}  \sigma_{gl} }{\eta_{ex} \sigma_{ex} (1+h_{\epsilon} )}=o(1). 
\end{equation}
However, for  $\mathrm{Na^+}$  in the extracellular space, the magnitude of electric drift flux are comparable to diffusive flux since (see Appendix \ref{Estimation_on_transport})
\begin{equation}
\label{R_ex_Na}
R_{ex}^{Na}=\frac{\eta_{gl}  \sigma_{gl}}{\eta_{e x}  \sigma_{e x}\left(1+h_{\epsilon}\right)} \frac{c_{ex}^{Na}}{c_{ex}^{K}}=O(1).
\end{equation}

In the next discussion, we estimate the scales of the glial transmembrane velocity, glial  radial velocity, and extracellular radial velocity. The variation in osmotic pressure in the stimulated region is the driving force for the water flow and  circulation. Our estimation is based on the equations governing fluid flow and the spatial variation of osmotic pressure.

From the conservation of mass in glial compartment, we have
\begin{equation}
\label{Glial_water}
\frac{\partial \eta_{gl}}{\partial t}+\mathcal{M}_{gl} U^{m}_{gl}+\nabla \cdot\left(\eta_{gl} \mathbf{u}_{gl}\right)=0.
\end{equation}
Based on Eq. (\ref{O_ex_nT}), at $t=T_{sti}$,  we know there is cumulative  osmosis variation $\Delta O_{ex}(T_{sti})$  in the stimulated extracellular region. Since the glial compartment volume fraction ($\eta_{gl}$) is larger than the extracellular volume fraction  ($\eta_{ex}$), we have
\begin{equation*}
\vert \Delta O_{gl} \vert <\vert \Delta O_{ex} \vert.
\end{equation*}
Therefore, we view the $\Delta O_{ex}$ is the driving force for hydrostatic pressure variation. At the resting state,  Eq. (\ref{Glial_water}) yields
\begin{equation*}
\mathcal{M}_{gl} L_{gl}^{m}\left(p_{gl}^{re}-p_{ex}^{r e}-\gamma_{gl} k_{B} T\left(O_{gl}^{re}-O_{ex}^{r e}\right)\right)+\nabla \cdot\left(\eta_{gl}^{re} \mathbf{u}_{gl}^{re}\right)=0,
\end{equation*}
and by Eq. (\ref{Glial_water}), we get
\begin{eqnarray}
\label{delta_P_gl}
&\frac{\partial \Delta \eta_{gl}}{\partial t}+\mathcal{M}_{gl} L_{gl}^{m}\left(\Delta p_{gl}-\Delta p_{ex}-\gamma_{gl} k_{B} T\left(\Delta O_{g l}-\Delta O_{ex}\right)\right) \nonumber \\
&+\nabla \cdot\left( \Delta\left(\eta_{gl} \mathbf{u}_{gl}\right)\right)=0.
\end{eqnarray}
Based on Eq. (\ref{fluid_nd_gl}), the scale of the second term in Eq. (\ref{delta_P_gl})  is much larger than the third term, since
\begin{equation*}
\frac{\delta_{2}}{\delta_{1}}=\frac{ \kappa_{gl} \tau_{gl} }{\mu (r^{*})^2\mathcal{M}_{gl} L_{gl}^{m}}= o\left(1\right).
\end{equation*}
where we choose 
\begin{equation*}
U^{*}_{gl}= k_{B}TO^{*}, \ \ u^{*}_{gl}=\frac{\kappa_{gl}\tau_{gl}k_{B}TO^*}{\mu r^*}.
\end{equation*}
Therefore, Eq. (\ref{delta_P_gl}) in the stimulated glial region  can be approximated as  
\begin{eqnarray}
\label{p_difference_eq}
\frac{\partial\left(\Delta p_{gl}-\Delta p_{e x}\right)}{K_{gl} \partial t}&&+\mathcal{M}_{gl} L_{g l}^{m}\left(\Delta p_{gl}-\Delta p_{e x}\right)\nonumber \\
&&+\mathcal{M}_{gl} L_{gl}^{m} \gamma_{gl} k_{B} T \Delta O_{ex}=0,
\end{eqnarray}
with the initial condition
\begin{equation*}
\Delta \eta_{gl}(0)=\frac{\Delta p_{gl}(0)-\Delta p_{ex}(0)}{K_{gl}}=0.
\end{equation*}
In Eq. (\ref{p_difference_eq}),  we have used the relationship  between hydraulic pressures $p_{l}, \ l=gl,ex$ and glial compartment volume fraction $\eta_{gl}$  in Eq. (\ref{hydro_relation}) 
\begin{equation}
\label{eta_p}
K_{gl} \Delta \eta_{gl} = \Delta p_{gl}- \Delta p_{ex}.
\end{equation}
By using a linear approximation of  extracellular osmotic concentration variation $\Delta O_{ex}$
\begin{equation*}
\Delta O_{ex}(t)=  \frac{ \Delta O_{ex} (T_{sti} )}{T_{sti}}t,     \ \  t\in [0,T_{sti}],
\end{equation*}
the solution of $\Delta \left(p_{gl}-p_{ex}\right)$ in Eq. (\ref{p_difference_eq})  can be written as 
\begin{eqnarray}
\Delta p_{gl}(t)-\Delta p_{e x}(t)=&&\left(\frac{B t}{A} \exp (A t)-\frac{B}{A^{2}}(\exp (A t)-1)\right) \nonumber \\
&&\exp (-At)
\end{eqnarray}
where
\begin{equation*}
A=\mathcal{M}_{gl} L_{gl}^{m} K_{gl}, \quad B=-K_{gl} \mathcal{M}_{gl} L_{gl}^{m} \gamma_{gl} k_{B} T \frac{\Delta O_{ex}\left(T_{sti}\right)}{T_{sti}}
\end{equation*}
Hence, we estimate the average glial transmembrane water velocity in the stimulated region  as 
\begin{equation}
\label{u_gl_formula}
U_{gl}^{m}(t)=L_{gl}^{m}\left(\Delta p_{gl}(t)-\Delta p_{ex}(t)+\gamma_{gl} k_{B}T\Delta O_{ex}(t)\right), 
\end{equation}
and  the scale of glial transmembrane velocity in the  stimulated region as
\begin{equation}
U_{gl}^{*}=\left|U_{gl}^{m}(T_{sti})\right|.
\end{equation}
In Eq. (\ref{u_gl_formula}), the hydrostatic pressure variations $\Delta p_{l}, l=gl,ex$  passively react to the  osmotic pressure variation $k_{B}T\cdot \Delta O_{ex}$ in the stimulated region. Therefore, the direction of this glial transmembrane water flow is determined by osmotic pressure variation $k_{B}T\cdot\Delta O_{ex}$.

In the next step, we estimate the glial radial velocity scale $u_{gl}^{r*}$ and  extracellular radial velocity scale $u_{ex}^{r*}$. By the incompressibility  condition, we have
\begin{equation}
\label{compression}
\nabla \cdot\left(\eta_{gl} \mathbf{u}_{g l}\right)+\nabla \cdot\left(\eta_{ex} \mathbf{u}_{ex}\right)+\frac{\partial\left(\eta_{ax} u_{ax}^{z}\right)}{\partial z}=0.
\end{equation}

In Eq. (\ref{compression}), the dominant terms are the gradients in radial direction, because the length scale difference between $r^*$ and $z^*$ and the osmotic pressure variation are both in the radial direction. Therefore, Eq. (\ref{compression}) can be approximated by  
\begin{equation}
\label{radial_U}
\frac{\partial\left(\eta_{gl} u_{gl}^{r}\right)}{\partial r}+\frac{\partial\left(\eta_{ex} u_{e x}^{r}\right)}{\partial r}=0,
\end{equation}
The velocity boundary conditions at $r=0$,
\begin{equation*}
u^{r}_{gl}=u^{r}_{ex} = 0,
\end{equation*}
and Eq. (\ref{radial_U}) yield
\begin{equation}
\label{u_ex_and_u_gl}
\eta_{gl} u_{gl}^{r}+\eta_{ex}u_{ex}^{r}=0.     
\end{equation}
With the help of Eq. (\ref{u_ex_and_u_gl}),  we can rewrite $u_{gl}^r$ in form of
\begin{equation}
\label{u_gl_relation}
u_{gl}^{r}=(1-\chi) u_{gl}^{r}-\chi \frac{\eta_{e x}}{\eta_{gl}} u_{ex}^{r},
\end{equation}
where the $\chi$ is defined as 
\begin{equation*}
\chi=\frac{\kappa_{gl} \tau_{gl}}{\frac{\eta_{ex}}{\eta_{gl}} \kappa_{ex} \tau_{ex}+\kappa_{gl} \tau_{gl}}.
\end{equation*}
By substituting  Eqs. (\ref{def_u_gl}), (\ref{def_u_ex})  into Eq.  (\ref{u_gl_relation}), we estimate the radial velocity scale in the glial compartment as
\begin{eqnarray}
\label{U_gl_scale}
u_{gl}^{r*}=&&\left|(1- \chi) \frac{\kappa_{gl} \tau_{gl}}{\mu} \frac{\Delta p_{gl}-\Delta p_{ex}}{r^{*}}-(1- \chi) \frac{\kappa_{gl} \tau_{gl}}{\mu} \gamma_{gl} k_{B} T \frac{\Delta O_{gl}}{r^{*}} \right. \nonumber \\
&&\left. - \chi\frac{\eta_{ex}}{\eta_{gl}} k_{e} \tau_{ex} \frac{\Delta \phi_{ex}}{r^{*}}\right|_{t=T_{sti}}
\end{eqnarray}
In Eq. (\ref{U_gl_scale}), the $\Delta O_{gl}$  is due to the changes of the volume fraction  of the glial compartment $\Delta \eta_{gl}$ (see Remark \ref{Remark_5}) can be estimated as
\begin{equation*}
\Delta O_{gl} \approx \frac{\eta_{gl}^{re} }{\eta_{gl}^{re}+\Delta \eta_{gl}}O_{gl}^{r e}-O_{gl}^{r e}=-\frac{\Delta \eta_{gl}}{\eta_{gl}^{re}+\Delta \eta_{gl}} O_{gl}^{re},
\end{equation*}
where $\Delta \eta_{gl}$ can be written by using the $\Delta p_{l}$ as in Eq. (\ref{eta_p})
\begin{equation*}
\Delta \eta_{gl}=\frac{\Delta p_{gl} -\Delta p_{ex}}{K_{gl}}. 
\end{equation*}
Furthermore, by Eq. (\ref{u_ex_and_u_gl}), the scale of radial direction extracellular region velocity scale $(u_{ex}^{*})$  given  by 
\begin{equation}
\label{U_ex_scale}
u_{ex}^{*}=\frac{\eta_{gl}}{\eta_{ex}}  u_{gl}^{*}.   
\end{equation}

\begin{figure}[h]
	\centering
	\hspace*{-0.6cm}
	\includegraphics[width=3.25in,height=5cm]{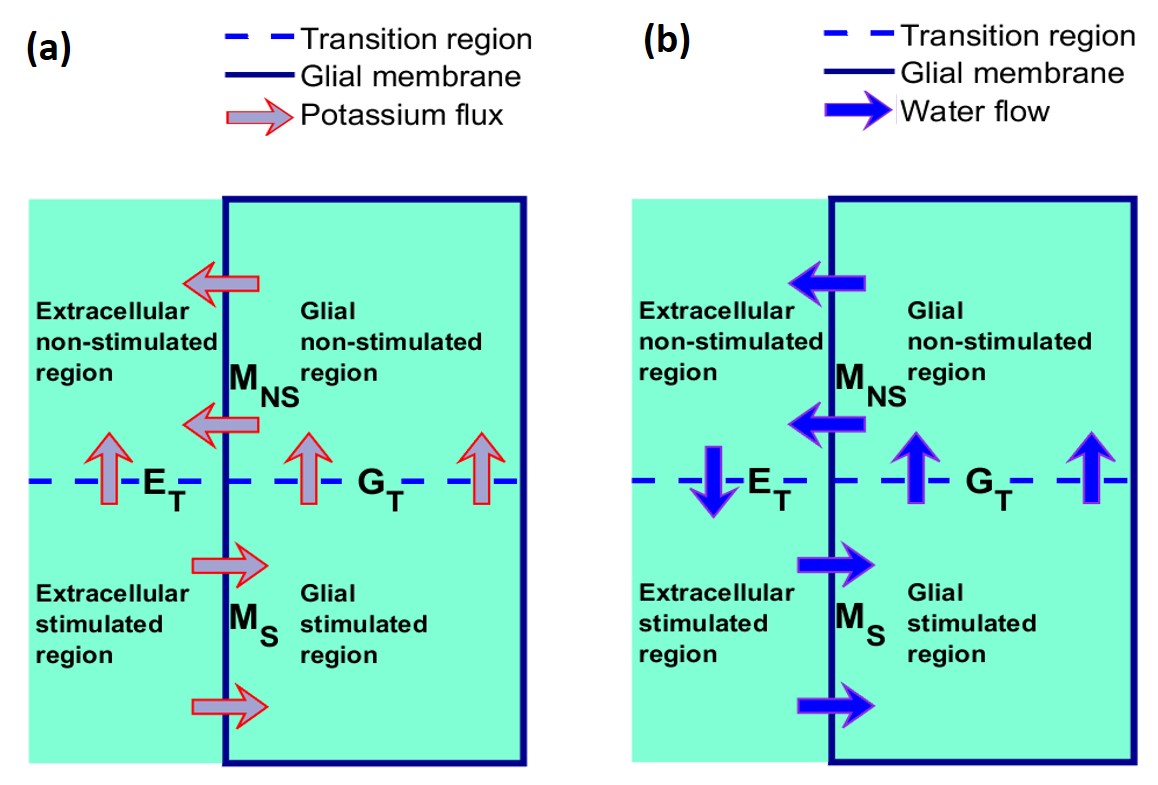}
	\caption{(a) Schematic graph of the potassium flux when inner part axon stimulated. In the stimulated region, the potassium takes the way of extracellular pathway and through the glial compartment via glial membrane. In the non- stimulated region, the potassium leaks out to the extracellular space through the glial membrane. (b) Schematic graph of the water circulation when inner part axon stimulated. In the stimulated region, the glial transmembrane water flow goes from extracellular space into glial compartment as the effect of osmosis difference. In the extracellular space, water goes from non-stimulated region to stimulated region in radial direction. In the glia compartment goes in the opposite direction. This compartment drawing is given only to aid qualitative understanding. }
	\label{fig::Patterns}
\end{figure}

Fig. \ref{fig::Patterns}b shows that the water flow exhibits circulation patterns between the extracellular space and glial compartment. The water flow in the glial compartment is from the stimulated region to the non-stimulated region in the radial direction. In extracellular space, the water flow in the radial direction is from the non-stimulated region to stimulated region.

\begin{remark}
	\label{Remark_5}
	\normalfont
	We assume the average total number of molecules (not concentration) in the stimulated glial region does not change since the major glial transmembrane ion flux in the stimulated region is  $\mathrm{K^+}$ flux  and  this $\mathrm{K^+}$ flux  from the stimulated extracellular space moves through the glial transition $S_t$  to the non-stimulated  extracellular space as Eq. (\ref{equal_current_2}). 
\end{remark}

\subsection{The relative importance  of ion flux components}
In this section, we discuss the relative importance of ion flux components, due to diffusion, convection, and electric drift in the glial and extracellular regions, respectively. Our discussion focuses on the radial direction since these are the dominant fluxes.

In the extracellular space, we characterize the relative importance of electric drift and diffusion (of potassium and sodium) in the extracellular space by the ratios $R_{ex}^{K}$ and $R_{ex}^{Na}$ analyzed in Eq. (\ref{R_ex_K}) and Eq. (\ref{R_ex_Na})
\begin{equation*}
\label{R_ex}
R_{ex}^{K}=\left| \frac{\eta_{gl}  \sigma_{gl}}{\eta_{ex}  \sigma_{ex}\left(1+h_{\epsilon}\right)}\right|, \quad R_{ex}^{Na}=\left| \frac{\eta_{gl}  \sigma_{gl}}{\eta_{ex} \sigma_{ex}\left(1+h_{\epsilon}\right)} \frac{c_{ex}^{N a}}{c_{ex}^{K}}\right|.
\end{equation*}
For radial direction flux, the  ratio between convection and diffusion in the extracellular space is estimated by the Peclet number shown in Eq. (\ref{Peclet}) 
\begin{equation}
\label{Pe_ex}
Pe_{e x}^{i}=\left|\frac{ c_{ex}^{i}u_{ex}^{*} r^{*}}{D_{ex}^{i} \tau_{ex} \Delta c_{ex}^{i}}\right|, \quad  i=\mathrm{Na}^{+}, \mathrm{K}^{+},
\end{equation}
where we approximate radial diffusion flux scale in the extracellular space as
\begin{equation*}
\left|D_{ex}^{*} \tau_{ex} \frac{\Delta c_{ex}^{i}}{r^{*}}\right|, \quad i=\mathrm{Na}^{+}, \mathrm{K}^{+}.
\end{equation*}
In a similar way, we estimate the Peclet numbers shown in Eq. (\ref{Peclet})  in the glial compartment as
\begin{equation}
\label{pe_gl}
Pe_{gl}^{i}=\left|\frac{c_{gl}^{i} u_{gl}^{*} r^{*}}{D_{gl}^{*} \tau_{gl} \Delta c_{gl}^{i}}\right|, \quad i=\mathrm{Na}^{+}, \mathrm{K}^{+}.
\end{equation}
Note that the Peclet numbers for $\mathrm{Na^+}$ and $\mathrm{K^+}$ are significantly different due to their different  concentrations as shown in Eqs. (\ref{Pe_ex}) and (\ref{pe_gl}).  In the glial compartment, the  ratio between electric drift and diffusion is  
\begin{equation}
\label{R_gl}
R_{gl}^{K}=\left|\frac{1}{1+h_{\epsilon}} \frac{c_{gl}^{K} \Delta c_{ex}^{K}}{c_{ex}^{K} \Delta c_{gl}^{K}}\right|, \quad R_{g l}^{Na}=\left|\frac{1}{1+h_{\epsilon}} \frac{c_{gl}^{Na} \Delta c_{ex}^{K}}{c_{ex}^{K} \Delta c_{gl}^{Na}}\right|. 
\end{equation}
where we have used Eqs. (\ref{Nernst_K_variation}) and  (\ref{deltaE_relation}).  In Eq. (\ref{R_gl}), we  estimate the $\mathrm{K}^{+}$ concentration change $(\Delta c_{gl}^{K})$  in the stimulated glial compartment as
\begin{equation}
\label{glial_K}
\Delta c_{gl}^{K} \approx\left(n c_{sti}-\Delta c_{ex}^{K}\right) \frac{\lambda_{gl}^{m, K}}{\lambda_{gl}^{m,K}+\lambda_{ex}^{K}} \frac{\eta_{ex}}{\eta_{gl}},
\end{equation}
where $\lambda_{gl}^{m,K}$ and $\lambda_{ex}^{K}$ are defined in Eq. (\ref{dynamic_K_ex}), and $n$ is the number of stimuli.\\
We estimate the $\Delta c_{gl}^{Na}$ in the stimulated glial  compartment as  
\begin{equation}
\label{glial_Na}
\Delta c_{gl}^{Na} \approx-\frac{3 \Delta I_{gl}}{g_{gl}^{K}\left(\Delta V_{gl}-\Delta E_{gl}^{K}\right)} \Delta c_{gl}^{K},
\end{equation}
where $\Delta I_{gl}$ are approximated by Taylor expansion as 
\begin{equation*}
\Delta I_{gl} \approx 2\left(\frac{\mathrm{K}_{\mathrm{K} 1} I_{gl}^{r e, 1}}{c_{ex}^{K,re}\left(c_{ex}^{K,re}+K_{K1}\right)}+\frac{\mathrm{K}_{\mathrm{K} 2} I_{gl}^{re,2}}{c_{ex}^{K, re}\left(c_{ex}^{K,re}+K_{K2}\right)}\right) \Delta c_{ex}^{K}.
\end{equation*}
In the next section, we carry out a numeric simulation as mentioned previously. Furthermore, we compare the results between the electrodiffusion model with the convection-electrodiffusion (full) model.

\section{Numerical simulation\label{sec: Numerical simulation}}
In this section,  numerical simulations are used to confirm our asymptotic estimations.   The comparison between electrodiffusion model and the full convection-electrodiffusion model is   conducted  to understand how the nervous (neuron-glia) system interacts with the extracellular space to create microcirculation.

A train of stimuli is applied to stimulate the axon membrane near the left boundary $(\{(z_0,r) \vert z_0=\textcolor{black}{1.875} \  \mathrm{mm}  \ \mathrm{and} \  r<r_{sti}= \frac{1}{2} r^{*}= 24 \  \mathrm{\mu m} \})$. Each single stimulus has current strength $I_{sti}=3\times 10^{-3} \ \mathrm{A/m^2} $ with duration  $3\ \mathrm{ms}$. The frequency of the stimuli is $50 \ \mathrm{Hz} \ (T=0.02 \ \mathrm{s})$  and  the duration is $T_{sti}=0.2 \ \mathrm{s}$.  
The obtained full model is solved by using Finite Volume Method with mesh size $h=1/20$ and temporal size $t =1/10$ in dimensionless.  The code is written in the Matlab environment. 
\subsection{Estimation of velocity scales}
We first estimate how large are the fluid velocities in extracellular space and glial compartment generated by a train of stimuli. From Eqs. (\ref{Na_solution}) and  (\ref{K_solution}), the estimated concentration variations in the stimulated extracellular region at $t=T_{sti}$ are
\begin{equation*}
\Delta c_{ex}^{Na} \approx-1.06\  \mathrm{mM}, \quad \Delta c_{ex}^{K} \approx 0.89\  \mathrm{mM}, \quad \Delta O_{ex} \approx-0.34 \ \mathrm{mM}.
\end{equation*}
The estimated glial transmembrane velocity by Eq. (\ref{U_gl_scale}) is
\begin{equation*}
U_{gl}^{ *} \approx 9.78 \times 10^{-2} \ \mathrm{nm} / \mathrm{s}.
\end{equation*}
From Eqs. (\ref{U_gl_scale}) and (\ref{U_ex_scale}), the estimated scale of radial water velocities inside glial compartment and extracellular space are 
\begin{equation*}
u_{ex}^{*} \approx 1.56 \times 10^{1} \ \mathrm{nm} / \mathrm{s}, \ \ \ u_{gl}^{*} \approx 3.90 \ \mathrm{nm} / \mathrm{s}.
\end{equation*}

\begin{figure}[h]
	\centering
	\includegraphics[width=3.25in,height=5cm]{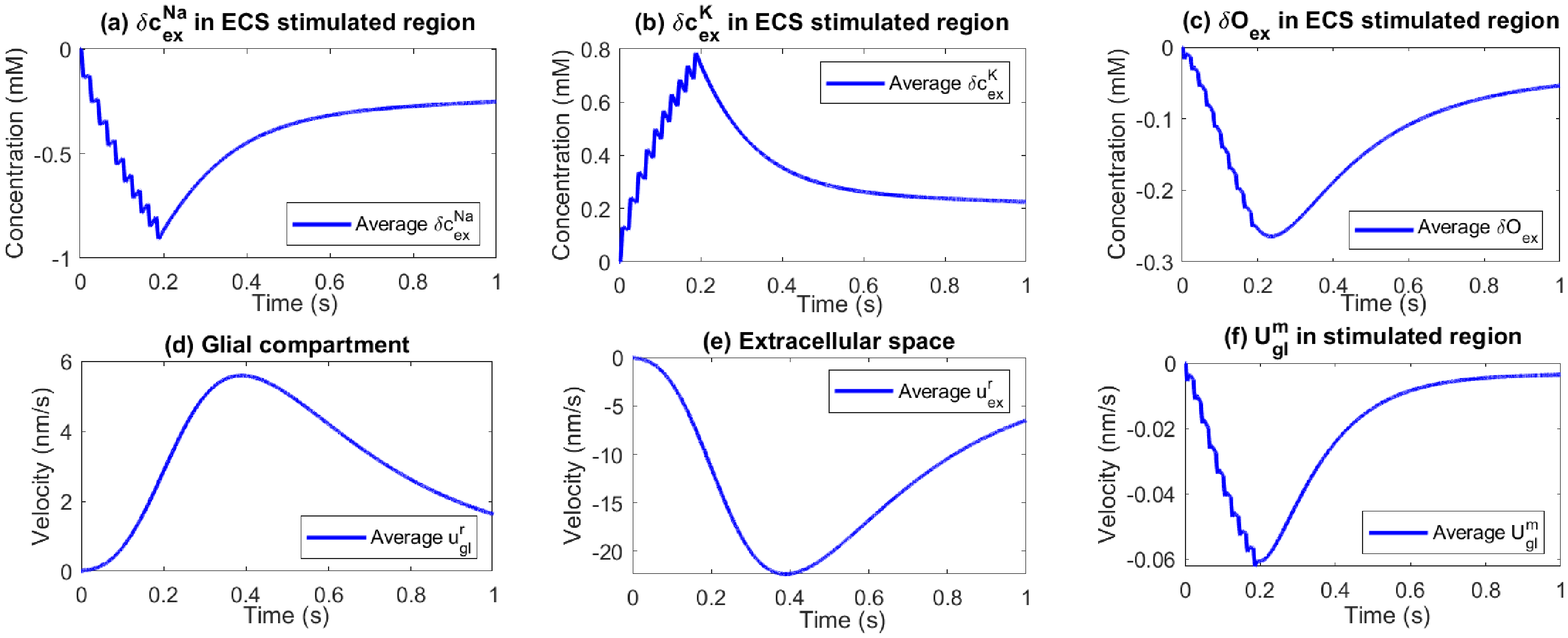}
	\caption{Numerical Results. (a-c) Average concentration variations in the stimulated extracellular region; (d-e) Average radial velocity in the intradomain; (f) Average glial transmembrane velocity in the stimulated region (with normal direction points to ECS).}
	\label{fig::Nermical_result}
\end{figure}

In Fig. \ref{fig::Nermical_result}a-c, we plot the computed average variation of concentrations in the stimulated extracellular region. These computed concentration changes are consistent with the estimates presented previously. The change of concentration reaches its peak at the end of the train of stimulus $(t=T_{sti})$ and quickly returns to its previous equilibrium value.  

In Fig.  \ref{fig::Nermical_result}f, we plot the computed average transmembrane water flow through the glial membrane in the stimulated region. We see Fig. \ref{fig::Patterns}b that water flows into the glial compartment from the extracellular space in the stimulated region. This transmembrane water flow generates the water circulation between the stimulated region and non-stimulated region in the radial direction. As in the Fig. \ref{fig::Patterns}b, in the extracellular compartment,  the water flow goes from the non-stimulated region to the stimulated region and in the glial compartment, water flows in the opposite (radial) direction. In the Fig.  \ref{fig::Nermical_result}d-e, we plot the computed average water velocity in the radial direction in the glial compartment and in the extracellular space. The computations are consistent with our estimation above.

\begin{figure}[h]
	\centering
	\includegraphics[width=3.25in,height=3cm]{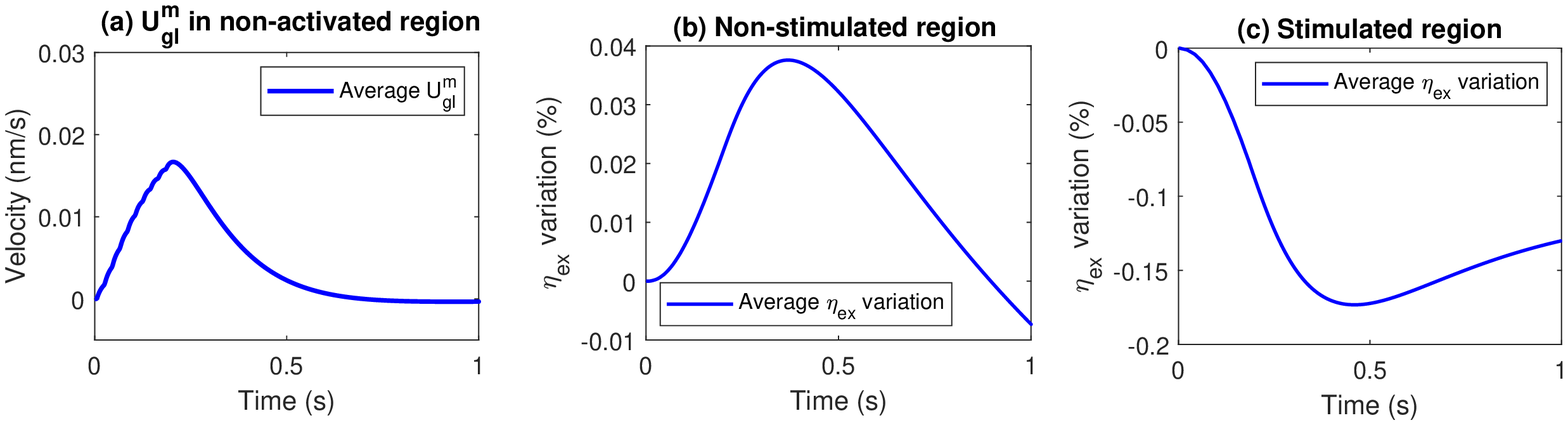}
	\caption{(a) Average glial transmembrane velocity in the non-stimulated region (the normal direction points from glial compartment to extracellular space.); (b-c) Average variation of the extracellular volume fraction in non- stimulated region and stimulated regions. }
	\label{fig::Nermical_result2}
\end{figure}

\begin{figure}[h]
	\centering
	\includegraphics[width=3.25in,height=9cm]{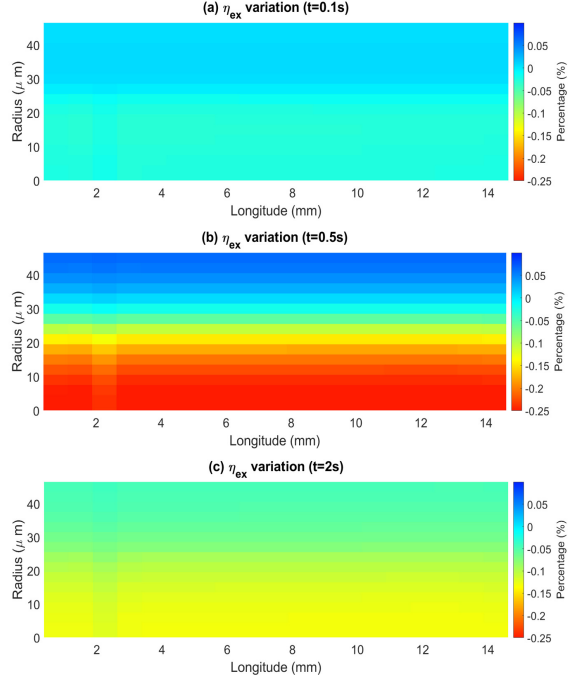}
	\caption{(a)-(c): Extracellular space volume fraction $(\eta_{ex})$ variation at time $t=0.1\mathrm{s},~0.5\mathrm{s},~2\mathrm{s}$. The blue is the enlarged region of extracellular space and red is the shrunken region of the extracellular space which is qualitatively  consistent with the results in Ref.~\cite{ holthoff2000directed,kofuji2004potassium}. \textcolor{black}{The stimulus current has been applied at $z_{0}=1.875 \mathrm{mm}$ as shown in Fig. \ref{fig:stimulated_region}, which induces ion concentration  and osmosis variation differ. The volume fraction changes depend on the hydrostatic pressure difference which involves the osmotic pressure (see Fig. \ref{fig:ox} in the Appendix).}}
	\label{fig::eta_ex_variation}
\end{figure}

\begin{figure}[!h]
	\centering
	\includegraphics[width=3.25in,height=9cm]{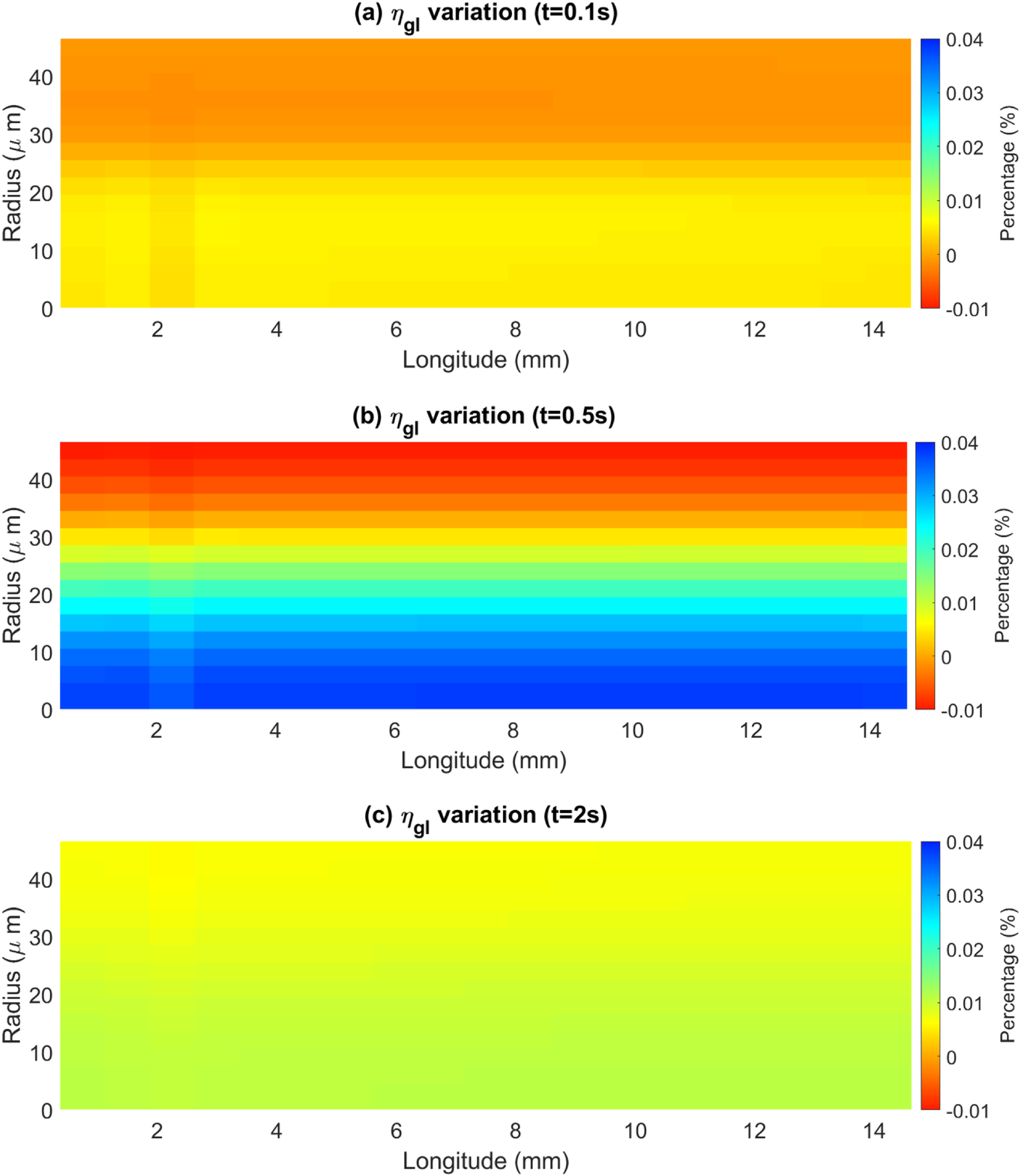}
	\caption{(a)-(c):Glial compartment volume fraction $(\eta_{gl})$ variation at time $t=0.1\mathrm{s},~0.5\mathrm{s},~2\mathrm{s}$.}
	\label{fig::eta_gl_variation}
\end{figure}

In the Fig. \ref{fig::Nermical_result2}a, we show the transmembrane water flow through the glial membrane in the non-stimulated region as in the Schematic Fig. \ref{fig::Patterns}b.  This water flow to the extracellular space produces widening of the extracellular space volume in the non-stimulated region, as shown in Fig. \ref{fig::Nermical_result2}b. At the same time, the extracellular space volume shrinks (in the stimulated region) as shown in Fig. \ref{fig::Nermical_result2}c. The shrinkage is produced by the inward water flow through the glial membrane in stimulated region, as in Fig. \ref{fig::Nermical_result2}f. In Fig. \ref{fig::eta_ex_variation} and Fig \ref{fig::eta_gl_variation},  the variations of volume fractions of the  extracellular space and glial compartment   in the whole domain are plotted at time $t = 0.1 \mathrm{s}$ (during the stimulus), $t=0.5 \mathrm{s}$ (maximum variations) and $t=2\mathrm{s}$ (back to resting state).    Our simulation is consistent with the experiments in references \cite{holthoff2000directed,kofuji2004potassium}, where the extracellular space becomes smaller in the middle cortical layers (where the stimulus is applied)  but widens in the most superficial and deep cortical layers (where no stimulus is applied).

\begin{remark}
	\normalfont
	\textcolor{black}{
		In Figs. \ref{fig::eta_ex_variation}-\ref{fig::eta_gl_variation}, it is an illusion that there are jumps in the contours of volume fractions for extracellular space and glial compartment. By checking  a line-plot at a fixed radius  $r=1.5\mathrm{\mu m}$, Fig.\ref{fig:z_direction_eta} in the Appendix  illustrates that  there are not jumps rather than local extreme values at  the $z_{0}=1.875 \mathrm{mm}$ where the stimuli are applied.  These stimuli result in the local potassium accumulation which decreases the osmosis variation in the extracellular space near $z_{0}$ (see Appendix Fig. \ref{fig:ox}). Therefore, less shrunken of the extracellular volume fraction near $z_{0}$ as Figs. \ref{fig::eta_ex_variation}-\ref{fig::eta_gl_variation} shown.}
\end{remark}

\subsection{Importance of convection}
In this section, we explore the importance of fluid convection during potassium clearance in each region. We first examine the estimated Peclet numbers for $\mathrm{Na^+}$ and $\mathrm{K^+}$ in the extracellular and glial compartments. 
By Eq. (\ref{Pe_ex}), the Peclet numbers (for the radial ion flux) in the extracellular space are
\begin{eqnarray*}
	Pe_{ex}^{K}&=\left|\frac{c_{ex}^{K} u_{ex}^{*} r^{*}}{D_{e x}^{K} \tau_{ex} \Delta c_{ex}^{K}}\right| \approx 1.0 \times 10^{-2},\\ \quad Pe_{ex}^{Na}&=\left|\frac{c_{ex}^{Na} u_{ex}^{*} r^{*}}{D_{ex}^{Na} \tau_{ex} \Delta c_{ex}^{Na}}\right| \approx 3.5 \times 10^{-1}.
\end{eqnarray*}
By Eqs. (\ref{R_ex_K}) and (\ref{R_ex_Na}), the  ratios between electric drift and diffusion (of the radial ion flux) in the extracellular space  are
\begin{eqnarray*}
	&R_{ex}^{K}= \left|\frac{\eta_{gl}  \sigma_{gl}}{\eta_{ex}  \sigma_{ex}\left(1+h_{\epsilon}\right)}\right|\approx 6.2 \times 10^{-2}, \\
	&R_{ex}^{Na}=\left|\frac{\eta_{gl}  \sigma_{gl}}{\eta_{ex} \sigma_{ex}\left(1+h_{\epsilon}\right)} \frac{c_{ex}^{N a}}{c_{ex}^{K}}\right|\approx 2.3.
\end{eqnarray*}

In the glial compartment, based on Eqs. (\ref{pe_gl}),  (\ref{glial_K}) and  (\ref{glial_Na}), we get the Peclet numbers (for the radial ion flux) in the glial compartment are
\begin{eqnarray*}
	&Pe_{gl}^{K}=\left|\frac{c_{gl}^{K} u_{gl}^{*} r^{*}}{D_{gl}^{K} \tau_{gl} \Delta c_{gl}^{K}}\right| \approx 2.9 \times 10^{1}, \\ &Pe_{gl}^{Na}=\left|\frac{c_{gl}^{Na} u_{gl}^{ *} r^{*}}{D_{gl}^{Na} \tau_{gl} \Delta c_{gl}^{N a}}\right| \approx 1.7 \times 10^{1}.
\end{eqnarray*}

By Eq. (\ref{R_gl}), the  ratios between electric drift and diffusion (of the radial ion flux) in the glial compartment  are
\begin{eqnarray*}
	&&R_{gl}^{K}=\left|\frac{1}{1+h_{\epsilon}} \frac{c_{gl}^{K} \Delta c_{ex}^{K}}{c_{ex}^{K} \Delta c_{gl}^{K}}\right| \approx 4.3 \times 10^{2}, \\ 
	&&R_{gl}^{N a}=\left|\frac{1}{1+h_{\epsilon}} \frac{c_{gl}^{N a} \Delta c_{ex}^{K}}{c_{ex}^{K} \Delta c_{gl}^{N a}}\right| \approx 1.7 \times 10^{2}.
\end{eqnarray*}

\begin{figure}[h]
	\centering
	\includegraphics[width=3.25in,height=6cm]{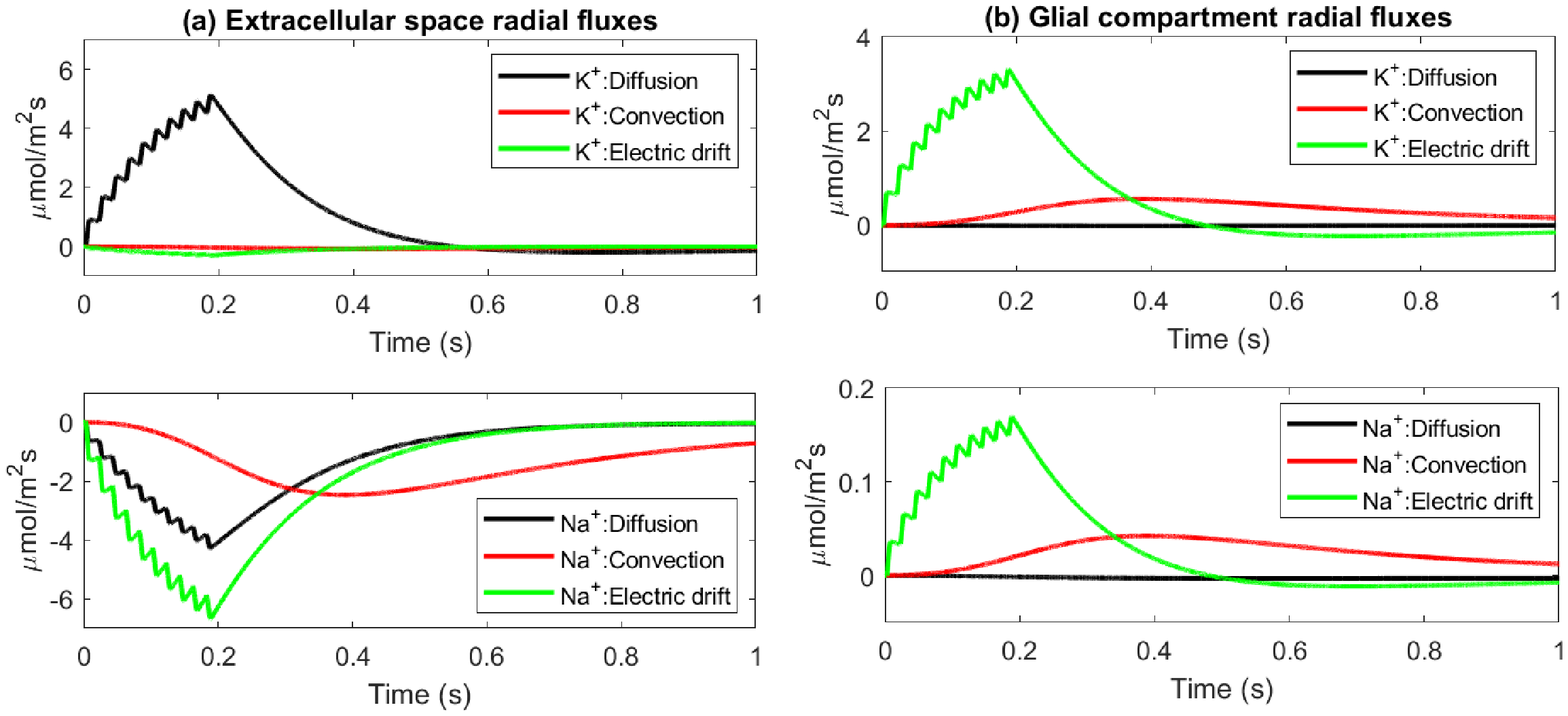}
	\caption{(a) Average radial direction fluxes components in the extracellular space; (b) Average radial direction fluxes components in the glial compartment (radial direction as normal direction).}
	\label{fig::Nermical_result3}
\end{figure}

In Fig. \ref{fig::Nermical_result3}, we plot the computed potassium and sodium fluxes (in the radial direction)  in the extracellular space and glial compartments .

In the extracellular space,  the importance of different fluxes are complicated because they depend on the ion species concentration as shown in Eq. (\ref{Pe_ex}). For potassium, the diffusion flux is dominant  as shown in Fig. \ref{fig::Nermical_result3}a upper panel. But for the sodium (Fig. \ref{fig::Nermical_result3}a lower panel),  the three fluxes, diffusion, convection, and electric drift, are comparable with the electric drift flux being somewhat larger. These simulation results agree with our estimations above. In the extracellular space, the potassium's Peclet  number $Pe_{ex}^{K}$ and the ratio $R_{ex}^{K}$ are in $O(10^{-2})$, while the sodium's Peclet  number  $Pe_{ex}^{Na}$ is order of $O(10^{-1})$ and the ratio $R_{ex}^{Na}$ is in $O(1)$.

In the glial compartments (Fig. \ref{fig::Nermical_result3}b), the situation is different from the extracellular space. The electric drift is dominant, and convection flux comes as second in importance for both sodium and potassium.  The water flow has a more important effect on potassium in the glial compartment than in the extracellular space. The maximum of the convection flux occurs after the stimuli, since it takes that long for osmotic pressure to accumulate.   Also, it lasts longer time when the effect of electric drift has diminished.

\begin{figure}[h]
	\centering
	\includegraphics[width=3.25in,height=3cm]{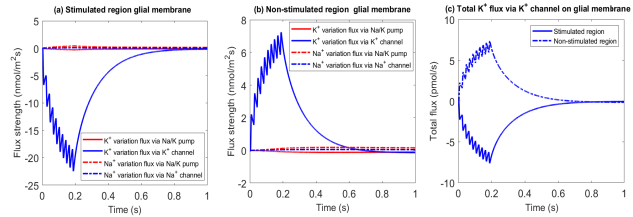}
	\caption{(a) Potassium and sodium flux variation through Na/K pump and ion channels on the glial membrane in the stimulated region; (b) Potassium and sodium flux variation through Na/K pump and ion channels on the glial membrane in the non-stimulated region. c: the total potassium flux through potassium channel on the glial membrane.}
	\label{fig::Nermical_result4}
\end{figure}

In the Fig. \ref{fig::Nermical_result4}a and \ref{fig::Nermical_result4}b, the potassium and sodium flux through the glial membrane are presented and the results are consistent with our estimates. The major current through the glial membrane is through the potassium channel in both stimulated region and non-stimulated region. Fig. \ref{fig::Nermical_result4}c compares the stimulated and non-stimulated region by showing the total potassium flux  through potassium channels (integrated over all the glial membrane). The total potassium flux has different direction in the stimulated region and non-stimulated region, as shown in our estimation in Eq. (\ref{equal_current_2}). The strength is the same, but the direction is different.

\begin{figure}[h]
	\centering
	\includegraphics[width=3.25in,height=4cm]{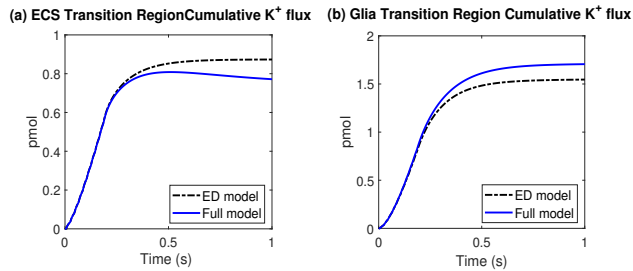}
	\caption{ (a) Cumulative $\mathrm{K^+}$ flux on extracellular transition region; (b) Cumulative $\mathrm{K^+}$ flux on glail transition region (radial direction as normal direction).}
	\label{fig::Nermical_result5}
\end{figure}

Fig. \ref{fig::Nermical_result5} compares the potassium flux in the electrodiffusion (ED) model and convection-electrodiffusion (full) model. In the full model, the water circulation between the stimulated and non-stimulated region in both extracellular and glial compartments have an important role in the circulation of potassium. The water circulation has an important role in buffering potassium in the optic nerve bundle. The water circulation increases the potassium flow through the glial compartment.

Fig. \ref{fig::Nermical_result5}b show  how water flow increases the potassium flux through the glia in the transition region between the stimulated and non-stimulated region. The potassium flux moves back to the stimulated extracellular region from non-stimulated extracellular region through the extracellular pathway, as shown in Fig. \ref{fig::Nermical_result5}a. The  time rate of change of the cumulative $\mathrm{K^+}$ flux through the extracellular transition region  decreases after stimulus.

\begin{figure}[h]
	\centering
	\includegraphics[width=3.25in,height=4cm]{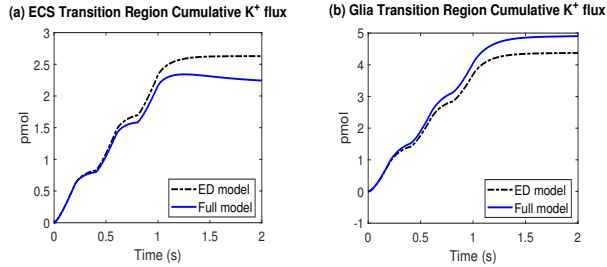}
	\caption{Multiple trains of action potentials. (a) Cumulative $\mathrm{K^+}$ flux on extracellular transition region; (b) Cumulative $\mathrm{K^+}$ flux on glail transition region (radial direction as normal direction).}
	\label{fig::Nermical_result6}
\end{figure}

Multiple trains of action potentials strengthen the effect of water flow on the transport through the glial compartment. In the Fig. \ref{fig::Nermical_result6}, three trains of action potentials occur with $0.2 \ \mathrm{s}$ resting period between each. Fig. \ref{fig::Nermical_result6}b shows that water flow increases $25\%$ of the amount of cumulative potassium flux through the transition region in the glial compartment, beyond the potassium flow in the electrodiffusion model. Consequently, the amount of cumulative potassium flux through the transition region in the extracellular space is around $15\%$ less than in the electrodiffusion model see Fig.  \ref{fig::Nermical_result6}a.

\section{Discussion\label{Discussion}}

Biological systems, like engineering systems, are complex, involving many components connected in specific structures, using a range of forces to perform specific functions, often that can be defined by quantitative measurements and relations. These systems are defined in textbooks of physiology and some in more mathematical detail elsewhere.

Many parameters are involved that need to be known if function is to be understood and predicted. What is not so well known is how these parameters are determined. In one extreme, the circuits of electronic devices all parameters—every one—are known by independent measurements. Curve fitting is not involved at all. Indeed, it is hard to imagine how a computer of some $10^{13}$   devices that interact with each other some $10^9$ times a second could function if parameters were not definite and known to the designer of the circuit. Thus, complexity in itself does not prevent definite understanding.

A crucial help in dealing with electronic circuits is the universal and exact nature of the Maxwell equations that govern electronic current flow in these structures. The same equations are true for biological systems for ions, but the mechanical response of the system to the charges and their movement when electric fields change (loosely called `polarization') is not so well known. Measurements of the physical and electrical structure of tissues is, however, sometimes possible giving some of the certainty to fortunate biological systems that the Maxwell-Kirchhoff equations bring to electronic systems.
It is natural to try to simplify the electrical and then the electrodiffusional and osmotic properties of biological tissues with compartment models, in which spatial variables and differential equations in space and time are replaced by compartments and ordinary differential equations in time. These compartments can be derived in some cases by well defined perturbation procedures (some of which we use here) but the accuracy of the perturbation scheme and reduced models is difficult to determine, to put it mildly, given the large number of parameters that affect that accuracy, particularly as conditions change. The compartments introduce a level of uncertainty that is hard to resolve and is likely to impede agreement among investigators and thus the progress of knowledge. 
In some fortunate cases, biological systems are known well. Then field equations can be written and solved that are general and quite independent of the choice of compartments, as we have tried to do here. The system of long cylindrical nerve fibers, ionic channels and membranes—particularly their capacitance—that conducts the signals (action potentials) of the nervous system is known quite well. Independent measurements of every component are available. Parameters can be measured of almost all components in several independent ways that give indistinguishable results. Thus action potential propagation can be computed with little ambiguity.

Some syncytial tissues are known almost this well. The lens of the eye has been studied by impedance spectroscopy and morphometry so the structure and structural parameters are well known. Flows have been directly measured and also pressure, sometimes with spatial dependence, in Mathias group more than anywhere else In the case of the lens, the biological system is nearly as well determined as the electronic system.
The optic nerve is not so well known. Here we have good structural information but limited knowledge of parameters. Membrane capacitance and extracellular and intracellular resistivities are known. Conductance of voltage activated channels and connexins is known but the spatial distribution of connexins and channels is not known, and even the identity of the channels is not known. Thus calibration of our optic nerve model is incomplete, as we have tried to explain in detail in the text. And so validation is limited as well. 
What is needed for calibration in the optic nerve more than anything else is experimental measurements of the type and spatial distribution of pumps and channels. What is needed for validation is experimental measurements of the spatial distribution of potentials, concentrations and pressures. The theory can easily be extended to compute those quantities not already included. 
Indeed, this process of calibration and validation is what is needed, in our view, to understand the role of water flow, ion migration and diffusion in other systems in the central nervous system. Understanding the glymphatic flows in the central nervous system requires a field theory in the spirit of that presented here. It requires calibration with the spatial distribution of pumps and channels. It requires validation by measurement of the spatial distribution of concentration, electrical potential and pressure. A validated and calibrated theory can then predict and understand the glymphatic flows so important in biological processes like sleep and pathological situations like migraine and epilepsy.

\section{Conclusion \label{Conclusion}}
This work provides a comprehensive set of estimates and computations, showing the water circulation in the optic nerve. The water flow is generated by the osmotic difference between the glial compartment and extracellular space. Through the estimation, we show that in the stimulated region, the extracellular osmotic changes are not induced by ion fluxes from the axon compartment when the axon is firing. Indeed, based on the analysis, we found that the leading order of potassium flux out and sodium flux into axon is the same during the action potential, which is consistent with the literature \cite{ostby2009astrocytic,keynes1951ionic}.  The osmotic difference is generated due to the sodium and potassium conductance difference in the glial membrane. In other words, more potassium leaks into the glial compartment, and less sodium leaks out.  As a result of this glial transmembrane water flow in the stimulated region,  it forms a water circulation in the radial direction between the stimulated region and the non-stimulated region.

Our estimation of the velocity scales in the glial compartment and extracellular space shows that this water flow has a considerable effect on potassium flux in the glial compartment. By comparing the full model (including water) with the electrodiffusion model (exclude water),  we validate that water circulation through the glial pathway helps clearance of potassium in the extracellular space and enhance the glial buffering effect.  With additional numerical simulations, we show that the repetitive activity of the nerve fibers further increases the importance of water flow, and the water flow contribution to glia buffering, which is likely to dramatically dominate pathological situations of repetitive activity.

Besides, through our analysis,  we show that the electrical syncytium property of the glial cells is critical for clearing potassium (from the extracellular space) when the neuron fires. Based on the governing equation of glial electric potential, we explain why the inward glial transmembrane potassium flux in the stimulated region is almost the same as the outward potassium flux out to the extracellular space in the non-stimulated region when axon firing. This is because the electric potential spreads through the connected cells in the glial compartment. The glial electric potential in the non-stimulated region becomes more positive in response to the depolarization of the glial electric potential in the stimulated region. This electric property for the glial compartment  is always exist as long as there exists two distinguish stimulated region and non-stimulated region. The glial wrap the axon like a faster potassium transporter, which quickly remove the extra potassium (in the extracellular space) from the stimulated region to the non-stimulated region.   

Finally, we’d like to point out that the coupling of ionic and water flows is not unique to optic nerve. It is ubiquitous in many parts of the mammalian body and other biological tissues. Our analysis of the model for the optic nerve is just a first small step towards the understanding of the mechanisms of various transport processes and the consequences of a disrupted process under pathological conditions.

\section*{Author Contributions}
Y.Z., S.X., and H.H. did the model derivations and carried out the numerical simulations. R.S.E. and H.H. designed the study, coordinated the study, and commented on the manuscript. All authors gave final approval for publication.

\section* {Acknowledgments}
	This research is supported in part by National Natural Science Foundation of China 12071190 (S.X), the Fields Institute for Research in mathematical Science (S.X., R.S.E., and H.H.) and the Natural Sciences and Engineering Research Council of Canada (H.H.). Authors also would like to thank 	 anonymous	 reviewers	 for their valuable suggestions on model calibration $\&$ validation.

\appendix
\section{Notations}

\begin{table}[h]
	\begin{tabular}{l}
		$c_{l}^{i}$: Ion $i$ concentration in the $l$ region,\\
		$\phi_{l}$: Electric potential in $l$ region,\\
		$p_{l}$: Hydrostatic pressure in $l$ region,\\
		$\mathbf{u}_{l}$: Fuild velocity inside of the $l$ region,\\
		$\eta_{l}$: Volume fraction of $l$ region,\\
		$O_{l}$: Osmotic concentration in $l$ region,\\
		$\mathcal{M}_{k}$: Membrane area $k$ in per unit control volume,\\
		$\kappa_{l}$: Water permeability of $l$ region,\\
		$L_{k}^{m}$: Membrane hydrostatic permeability of $k$ membrane,\\
		$\mu$: Fluid viscosity.\\
		$K_k$: Stiffness constant of $k$ membrane,\\
		$\tau_{l}$: Tortuosity of $l$ region,\\
		$z^i$: Valence of the ion $i$,\\
		$A_{l}$: Negative charged protein density in $l$ region,\\
		$J_{p,k}^i$: Active ATP based ion $i$ pump on $k$ membrane,\\
		$J_{c,k}^i$: Passive transmembrane source of $k$ membrane,\\
		$g_{k}^{i}$: Conductance of k membrane for ion $i$,\\
		$\bar{g}^i$: Maximum conductance of axon membrane for ion $i$,\\
		$g_{leak}^i$: Leak conductance of axon membrane for ion $i$,
	\end{tabular}
\end{table}

\section{Comparison between membrane potential and Nernst potential on axon membrane}\label{nerst_axon}

The classical Hodgkin Huxley analysis of a single action potential \cite{chiu1979quantitative}  assumes that changes in concentration of ions are much less important than current flow in determining the shape of the action potential. In other words, the change in the Nernst (i.e., equilibrium) potential is much less than the change in the membrane potential.  In this section,  we show that  the variation of the Nernst potential for $\mathrm{Na^+}$, $\mathrm{K^+}$ and $\mathrm{Cl^-}$ on the axon membrane is much smaller than the axon membrane potential changes during action potentials,
\begin{equation*}
\Delta E_{ax}^{i}= o \left(\Delta V^{*}_{ax}\right), \quad i=\mathrm{Na^{+}}, \mathrm{K^{+}}, \mathrm{Cl^{-}}.
\end{equation*}
During action potentials,the scale of the  $\Delta V_{\mathrm{ax}}$  can be approximated by the $\mathrm{Na}^{+}$ and $\mathrm{K}^{+}$ Nernst potential difference at the resting state, 
\begin{equation}
\label{axon_membrane_approx}
\Delta V^{*}_{\mathrm{ax}} = O\left(E^{Na,re}_{ax} - E^{K,re}_{ax} \right).
\end{equation}
We take the $\mathrm{Cl^-}$ Nernst potential for example. By the charge neutrality condition in  Eq. (\ref{Charge_nuetrality}), we have 
\begin{equation}
\label{approx_Cl}
\Delta c_{ax}^{Cl} \approx -\frac{\eta_{ex}}{\eta_{ax}} \Delta c_{ex}^{Cl}.
\end{equation}
Therefore, the variation of $\mathrm{Cl^{-}}$ Nernst potential on axon membrane yields
\begin{equation}
\label{Cl_nernst}
\begin{aligned}
\Delta E_{ax}^{Cl}&=V^{*}\left(\log \left(\frac{c_{ex}^{Cl, re}+\Delta c_{ex}^{Cl}}{c_{ax}^{Cl, re}+\Delta c_{ax}^{C l}}\right)-\log \left(\frac{c_{ex}^{Cl, re}}{c_{ax}^{Cl, re}}\right)\right) \\
&\approx V^{*}\left(\log \left(1+\frac{\Delta c_{ex}^{Cl}}{c_{ex}^{Cl, re}}\right)-\log \left(1-\frac{\eta_{ex} \Delta c_{ex}^{Cl}}{\eta_{ax} c_{ax}^{Cl,re}}\right)\right),
\end{aligned}
\end{equation}
where
\begin{equation*}
V^{*}=\frac{k_BT}{e}, \quad \frac{1}{c_{ex}^{Cl,re}} = O\left( 10^{-2} \right), \quad  \frac{\eta_{ex}}{\eta_{ax} c_{ax}^{Cl,re}} = O\left(10^{-2}\right).
\end{equation*} 
In addition, the characteristic time for a single action potential  $T_{ax}^{*}$ is in millisecond level $(O\left( 10^{-3} \right))$, so the scale of $\Delta c_{ex}^{Cl}$ in the stimulated region is
\begin{equation}
\label{Cl_variation}
\Delta c_{ex}^{Cl,*} = \Delta c_{ex}^{Na,*}+ \Delta c_{ex}^{K,*}   < O\left( \frac{T_{ax}^{*} \mathcal{M}_{ax} \bar{g}^{Na} \Delta V^{*}_{ax}}{e \eta_{ex}} \right) = O(1),
\end{equation} 
where we use charge neutrality condition and maximum conductance of the voltage-gated $\mathrm{Na}^{+}$ channel. Therefore,  Eq. (\ref{Cl_nernst}) yields
\begin{equation}
\label{Cl_nernst2}
\Delta E_{ax}^{Cl}\approx V^{*}\left(\frac{1}{c_{ex}^{Cl, re}}+\frac{\eta_{e x}}{\eta_{ax} c_{ax}^{Cl, re}}\right) \Delta c_{ex}^{Cl},
\end{equation}
Based on Eqs. (\ref{axon_membrane_approx}), (\ref{Cl_nernst2}) and (\ref{Cl_variation}), and the fact that $\frac{V^{*}}{\Delta V^*_{ax}}=o(1)$, we have $\Delta E_{ax}^{Cl} = o \left(\Delta V^{*}_{ax}\right)$. In a similar way, we can get 
\begin{equation}
\Delta E_{ax}^{i} = o \left(\Delta V^{*}_{ax}\right), \quad i=\mathrm{Na^{+},K^{+}}.
\end{equation}

\section{Estimations of $t_{m1}$ and $t_{m2}$}\label{robust} 
In this section, we provide estimations on $t_{m1}$ and $t_{m2}$. For the first time interval parameter  $t_{m1}$,
by substituting  Eq. (\ref{alpha_m}),  Eq. (\ref{V_ax_app}) into  Eq. (\ref{m_1_solution}), we obtain

	\begin{equation}
	\label{m_eq_1}
	\begin{aligned}
	m^{dy}(t_{m1})=& m_{0}\exp\left(\frac{18t_{m1}}{35}\left( \exp \left(\frac{-70}{9}\right) -1 \right) + \frac{t_{m1}}{14} \left[ \mathrm{Li}_{2} \left(\exp(x) \right) + x\ln \left(1 -\exp(x) \right)-\frac{1}{2}x^{2}  \right] \bigg{\vert}_{2.5}^{-11.5}  \right) -\frac{t_{m1}}{14} \int_{2.5}^{-11.5} \frac{s}{\exp(s)-1}  \\
	&\exp\left( \frac{18t_{m1}}{35}\left( \exp\left(-\frac{70}{9}\right)-\exp\left(-\frac{25-10s}{18}\right)
	\right) +\frac{t_{m1}}{14} \left[ \mathrm{Li}_{2}(\exp(x)) + x\ln( 1-\exp(x)) -\frac{1}{2}x^{2}  \right] \bigg{\vert}_{s}^{-11.5}  \right)ds,
	\end{aligned}
	\end{equation}

Based on  Eq. (\ref{m_eq_1}), we present the estimations of $t_{m1}$ by choosing different open probabilities value for $m^{dy}(t_{m1})$ in Table \ref{Table_1} below.
\begin{table}[!htbp]
	\centering
	\caption{\label{Table_1} Estimation of $t_{m1}$ }
		\begin{tabular}{cccc}
			$m^{d y}\left(t_{m 1}\right)$ & 0.93 & 0.95 & 0.97 \\
			\hline$t_{m 1}$ & $0.57 \mathrm{~ms}$ & $0.67 \mathrm{~ms}$ & $0.92 \mathrm{~ms}$ \\
		\end{tabular}
\end{table}
Table \ref{Table_1}  shows that the estimation of $t_{m1}$ through  Eq. (\ref{m_eq_1}) has consistent results. In the similar way, for the second time interval parameter $t_{m2}$, by substituting  Eq. (\ref{alpha_m}),  Eq.  (\ref{V_ax_app2}) into  Eq. (\ref{m_1_solution}), we obtain

	\begin{equation}
	\begin{tiny}
	\label{m_eq_2}
	\begin{aligned}
	&m^{dy}(t_{m2}) =m_{0}\exp\left( \frac{36t_{m2}}{75}  \left( \exp\left(\frac{-70}{9}\right)-\exp\left(\frac{5}{9}\right)\right)  + \frac{t_{m2}}{15} \left[  \mathrm{Li}_{2} ( \exp(x))+x\ln(1-\exp(x))-  \frac{1}{2} x^{2}  \right]  \bigg{\vert}_{3.5}^{-11.5}  \right) +\frac{t_{m2}}{15}\\
	&\int_{-11.5}^{3.5} \frac{s}{\exp(s)-1} \exp\left(\frac{36t_{m2}}{75}\left(\exp\left(\frac{-(35-10s)}{18}\right) -\exp\left(\frac{5}{9}\right)  \right)  +\frac{t_{m2}}{15}\left[ \mathrm{Li}_{2}(\exp(x) )+x\ln(1-\exp(x))-\frac{1}{2}x^{2} \right] \bigg{\vert}_{3.5}^{s}  \right) ds.
	\end{aligned}
	\end{tiny}
	\end{equation}

In the second time interval, we choose $m^{dy}(t_{m1})=0.95$ as the initial value $m_{0}$ in  Eq. (\ref{m_eq_2}). Table \ref{Table_2} shows  consistent estimation of the $t_{m2}$ when different value for $m^{dy}(t_{m2})$  has been chosen.    
\begin{table}[!htbp]
	\centering
	\caption{\label{Table_2} Estimation of $t_{m 2}$}
		\begin{tabular}{cccc}
			$m^{d y}\left(t_{m 2}\right)$ & 0.15 & 0.1 & 0.05 \\
			\hline$t_{m 2}$ & $2.44 \mathrm{~ms}$ & $3.00 \mathrm{~ms}$ & $4.01 \mathrm{~ms}$
		\end{tabular}
\end{table}

In sum, based on the results in Table \ref{Table_1}-\ref{Table_2}, we confirm that by using  Eq. (\ref{m_eq_1}) and  Eq. (\ref{m_eq_2}) to estimate the time parameter $t_{m1}$ and $t_{m2}$ for $\Delta V_{ax}$ have robust results.

\section{Estimation of transmembrane currents}\label{pump_increment}  

After the axon stop firing,  we assume that voltage-gated $\mathrm{Na}^{+}$ and $\mathrm{K}^{+}$  channel's conductance on axon membrane have returned to their resting state in the stimulated region, 
\begin{equation*}
g_{ax}^{i,dy}\approx g_{ax}^{i,re}, \ i=\mathrm{Na^+,K^+}.
\end{equation*}
At this stage, we have ion channel conductance on the glial and axon membrane as
\begin{equation}
\label{conductance_1}
\{g_{ax}^{Na,re},\ g_{ax}^{K,re},\ g_{ax}^{Cl},\ g_{gl}^{Cl}, \ g_{gl}^{Na}\}  \subset o \left( g_{gl}^{K} \right).
\end{equation}
Similar to  Eq. (\ref{Nernst_K_variation}), we claim in the stimulated region
\begin{equation}
\label{Nernst_1}
\Delta E_{k}^{i} = o\left( \Delta E_{gl}^{K} \right), \  i=\mathrm{Na^+,Cl^-}, \  k=gl,ax,
\end{equation}
since   Eq. (\ref{order_C}) and
\begin{equation*}
c_{ex}^{K,re}= o\left(c_{ex}^{i,re} \right), \quad i=\mathrm{Na}^{+},\mathrm{Cl}^{-}.
\end{equation*}
In addition, for the increase current through $\mathrm{Na/K}$ pump  in   Eq. (\ref{phi_ex_governing}), we have
\begin{equation*}
z^{Na}e\Delta J_{p,k}^{Na}+ z^{K}e\Delta J_{p,k}^{K}=\Delta I_{k},    \  k=gl,ax.
\end{equation*}
By the Taylor expansion, we approximate the increase current through the $\mathrm{Na/K}$ pump due to the extracellular $\mathrm{K^+}$ concentration changes as
\begin{equation}
\label{delta_I}
\Delta I_{k} \approx 2\left( \frac{K_{K1}I_{k}^{re,1}}{c_{ex}^{K,re}   (c_{ex}^{K,re}+K_{K1}) }+ \frac{K_{K2} I_{k}^{re,2}}{c_{ex}^{K,re} (c_{ex}^{K,re}+K_{K2}) }\right)\Delta c_{ex}^{K},      
\end{equation}
where $I_{k}^{re,1}$ and $I_{k}^{re,2}$ are the resting state current through $\alpha_{1}-$ and $\alpha_{2}-$ isoform of the Na/K pump on glial membrane $(k=gl)$ or axon membrane $(k=ax)$.\\
By comparison between   Eq. (\ref{Nernst_K_variation})  and   Eq. (\ref{delta_I}), we have 
\begin{equation}
\label{pump_1}
\Delta I_{k}= o\left( g_{gl}^{K} \Delta E_{gl}^{K}\right), \quad  \ k=gl,ax.
\end{equation}
In all, based on the estimations in  Eqs. (\ref{conductance_1}), (\ref{Nernst_1}) and (\ref{pump_1}), we claim the dominated term in the right-hand side of   Eq. (\ref{phi_ex_governing}) is 
\begin{eqnarray*}
	&\sum_{i} z^{i}e \mathcal{M}_{gl} \left( J_{p,gl}^{i}+ J_{c,gl}^{i} \right)+\sum_{i}z^{i}e \mathcal{M}_{ax} \left( J_{p,ax}^{i}+ J_{c,ax}^{i}\right) \\
	&\approx \mathcal{M}_{gl} g_{gl}^{K} \left(\Delta V_{gl} -\Delta E_{gl}^{K} \right),
\end{eqnarray*}
where we use the fact that at the resting state, the transmembrane currents in both axon membrane and glial membrane are negligible in compare to the source term $g_{gl}^{K}\Delta E_{gl}^{K}$.

\section{Comparison between $\Delta \phi_{gl}$ and $\Delta \phi_{ex}$}
\label{phi_ex_and_phi_gl} 
In this section, we show that the  scale of the glial electric potential variation $\Delta \phi_{gl}$ is much larger than the scale of the extracellular electric variation $\Delta \phi_{ex}$ in the stimulated region. Based on   Eq. (\ref{relation_phi}),  we know 
\begin{equation}
\label{order1}
O\left( \frac{\eta_{gl} \sigma_{gl} }{\eta_{ex}\sigma_{ex} }\right)= 10^{-2}, \quad \   O\left(\frac{ \tau_{ex}eD_{ex}^{\mathrm{diff}} }{\sigma_{ex} } \Delta c_{sti}\right) = 10^{-6}.
\end{equation}
If the $\Delta \phi_{ex} \neq o(\Delta \phi_{gl}) $, then based on  Eqs. (\ref{relation_phi})  and (\ref{order1}), we should have 
\begin{equation*}
O\left(\Delta \phi_{gl} \right) < 10^{-5}.
\end{equation*}
Therefore, the right-hand side of  Eq. (\ref{glia_phi_approx}) becomes
\begin{equation}
\label{RIGHT}
\left|\frac{g_{gl}^{K}}{e}\left(\Delta V_{gl}-\Delta E_{g l}^{K}\right)\right| \approx\left|\frac{g_{gl}^{K}}{e} \Delta E_{gl}^{K}\right| = O\left( 10^{-8} \right).
\end{equation}
where we use the estimation of $\Delta E_{gl}^{K} \ \left( =O\left(10^{-3}\right)  \right)$ in  Eqs. (\ref{Nernst_K_variation}) and (\ref{C_variation}),  and
\begin{equation*}
O\left(\Delta V_{gl} \right) = O\left(\Delta \phi_{gl} -\Delta \phi_{ex} \right) < 10^{-5}.
\end{equation*}
At the same time,  the left-hand side of   Eq. (\ref{glia_phi_approx}) gives
\begin{equation}
\label{LEFT}
\left|\frac{2}{r_{sti}} \frac{\eta_{gl} \sigma_{gl} }{\mathcal{M}_{gl}} \frac{\Delta \phi_{gl}}{r^{*}}\right| < O\left( 10^{-11} \right). 
\end{equation}
In   Eq. (\ref{glia_phi_approx}), based on   Eqs. (\ref{LEFT})  and (\ref{RIGHT}), the order of right-hand side does not match with the order of left-hand side. Therefore, we conclude that 
\begin{equation*}
\Delta \phi_{ex} = o(\Delta \phi_{gl}).
\end{equation*}

\section{Estimation of extracellular $\mathrm{Na^+}$ and $\mathrm{K^+}$ transport}
\label{Estimation_on_transport}
For the $\mathrm{K^+}$ clearance in the stimulated extracellular region in   Eq. (\ref{dynamic_K_ex}),  based on Eqs. (\ref{Nernst_K_variation}) and  (\ref{K_current_eq}),  the effect of average glial transmembrane $\mathrm{K}^{+}$ flux in the stimulated region is
\begin{equation}
\lambda_{gl}^{m,K} = \frac{\mathcal{M}_{gl} g_{gl}^{K} h_{\epsilon} k_{B} T}{z^{K}\left(1+h_{\epsilon}\right) e^{2} c_{ex}^{K,re}}.
\end{equation}
For $\mathrm{K^+}$ flux through the extracellular pathway, we only consider the effects from diffusion and electric drift terms in the radial $\mathrm{K^+}$ flux. The fluid flows in the extracellular space from the non-stimulated region to the stimulated region. So, the convection flux in the extracellular is a consequence of the osmosis and flattens the variation of osmotic pressure in the stimulated region.

The scale of the radial diffusive $\mathrm{K^+}$ flux in the extracellular space can be approximated as
\begin{equation}
\label{K_diff_ex}
O\left(-D_{ex}^{K} \tau_{ex} \frac{d c_{ex}^{K}}{d r}\right) =  \frac{D_{ex}^{K} \tau_{ex}}{r^{*}} \Delta c_{ex}^{K}.
\end{equation}
The scale of the radial electric drift  $\mathrm{K^+}$ flux in the extracellular space is
\begin{equation}
\begin{aligned}
\label{K_electric_ex}
O\left(-\frac{D_{ex}^{K} \tau_{ex} e}{k_{B} T} c_{ex}^{K} \frac{d \phi_{ex}}{d r}\right) & = \frac{D_{ex}^{K} \tau_{ex} e}{k_{B} T} c_{ex}^{K} \frac{\Delta \phi_{ex}}{r^{*}} \\
& \approx- \frac{\eta_{gl}\sigma_{gl} D_{e x}^{K} \tau_{ex} }{\eta_{ex}  \sigma_{ex}\left(1+h_{\epsilon}\right)r^{*}}  \Delta c_{ex}^{K},
\end{aligned}
\end{equation}
where  $\Delta \phi_{ex}$ used the estimation from   Eq. (\ref{phi_ex_scale}).\\
Based on   Eqs. (\ref{K_diff_ex}) and  (\ref{K_electric_ex}),  we note that the electric drift   $\mathrm{K^+}$ flux is in the opposite radial direction to the diffusive  $\mathrm{K^+}$ flux in the extracellular space. At the same time, the electric drift $\mathrm{K^+}$ flux  has a much smaller magnitude than the diffusive $\mathrm{K^+}$ flux because the  ratio $R_{ex}^{K}$  between the electric drift and  diffusion terms is 
\begin{equation}
R_{ex}^{K}=\frac{\eta_{gl}\sigma_{gl} }{\eta_{ex} \sigma_{ex} (1+h_{\epsilon} )}=o(1). 
\end{equation}
Therefore, in   Eq. (\ref{dynamic_K_ex}), the average  effect of the $\mathrm{K^+}$ transport through extracellular pathway can be approximated as 
\begin{equation}
\lambda_{ex}^{K}=\frac{2\eta_{ex} D_{ex}^{K} \tau_{ex} }{ r_{sti} r^{*} },                                                   
\end{equation}
where we used the ratio between volume $V_{S}$ and the effective radial surface.

In   Eq. (\ref{Na_dynimic}), we first look for the effect of  $\mathrm{Na}^+$ fluxes through the extracellular pathway.  
Similar to   Eq. (\ref{K_diff_ex}),  the scale of the radial diffusive $\mathrm{Na^+}$ flux  in the extracellular space is 
\begin{equation}
\label{Na_diff_approx}
O\left(-D_{ex}^{Na} \tau_{ex} \frac{d c_{ex}^{Na}}{dr}\right) = \frac{D_{ex}^{Na} \tau_{ex}}{r^{*}} \Delta c_{ex}^{Na} .
\end{equation}
The scale of the radial electric drift flux for $\mathrm{Na^+}$ in in the extracellular space is
\begin{equation}
\label{Na_eletric_approx}
\begin{aligned}
O\left(-\frac{D_{ex}^{N a} \tau_{ex} e}{k_{B} T} c_{ex}^{Na} \frac{d \phi_{ex}}{d r}\right) & = \frac{D_{ex}^{Na} \tau_{ex} e}{k_{B} T} c_{ex}^{Na} \frac{\Delta \phi_{ex}}{r^{*}} \\
& \approx- \frac{\eta_{gl}  \sigma_{gl} D_{ex}^{Na} \tau_{ex} }{\eta_{ex} \sigma_{e x}\left(1+h_{\epsilon}\right)r^{*}} \frac{c_{ex}^{Na}}{c_{ex}^{K}}\Delta c_{ex}^{K}
\end{aligned}
\end{equation}
For $\mathrm{Na^+}$ in the extracellular space, the radial electric drift $\mathrm{Na^+}$ flux is in the same direction as the radial diffusive $\mathrm{K^+}$ flux since $\Delta c_{ex}^{Na}$ is negative in the stimulated region.\\
The scale of the radial diffusive $ \mathrm{Na^+}$ flux  is at same level as the radial electric drift $ \mathrm{Na^+}$ flux in the extracellular space. From   Eqs. (\ref{Na_diff_approx}) and (\ref{Na_eletric_approx}), the  ratio $R_{ex}^{Na}$ is 
\begin{equation}
R_{ex}^{Na}=\frac{\eta_{gl}  \sigma_{gl}}{\eta_{e x} \sigma_{e x}\left(1+h_{\epsilon}\right)} \frac{c_{ex}^{Na}}{c_{ex}^{K}}=O(1),
\end{equation}
since $\Delta c_{ex}^{Na}$ and $\Delta c_{ex}^{K}$ is at the same leading order. 
The $\mathrm{Na^+}$ flux through glial transmembrane   is much smaller than the $\mathrm{K^+}$ flux such that
\begin{equation}
\label{lambd_m_Na}
\lambda_{gl}^{m,Na}=o\left(\lambda_{gl}^{m,K} \right). 
\end{equation}
This is because  the conductance on the glial membrane $g_{gl}^{Na} = o\left( g_{gl}^{K}\right)$.  The effect of  $\mathrm{Na^{+}}$  flux through glial transmembrane can be neglected in   Eq. (\ref{Na_dynimic}), since    Eq. (\ref{lambd_m_Na}),  and  the diffusive fluxes in  Eqs. (\ref{Na_diff_approx}) and  (\ref{K_diff_ex}) are in the same magnitude.
In sum, for  Eq. (\ref{Na_dynimic}), we get
\begin{equation*}
\lambda_{ex}^{Na,1}=\frac{2 \eta_{ex}D_{ex}^{Na} \tau_{ex} }{r_{sti}r^{*} }, \quad \lambda_{ex}^{Na,2}=\frac{2 \eta_{gl}\sigma_{gl} D_{ex}^{Na}\tau_{ex}  c_{ex}^{Na, re} }{ r_{sti}\sigma_{ex}\left(1+h_{\epsilon}\right)  r^{*}c_{ex}^{K,re}}.
\end{equation*}
where we used the ratio between volume $V_{S}$ and the effective radial surface.

In the end of this section, we consider the solution for the coupled dynamical system of  (\ref{dynamic_K_ex}) and  (\ref{Na_dynimic})
\begin{equation}
\label{na_k_system}
\frac{d}{dt}
\left(
\begin{aligned}
&\Delta c_{ex}^{K} \\
&\Delta c_{ex}^{Na}
\end{aligned}
\right)
=A
\left(
\begin{aligned}
&\Delta c_{ex}^{K} \\
&\Delta c_{ex}^{Na}
\end{aligned}
\right),
\end{equation}
where
\begin{equation}
A=\left[
\begin{array}{cc}
A_{11} & 0 \\
A_{21} & A_{22}
\end{array}
\right]
=\left
[\begin{array}{cc}
-\left(\lambda_{gl}^{m, K}+\lambda_{ex}^{K}\right) / \eta_{e x}^{re} & 0 \\
\lambda_{e x}^{Na, 2} / \eta_{ex}^{re} & -\lambda_{ex}^{Na, 1} / \eta_{ex}^{re}
\end{array}
\right].
\end{equation}
In the system (\ref{na_k_system}), we assume that  $\eta_{ex}$ keeps at its resting state $(\eta_{ex}^{re})$ and the initial condition is
\begin{equation}
\left(
\begin{aligned}
\Delta c_{ex}^{K,0} \\
\Delta c_{ex}^{Na,0}
\end{aligned}
\right)
=\left(
\begin{aligned}
\Delta c_{sti} \\
-\Delta c_{sti}
\end{aligned}
\right).
\end{equation}
The solution for System (\ref{na_k_system}) in the time interval $t\in [0,T]$ is
\begin{equation}
\label{solution_T}
\left\{
\begin{aligned}
\Delta c_{ex}^{K}(t)=&\Delta c_{sti} \exp \left(A_{11} t\right), \\
\Delta c_{ex}^{Na}(t)=&\frac{A_{21} \Delta c_{sti}}{A_{11}-A_{22}}\left(\exp \left(A_{11} t\right)-\exp \left(A_{22} t\right)\right)\\
&-\Delta c_{sti} \exp \left(A_{22} t \right),
\end{aligned}
\right.
\end{equation}
where $T$ is the time interval between each single action potential in the axon compartment.
There are $n \ (=\frac{T_{sti}}{f_{m}})$ stimuli in the time interval $[0,T_{sti}=nT]$, we have 
\begin{eqnarray*}
	\Delta c_{ex}^{K}(iT)=\Delta c_{ex}^{K} (iT)+\Delta c_{sti}, \ \  \Delta c_{ex}^{Na}(iT)=&\Delta c_{ex}^{Na} (iT)-\Delta c_{sti}. \\ 
	&i=1\dots n-1,
\end{eqnarray*}
In the above, we view the extracellular $\mathrm{K^{+}}$ and  $\mathrm{Na^{+}}$ concentration immediately changes due to axon firing.
By using   Eq. (\ref{solution_T}), we have 
\begin{equation}
\label{K_solution}
\Delta c_{ex}^{K}(nT)=\Delta c_{sti} \frac{\exp \left(A_{11} T\right)-\exp \left((n+1) A_{11} T\right)}{1-\exp \left(A_{11} T\right)},
\end{equation}
and 
\begin{eqnarray}
\label{Na_solution}
\Delta c_{ex}^{Na}(nT)=&&
\sum_{i=1}^{n} \frac{A_{21} \Delta c_{ex}^{K}( (i-1) T)}{4}\left( \exp \left(A_{11} T\right)-\exp \left(A_{22} T\right) \right) \nonumber \\
&&\exp\left((n-i) A_{22} T\right)-\Delta c_{sti} \sum_{i=1}^{n} \exp \left(i A_{22} T\right),
\end{eqnarray}
where 
\begin{equation*}
\Delta c_{ex}^{K}(jT)=\Delta c_{sti} \frac{1-\exp \left((j+1) A_{11} T\right)}{1-\exp \left(A_{11} T\right)}, \ \  j=0,1, \ldots n-1.
\end{equation*}

\section{Spatial Distribution of velocity and osmotic pressure}
\begin{figure}[htp]
	\centering
	\includegraphics[width=3.25in,height=3cm]{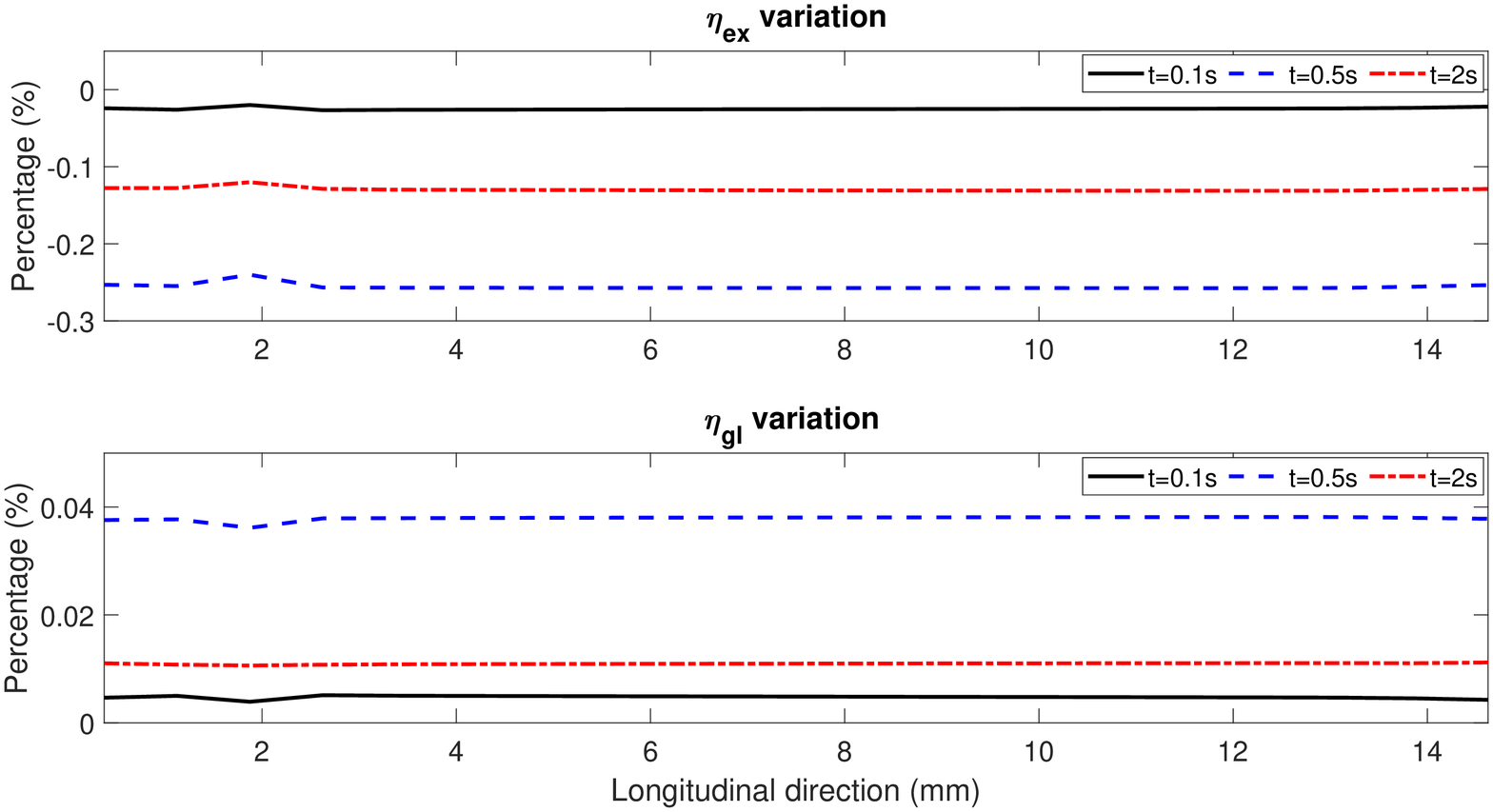}
	\caption{\textcolor{black}{Longitudinal direction changes of $\eta_{ex}$ and $\eta_{gl}$ at $r=1.5 \mathrm{\mu m}$ at $t=0.1 \mathrm{s},0.5 \mathrm{s},2 \mathrm{s}$.}}
	\label{fig:z_direction_eta}
\end{figure}

\begin{figure}[htp]
	\centering
	\includegraphics[width=3.25in,height=4cm]{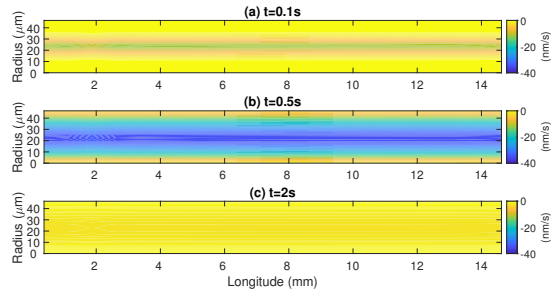}
	\caption{\textcolor{black}{Spatial distribution of velocity in radius direction during and after a train of stimuli.}}
	\label{fig: velocityr}
\end{figure}

\begin{figure}[htp]
	\centering
	\includegraphics[width=3.25in,height=3cm]{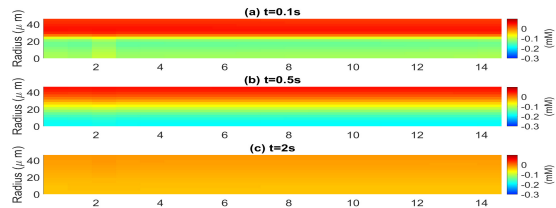}
	\caption{\textcolor{black}{Spatial distribution of osmotic pressure changes from resting state during and after a train of stimuli.}}
	\label{fig:ox}
\end{figure}

\begin{table}
	\begin{scriptsize}
		\caption{Parameters in Optic Nerve Model}
			\begin{tabular}{cccc}
				Parameters  &  Value   & Parameters  &  Value  
				\\
				\hline
				$R_{a}$ & $4.8 \times 10^{-5} \mathrm{~m}$ \ (Ref.\cite{kuffler1966physiological,bracho1975further}) & $\mu$ & $7 \times 10^{-4} \mathrm{~Pa} \cdot \mathrm{s}$ \ (Ref.\cite{mathias1985steady})\\
				\hline 
				$R_{b}$	 & $6\times 10^{-5}  \mathrm{m} $ \ (Ref.\cite{wang2019intraocular}) &	$c_{csf,eye}^{Na}$ &	$111 \ \mathrm{mM}$\ (Ref.\cite{kuffler1966physiological}) \\
				\hline  
				$L$ & $1.5 \times 10^{-2} \mathrm{~m}$ (Ref.\cite{kuffler1966physiological}) & $c_{\text {csf,eye }}^{\text {K }}$ & $3 \ \mathrm{mM}$\ (Ref.\cite{kuffler1966physiological})\\
				\hline 
				$e$ & $1.69 \times 10^{-19} \mathrm{~A} \cdot \mathrm{s}$ & $c_{gl}^{\text {Na,re }}$ & $7.57\ \mathrm{mM}$ (*) \\
				\hline
				$k_{B}$ & $1.38 \times 10^{-23} \mathrm{~J} / \mathrm{K}$ &  $c_{gl}^{K, re}$ & $100.84\  \mathrm{mM}$ (*,Ref.\cite{kuffler1966physiological}) \\
				\hline
				$T $& $296.15 \mathrm{~K}$ (Ref.\cite{kuffler1966physiological}) & $c_{ax}^{\text {Na,re}}$ & $10.17\  \mathrm{mM}$ (*) \\
				\hline
				$\eta_{ax}^{re}$ & $5 \times 10^{-1}$  (Ref.\cite{kuffler1966physiological}) & $c_{ax}^{K, re}$ & $100.04 \  \mathrm{mM}$ (*)\\
				\hline 
				$\eta_{gl}^{re}$ & $4 \times 10^{-1}$ \ (Ref.\cite{kuffler1966physiological}) & $A^{re}_{ax, gl}$ & $105 \ \mathrm{mM}$ (*)\\
				\hline
				$\eta_{ex}^{re}$ & $1 \times 10^{-1}$ (Ref.\cite{kuffler1966physiological}) &  $\tau_{ex}^{OP}$ & $0.16$ \ (Ref.\cite{mathias1985steady,malcolm2006computational})  \\
				\hline 
				$\mathcal{M}_{ax}$ & $5.9 \times 10^{6} \mathrm{~m}^{-1}$ \ (Ref.\cite{pilgrim1982volume}) & $\tau_{ex}^{SAS}$ & $1$ (*)  \\
				\hline 
				$\mathcal{M}_{gl}$ & $1.25 \times 10^{7} \mathrm{~m}^{-1}$\  (Ref.\cite{pilgrim1982volume}) & $\tau_{gl}$ & $0.5$ (*)  \\
				\hline
				$z^{Na, K}$ & $1$ & $p_{CSF}$ & $1.3 \times 10^{3} \mathrm{~Pa}$ \ (Ref.\cite{band2009intracellular}) \\
				\hline 
				$z^{Cl}$ & $-1$ & $p_{ICP}$ & $4 \times 10^{3} \mathrm{~Pa}$ \ (Ref.\cite{band2009intracellular}) \\
				\hline
				$z^{ax, gl}$ & $-1$\ (*) & $p_{OBP}$ & $0 \mathrm{~Pa}$\  \ (Ref.\cite{band2009intracellular}) \\
				\hline

				$ \gamma_{\text {ax,gl}}$ & $1$ \ (Ref.\cite{mathias1985steady,malcolm2006computational})&$ D_{ex,ax}^{Na}$ & $1.39 \times 10^{-9} \mathrm{~m}^{2} / \mathrm{s}$ (Ref.\cite{mathias1985steady})\\
				\hline
				$\gamma_{pia}$ & $1$ (Ref.\cite{mathias1985steady,malcolm2006computational}) & $D_{ex, ax}^{K}$ & $2.04 \times 10^{-9} \mathrm{~m}^{2} / \mathrm{s} $ \ (Ref.\cite{mathias1985steady}) \\
				\hline 
				$K_{\text {Na1,Na2}}$ & $2.3393 \mathrm{mM}$ (Ref.\cite{zhu2019bidomain})& $D_{ex,ax}^{Cl}$ & $2.12 \times 10^{-9} \mathrm{~m}^{2} / \mathrm{s}$ (Ref.\cite{mathias1985steady}) \\
				\hline
				$K_{K1}$ & $1.6154 \mathrm{mM}$ (Ref.\cite{zhu2019bidomain}) & $D_{gl}^{N a}$ & $1.39 \times 10^{-11} \mathrm{~m}^{2} / \mathrm{s}$ (Ref.\cite{mathias1985steady}) \\
				\hline 
				$K_{K 2}$ & $0.1657 \mathrm{mM}$ (Ref.\cite{zhu2019bidomain})& $D_{gl}^{K}$ & $2.04 \times 10^{-11} \mathrm{~m}^{2} / \mathrm{s}$ (Ref.\cite{mathias1985steady}) \\
				\hline
				$I_{gl,1}$ & $4.78 \times 10^{-4} \mathrm{~A} / \mathrm{m}^{2}$ (**,Ref.\cite{zhu2019bidomain})  & $D_{gl}^{Cl}$ & $2.12 \times 10^{-11} \mathrm{~m}^{2} / \mathrm{s}$ (Ref.\cite{mathias1985steady}) \\
				\hline 
				$I_{gl, 2}$ & $6.5 \times 10^{-5} \mathrm{~A} / \mathrm{m}^{2}$ (**,Ref.\cite{zhu2019bidomain}) & $k_{ex}^{OP}$  & $1.3729 \times 10^{-8} \mathrm{~m}^{2} / \cdot \mathrm{s}$ (Ref.\cite{malcolm2006computational}) \\
				\hline
				$I_{ax, 1}$ & $9.56 \times 10^{-4} \mathrm{~A} / \mathrm{m}^{2}$ (**,Ref.\cite{zhu2019bidomain})& $k_{ex}^{SAS}$ & $0 \mathrm{~m}^{2} / \mathrm{V} \cdot \mathrm{s}$\ (*) \\
				\hline 
				$I_{ax, 2}$ & $1.3 \times 10^{-4} \mathrm{~A} / \mathrm{m}^{2}$ (**,Ref.\cite{zhu2019bidomain})& $K_{ax}$ & $1.67 \times 10^{6} \mathrm{~Pa}$\ (Ref.\cite{hua2018cerebrospinal,lu2006viscoelastic})  \\
				\hline
				$g_{gl}^{N a}$ & $2.2 \times 10^{-3} \mathrm{~S} / \mathrm{m}^{2}$ (Ref.\cite{mathias1985steady})& $K_{gl}$ & $8.33 \times 10^{5} \mathrm{~Pa}$\ (Ref.\cite{hua2018cerebrospinal,lu2006viscoelastic}) \\
				\hline 
				$g_{gl}^{K}$ & $2.1 \mathrm{~S} / \mathrm{m}^{2}$ (Ref.\cite{mathias1985steady}) & $L_{dr}^{m}$ & $8.89 \times 10^{-13} \mathrm{~m} / \mathrm{Pa} \cdot \mathrm{s}$ (Ref.\cite{malcolm2006computational,zhu2019bidomain}) \\
				\hline 
				$g_{gl}^{Cl}$ & $2.2 \times 10^{-3} \mathrm{~S} / \mathrm{m}^{2}$ (Ref.\cite{mathias1985steady}) & $L_{pia}^{m}$ & $8.89 \times 10^{-13} \mathrm{~m} / \mathrm{Pa} \cdot \mathrm{s}$ (Ref.\cite{malcolm2006computational,zhu2019bidomain})\\
				\hline
				$g_{leak}^{Na}$ & $4.8 \times 10^{-3} \mathrm{~S} / \mathrm{m}^{2}$ (**,Ref.\cite{song2018electroneutral})& $L_{gl}^{m}$ & $1.34 \times 10^{-13} \mathrm{~m} / \mathrm{Pa} \cdot \mathrm{s}$ (Ref.\cite{malcolm2006computational,zhu2019bidomain}) \\
				\hline 
				$g_{leak}^{K}$ & $2.2 \times 10^{-2} \mathrm{~S} / \mathrm{m}^{2}$ (**,Ref.\cite{song2018electroneutral})& $L_{ax}^{m}$ & $7.954 \times 10^{-14} \mathrm{~m} / \mathrm{Pa} \cdot \mathrm{s}$ (Ref.\cite{villegas1960characterization}) \\
				\hline 
				$\bar{g}^{Na}$ & $1.357 \times 10^{1} \mathrm{~S} / \mathrm{m}^{2}$ (**,Ref.\cite{song2018electroneutral})& $\kappa_{g l}$ & $9.366 \times 10^{-19} \mathrm{~m}^{2}$ (Ref.\cite{malcolm2006computational,zhu2019bidomain}) \\
				\hline
				$\bar{g}^{K}$ & $2.945 \mathrm{~S} / \mathrm{m}^{2}$ (**,Ref.\cite{song2018electroneutral})& $\kappa_{ax}$ & $1.33 \times 10^{-16} \mathrm{~m}^{2}$ (Ref.\cite{malcolm2006computational,zhu2019bidomain})  \\
				\hline 
				$g_{ax}^{Cl}$ & $1.5 \times 10^{-1} \mathrm{~S} / \mathrm{m}^{2}$ (*) & $\kappa_{ex}^{OP}$ & $3.99 \times 10^{-16} \mathrm{~m}^{2}$ (**,Ref.\cite{malcolm2006computational,zhu2019bidomain})\\
				\hline 
				$G_{pia}^{Na,K,Cl}$ & $3 \mathrm{~S} / \mathrm{m}^{2}$ (*)& $\kappa_{ex}^{SAS}$ & $1.33 \times 10^{-14} \mathrm{~m}^{2}$ (**,Ref.\cite{malcolm2006computational,zhu2019bidomain}) 
			\end{tabular}
		\begin{tablenotes}
			\item[a] Note: the `*' estimated or induced from the concentration balance.
			\item[b] Note: the `**' deducted proportional from reference.
		\end{tablenotes}
	\end{scriptsize}
\end{table}

 
\bibliographystyle{plain}
\bibliography{Mybib}



\newpage
\appendix

\end{document}